\documentclass[twocolumn, numberedappendix,twocolappendix]{aastex631mod}

\usepackage[encapsulated]{CJK}
\usepackage{graphicx}
\usepackage{bm}
\usepackage{booktabs}
\usepackage{colortbl}
\usepackage{xcolor}
\usepackage{amsmath}
\usepackage{natbib}
\usepackage{microtype}
\usepackage{multirow}

\usepackage[all]{hypcap}
\newcommand{\Planck}{\rm Planck}

\newcommand{\cntext}[1]{\begin{CJK}{UTF8}{gbsn}#1\end{CJK}}

\newcommand{\bs}{\beta_s}
\newcommand{\GHz}{\,\mathrm{GHz}}
\newcommand{\refpte}{Figure \ref{fig:null-pte}} 
\definecolor{citecolor}{rgb}{0.08,0.30,0.85}
\newcommand{\inprep}[1]{{\color{citecolor} #1 in preparation}}
\newcommand{\Linprep}{{\hyperlink{cite.Li23}{L23}}}
\defcitealias{Li23}{L23}

\newcommand{\lcdm}{$\Lambda$CDM }

\newcommand{\jhu}{William H. Miller III Department of Physics and Astronomy, Johns Hopkins University, Baltimore, MD 21218, USA}

\newcommand{\goddard}{NASA Goddard Space Flight Center, 8800 Greenbelt Road, Greenbelt, MD 20771, USA}
\newcommand{\upenn}{Department of Physics and Astronomy, University of Pennsylvania, 209 South 33rd Street, Philadelphia, PA 19104, USA}
\newcommand{\oslo}{Institute of Theoretical Astrophysics, University of Oslo, P.O. Box 1029 Blindern, N-0315 Oslo, Norway}
\newcommand{\cfa}{Center for Astrophysics, Harvard \& Smithsonian, 60 Garden Street, Cambridge, MA 02138, USA}
\newcommand{\nist}{National Institute of Standards and Technology, Boulder, CO 80305, USA}
\newcommand{\ucsc}{Departamento de Ingenier\'{i}a El\'{e}ctrica, Universidad Cat\'{o}lica de la Sant\'{i}sima Concepci\'{o}n, Alonso de Ribera 2850, Concepci\'{o}n, Chile}
\newcommand{\pucc}{Centro de Astro-Ingenier\'ia, Facultad de F\'isica, Pontificia Universidad Cat\'olica de Chile, Avenida Vicu\~na Mackenna 4860, 7820436, Chile}
\newcommand{\puci}{Instituto de Astrof\'isica, Facultad de F\'isica, Pontificia Universidad Cat\'olica de Chile, Avenida Vicu\~na Mackenna 4860, 7820436, Chile}
\newcommand{\argonne}{High Energy Physics Division, Argonne National Laboratory, 9700 S. Cass Avenue, Lemont, IL 60439, USA}
\newcommand{\uchicago}{Department of Astronomy and Astrophysics, University of Chicago, 5640 South Ellis Avenue, Chicago, IL 60637, USA}
\newcommand{\cepia}{CePIA, Astronomy Department, Universidad de Concepción, Casilla 160-C, Concepción, Chile}
\newcommand{\MIT}{MIT Kavli Institute, Massachusetts Institute of Technology, 77 Massachusetts Avenue, Cambridge, MA 02139, USA}
\newcommand{\lanl}{Space Remote Sensing and Data Science, Los Alamos National Lab, Los Alamos, NM 87545, USA}

\definecolor{s3_c}{HTML}{8D39C5}
\definecolor{s7_c}{HTML}{55AFA9}
\definecolor{s9_c}{HTML}{D2A741}
\definecolor{ps_c}{HTML}{CA3142}
\definecolor{bpka}{HTML}{ff8c00}
\definecolor{bpq}{HTML}{4169e1}
\definecolor{p143}{HTML}{BA55D3}

\newcommand{\cbox}[1]{\fcolorbox{black}{#1}{\rule{0pt}{0pt}\rule{0pt}{0pt}}}

\graphicspath{{./}{figures/}}

\begin{document}

\title{CLASS Angular Power Spectra and Map-Component Analysis for 40~GHz Observations through 2022}

\author[0000-0001-6976-180X]{Joseph R. Eimer}
\correspondingauthor{Joseph R. Eimer}
\email{jeimer1@jhu.edu}
\affiliation{\jhu}
\author[0000-0002-4820-1122]{Yunyang Li (\cntext{李云炀}\!\!)}
\affiliation{\jhu}

\author{Michael~K. Brewer}\affiliation{\jhu}
\author[0000-0001-7458-6946]{Rui Shi (\cntext{时瑞}\!\!)}\affiliation{\jhu}

\author[0000-0001-7941-9602]{Aamir Ali}\affiliation{\jhu}
\author[0000-0002-8412-630X]{John~W. Appel}\affiliation{\jhu}
\author[0000-0001-8839-7206]{Charles L. Bennett}\affiliation{\jhu}
\author[0000-0003-2682-7498]{Sarah~Marie Bruno}\affiliation{\jhu}
\author[0000-0001-8468-9391]{Ricardo Bustos}\affiliation{\ucsc}
\author[0000-0003-0016-0533]{David T.~Chuss}
\affiliation{Department of Physics, Villanova University, 800 Lancaster Avenue, Villanova, PA 19085, USA}
\author[0000-0002-7271-0525]{Joseph~Cleary}\affiliation{\jhu}
\author[0000-0002-1708-5464]{Sumit Dahal}\affiliation{\goddard}\affiliation{\jhu}
\author[0000-0003-3853-8757]{Rahul Datta}\affiliation{\uchicago}\affiliation{\jhu}
\author[0000-0002-0552-3754]{Jullianna Denes~Couto}\affiliation{\jhu}
\author[0000-0002-3592-5703]{Kevin~L. Denis}\affiliation{\goddard}
\author[0000-0003-3892-1860]{Rolando D\"unner}\affiliation{\puci}\affiliation{\pucc}
\author[0000-0002-4782-3851]{Thomas~Essinger-Hileman}\affiliation{\goddard}
\author{Pedro Flux\'{a}}\affiliation{\pucc}

\author[0000-0002-2781-9302]{Johannes Hubmayer}\affiliation{\nist}
\author[0000-0003-1248-9563]{Kathleen Harrington}\affiliation{\argonne}\affiliation{\uchicago}
\author[0000-0001-7466-0317]{Jeffrey Iuliano}\affiliation{\upenn}\affiliation{\jhu}
\author{John Karakla}\affiliation{\jhu}
\author[0000-0003-4496-6520]{Tobias~A. Marriage}\affiliation{\jhu}
\author[0000-0002-5247-2523]{Carolina N\'{u}\~{n}ez}\affiliation{\jhu}
\author[0000-0002-8224-859X]{Lucas Parker}\affiliation{\lanl}
\author[0000-0002-4436-4215]{Matthew A. Petroff}\affiliation{\cfa}
\author[0000-0001-5704-271X]{Rodrigo A. Reeves}\affiliation{\cepia}
\author[0000-0003-4189-0700]{Karwan Rostem}\affiliation{\goddard}
\author[0000-0003-3487-2811]{Deniz A. N. Valle}\affiliation{\jhu}
\author[0000-0002-5437-6121]{Duncan J. Watts}\affiliation{\oslo}
\author[0000-0003-3017-3474]{Janet L. Weiland}\affiliation{\jhu}
\author[0000-0002-7567-4451]{Edward J. Wollack}\affiliation{\goddard}
\author[0000-0001-5112-2567]{Zhilei Xu (\cntext{徐智磊}\!\!)}\affiliation{\MIT}
\author[0000-0001-6924-9072]{Lingzhen Zeng}\affiliation{\cfa}



\begin{abstract}

Measurement of the largest angular scale ($\ell < 30$) features of the cosmic microwave background (CMB) polarization is a powerful way to constrain the optical depth to reionization and search for the signature of inflation through the detection of primordial $B$-modes. We present an analysis of maps covering 73.6\% of the sky made from the $40\,\mathrm{GHz}$ channel of the Cosmology Large Angular Scale Surveyor (CLASS) from 2016 August to 2022 May. Taking advantage of the measurement stability enabled by front-end polarization modulation and excellent conditions from the Atacama Desert, we show this channel achieves higher sensitivity than the analogous frequencies from satellite measurements in the range $10 < \ell < 100$. Simulations show the CLASS linear (circular) polarization maps have a white noise level of $125 \,(130)\,\mathrm{\mu K\, arcmin}$. We measure the Galaxy-masked $EE$ and $BB$ spectra of diffuse synchrotron radiation and compare to space-based measurements at similar frequencies. In combination with external data, we expand measurements of the spatial variations of the synchrotron spectral energy density (SED) to include new sky regions and measure the diffuse SED in the harmonic domain. We place a new upper limit on a background of circular polarization in the range $5 < \ell < 125$ with the first bin showing $D_\ell < 0.023$ $\mathrm{\mu K^2_{CMB}}$ at 95\% confidence. These results establish a new standard for recovery of the largest-scale CMB polarization from the ground and signal exciting possibilities when the higher sensitivity and higher-frequency CLASS channels are included in the analysis.

\end{abstract}

\keywords{\href{http://astrothesaurus.org/uat/435}{Early Universe (435)}; 
    \href{http://astrothesaurus.org/uat/322}{Cosmic microwave background radiation (322)};  
    \href{http://astrothesaurus.org/uat/1146}{Observational Cosmology (1146)}; 
    \href{http://astrothesaurus.org/uat/1277}{Polarimeters (1277)}; 
    \href{http://astrothesaurus.org/uat/1858}{Astronomy Data Analysis (1858)}}


\section{Introduction}
\label{sec:intro}
Measurements of the cosmic microwave background temperature and polarization angular power spectra \citep[e.g.,][]{hinshaw13, aiola20, planck18VI, spt3g21}, in combination with surveys of large-scale structure \cite[e.g.,][]{ spt19clustercosmology,kids20shear, des22shear, hsc23shear,madhavacheril23},   baryon acoustic oscillations \cite[e.g.,][]{eboss21baocosmology,des22baocosmology}, and cosmic expansion \cite[e.g.,][]{pantheon22cosmology}, have led to a well-constrained cosmological model, which includes a cold dark matter (CDM) component and parameterizes dark energy as a cosmological constant, $\Lambda$. The resultant \lcdm model describes the matter and energy content of the Universe, as well as its age, ionization history, and a statistical description of its initial conditions. 

A hypothesized early period of accelerating expansion, termed inflation, provides an explanation for the nearly flat, homogeneous, and isotropic Universe we observe, as well as a mechanism for producing the nearly scale-invariant spectrum of density perturbations needed to seed large-scale structure within \lcdm cosmology \citep{starobinskii79, guth81, linde82}. 
If such an inflationary period occurred, a background of primordial gravitational waves (tensor modes) is expected to have imprinted an odd-parity, ``$B$-mode,'' component in the CMB polarization, along with even-parity, ``$E$-mode,'' polarization and temperature anisotropy \citep{kamionkowski97, seljak97}.
While anisotropy in the temperature and $E$-mode polarization are expected and observed from density (scalar) perturbations, primordial $B$-mode polarization is only expected from tensor modes. 
The $BB$ angular power spectrum amplitude can be related to the energy scale of the processes driving the inflationary expansion through the tensor-to-scalar ratio, $r$. 
The tightest constraint ($r<0.036$) comes from the BICEP/Keck $BB$ measurement at $\ell>30$ supported by \Planck\ and WMAP data \citep{BK21}. 
Other recent  $B$-mode measurements at $\ell>30$ include \cite{abs18}, \cite{pb20bmode}, and \cite{spider21}. $BB$ measurements on the largest angular scales ($\ell<30$) probe a distinctive CMB polarization enhancement due to Thomson scattering since recombination by the ionized intergalactic medium during and after reionization. The tightest constraint on the tensor-sourced $BB$ spectrum at $\ell<30$ is an upper limit from \Planck\ \citep[$r<0.274$;][]{debelsunce23}. 

Tensor modes from inflation also source low-$\ell$ $E$-mode polarization, but, given current limits on $r$, the tensor contribution to $E$-mode signal is dominated by contributions from scalar perturbations. 
Like the $BB$ spectrum, the $EE$ spectrum at $\ell<30$ is enhanced by Thomson scattering of CMB photons by the ionized intergalactic medium, and therefore contains information regarding the processes by which the first stars converted neutral hydrogen into its current ionized state \citep{hu03reionization, heinrich17reionization, watts20}. 
This so-called ``reionization peak'' has an amplitude proportional to the square of the optical depth to reionization \citep[$\tau^2$;][]{zaldarriaga97}. 
Of the six standard \lcdm parameters, the least well constrained is $\tau$. While other parameters are measured with subpercent precision, the reionization optical depth uncertainty is 10\% \citep{pagano19}. 
Improvement in the measurement precision of $\tau$ will break a critical degeneracy with the amplitude of the initial scalar curvature fluctuations $A_s$. 
Improved knowledge of $\tau$, $A_s$, and upcoming measurements of large-scale structure will provide constraints on the sum of the neutrino masses at a level sufficient to address the fundamental question of the mass eigenstate hierarchy of neutrinos \citep{allison15neutrinos,liu16neutrinos}.

The Cosmology Large Angular Scale Surveyor (CLASS) is a project to measure the largest angular scale CMB polarization from the ground \citep{essinger-hileman14spie,harrington16spie}. 
Located at 5140~m elevation in the Atacama Desert of northern Chile, the CLASS telescopes observe in frequency bands near 40, 90, 150, and 220~$\mathrm{GHz}$ \citep{dahal22}. 
CLASS was designed to address sources of systematic error, such as instrumental polarization and ground pickup, which traditionally limit measurements at the largest angular scales. 
Design elements include front-end polarization modulation \citep{chus12vpm,harrington18spie,harrington21}, a low-spill optical design with entrance pupil at the modulator \citep{eimer12spie,xu20}, and high-efficiency detectors with well-defined frequency/mode acceptance \citep{rostem12spie,appel14spie,appel19}. 
A unique aspect of CLASS is that its polarization modulator measures both linear and circular polarization \citep{padilla20,petroff20}. 
Other projects targeting the largest angular scale microwave polarization include ground-based (GroundBIRD, \citep{groundbird20}; LSPE-strip, \citep{lspe20}; C-BASS \citep{cbass18}; S-PASS \citep{spass19}; and QUIJOTE \citep{quijote23}), balloon-borne (LSPE-SWIPE \citep{lspe20}; and Taurus \citep{taurus20apra}), and space-based (LiteBIRD \citep{litebird22}) efforts.

In this paper, we present polarization maps, angular power spectra, and analyses of Galactic and cosmological components based on 40 $\mathrm{GHz}$\footnote{The moniker ``40 GHz" is adopted as the name of this frequency channel---it is not the band center.} observations from 2016 August to 2022 May. 
In a companion paper, we describe the 40 $\mathrm{GHz}$ instrument, observations, and data reduction to maps \citep[][hereafter \Linprep]{Li23}, and we include an overview of these topics in Sections~\ref{sec:instrument} and \ref{sec:maps}. 
In Section~\ref{sec:spec_est}, we describe the method of angular power spectrum estimation for data and simulations. Internal consistency tests of the maps, comprising a battery of jackknife ``null'' tests, are presented in Section~\ref{sec:internal-consistency}. 
In Section~\ref{sec:compare}, we present comparisons to other data sets, including a final calibration from comparison to the bright synchrotron signal near the Galactic plane established by WMAP. Galactic synchrotron and cosmological component analyses are given in Sections~\ref{sec:galactic} and \ref{sec:cosmo}. We conclude in Section~\ref{sec:conc}.

\section{CLASS overview}
\label{sec:instrument}

\subsection{Instrument Description}
\label{ssec:instrument}

The CLASS telescope array resides in the Atacama Desert of northern Chile and surveys the sky with single-frequency telescopes at 40 and 90~$\mathrm{GHz}$, as well as a dual-frequency 150/220~$\mathrm{GHz}$ telescope. 
With its observing site at latitude $23^\circ$S, large instantaneous field of view (FOV), and ability to perform constant-elevation scans over $720^{\circ}$ in azimuth, the array can efficiently survey 73.6\% of the sky. 
This full area is mapped daily, apart from a Sun avoidance region that changes throughout the year, with boresight angle changed each day to improve polarization angle coverage and check for systematic errors, e.g., due to unwanted ground pickup. The CLASS frequency coverage enables characterization of both the CMB and polarized Galactic emission from synchrotron and dust with sensitivity centered near the polarized foreground minimum. The foreground minimum depends upon the exact sky region considered, and tends to be near 80--90 GHz when large fractions of the sky are considered \citep{planck18IV}. 

The telescopes share a common architecture. Key to enabling the stability required to recover large angular scale features on the sky, the first optical element in each telescope is a variable-delay polarization modulator (VPM). The VPM consists of a wire grid polarizer in front of a moving mirror, which modulates one linear Stokes parameter into circular polarization and vice versa, which provides CLASS sensitivity to both linear and circular polarization \citep{chus12vpm, harrington18spie}. Each VPM has a $60\,\mathrm{cm}$ diameter clear aperture.
The fast (10 Hz) front-end polarization modulation elevates the signal band out of the dominant noise at low frequencies, e.g., $1/f$ noise from bright atmospheric emission partially polarized after the VPM, while mitigating systematic errors associated with polarization induced by other optical elements in the system \citep[][\inprep{Cleary et al.}]{harrington21}.

Following the VPM, the signal is guided via two ambient-temperature mirrors into a cryogenic receiver where cold ($< 4$ K) dielectric lenses focus the signal onto a focal plane array at $\sim 50\,\mathrm{mK}$ \citep{iuliano2018spie}. 
The focal planes use custom profiled, smooth-walled feedhorns \citep{zeng10} to guide the signal onto planar microwave circuits containing band-defining and out-of-band rejection filters before terminating the signal on a pair of superconducting transition edge sensor (TES) bolometers---one TES for each linear polarization \citep{rostem12spie,appel14spie,appel19,dahal18spie,dahal22}. The detectors are read out through a series of cryogenic time-division multiplexing amplifiers with ambient-temperature multichannel electronics (MCE) \citep{reintsema03,battistelli08,nist_tdm_mux13b}.

The telescopes are housed within a comoving ground screen with a flared forebaffle extension opening toward the sky. For most of the survey described in this paper, a thin polypropylene sheet was held taut at the base of the forebaffle extension to environmentally shield the telescope from dust and potentially damaging weather events. Identified as a source of wind-induced systematic error, the closeout film was removed for periods of good weather after 2021 September.

Pointing is achieved through a three-axis mount---boresight over elevation over azimuth. During normal operation, the boresight and elevation pointing are held fixed while the azimuth is scanned. 

This paper uses data from the 40~$\mathrm{GHz}$ telescope observations beginning 2016 August 31, after initial commissioning, and extending to 2022 May 19, when the telescopes were shut down for maintenance and instrument upgrades. We define the period before 2018 February 22 as \emph{Era 1} and after 2018 June 22 as \emph{Era 2}---various instrument changes and upgrades were performed in the intervening time.

The 40~$\mathrm{GHz}$ telescope includes 36 feedhorns (72 TES bolometers) with array average $1.56^\circ$ FWHM beam (using a Gaussian beam approximation). The FOV is $20^\circ$ wide in azimuth and $15^\circ$ wide in elevation at zero boresight \citep{eimer12spie, xu20}. More details can be found in \citetalias{Li23}.

\subsection{Observing Strategy}
\label{sec:obs_strat}
Observations were organized and scheduled on a 24 hr cycle using custom software \citep{petroff20spie}. 
Usually, the daily observing schedule began at midday when the boresight is set to one of seven values, one for each day of the week, from $-45^{\circ}$ to $+45^{\circ}$ relative to vertical in $15^{\circ}$ increments. 
After the cryogenic system stabilized following the boresight adjustment, detector loading was measured by performing a current--voltage (\emph{IV}) sweep, and optimal detector bias values were selected \citep{appel22}. 
The set of data collected after the \emph{IV} measurement until the next \emph{IV} or until the schedule terminated is called a \emph{span}. 
At nighttime, the telescope nominally scanned $\pm 360^{\circ}$ in azimuth at $45^{\circ}$ elevation and constant boresight. Over the course of the scan, the FOV of the telescope would sweep out a swath roughly $15^\circ$--$20^\circ$ in elevation, depending on the boresight angle, centered on the $45^{\circ}$ boresight direction.
The azimuth scanning direction typically reversed at $180^{\circ}$ azimuth (due south) during the night. In the daytime, the azimuth range was reduced to keep the array boresight center $> 20^\circ$ from the Sun to avoid overheating and damaging the instrument. No other celestial objects were avoided by the telescope, and bright objects, e.g., the Moon, are removed in data cuts. More details regarding data selection are discussed in \citetalias{Li23}.  For a schedule dedicated to the CMB survey, observations spanned nearly 24 hr, ultimately terminating in preparation for the next day's boresight adjustment. At times, however, the CMB survey was interrupted by weather, instrument malfunction, and dedicated observations of the Moon or planets conducted for calibration purposes \citep{xu20,dahal21venus}. A summary of key properties of the survey and instrument are presented in Table \ref{tab:summary}.
Further details regarding the survey and instrumental configuration changes can be found in \citetalias{Li23}. 

\begin{deluxetable}{cc}
\tablecaption{CLASS 40 GHz instrument and survey summary. \label{tab:summary}}
\tablewidth{0pt}
\tablehead{} 

\startdata
Site location (lat/lon) & $-22.96^\circ/-67.79^\circ$ \\
Site elevation & 5140 m \\
Entrance pupil diameter & 30 cm \\
Beam size (FWHM) & $1.56^\circ$ \\
Frequency range & 33--43 GHz \\ \hline
Sky Area (sq-deg) & 30362 \\
Map decl.~ limits & $-76^\circ$ to $30^\circ$ \\
Mount pointing elevation &  $45^\circ$\\
Mount azimuth range & $\pm 360^\circ$ \\
Linear (circular) map sensitivity & $125 \,(130)\,\mathrm{\mu K\, arcmin}$\\
\enddata
\end{deluxetable}

\section{Data processing and maps}
\label{sec:maps}

The data collected by CLASS are unique in the field of CMB surveys. These are the first and only data collected by telescopes employing a front-end VPM to stabilize the polarization measurement and guard against certain systematic errors. In this Section we summarize the collection and processing of these data. A more complete description is given in \citetalias{Li23}. 

\begin{figure*}[t!]
\centering
\includegraphics[width=\linewidth]{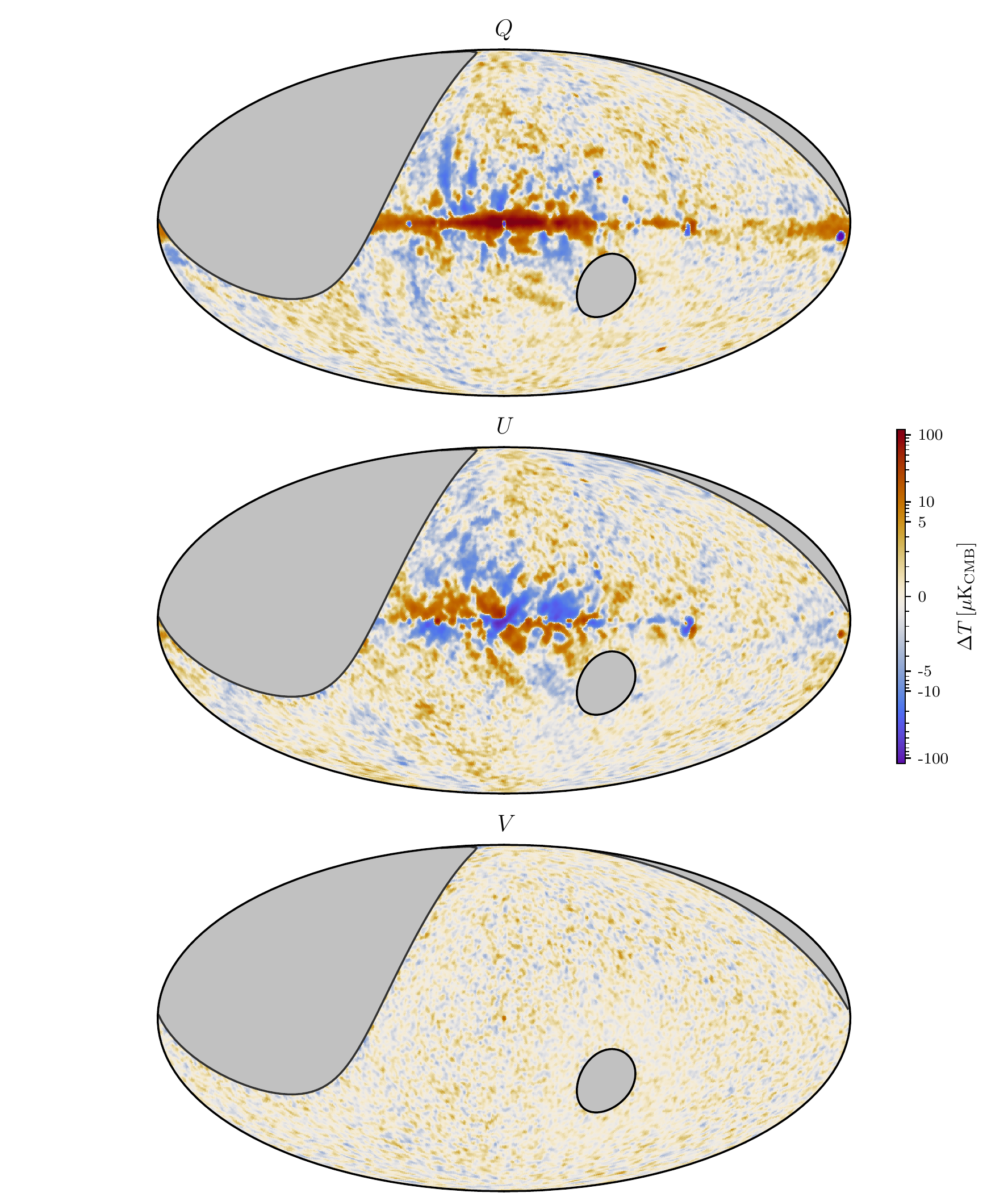}
    \caption{The $Q$/$U$/$V$ Stokes parameter maps for the CLASS 40 $\mathrm{GHz}$ channel are shown in Galactic coordinates using a Mollweide projection and $(l,b) = (0,0)$ placed at the center of each plot. Gray regions indicate portions of the sky not observed by CLASS or included too few observations to be well constrained. The maps have been smoothed with a 1.5$^\circ$ Gaussian beam for visualization purposes.
    To show both bright synchrotron features within the Galactic plane and the faint regions elsewhere in the map, the color scale is linear within $\pm5\,\mu$K and logarithmic beyond this range. The dipolar atmospheric $V$ signal has been removed from the $V$ map through the mapmaking filters, and no remaining Stokes $V$ signal is detected---the map is consistent with noise. The survey coverage is nonuniform; see \Linprep{} for hit maps. In addition to anisotropy in the signal, the noise also changes over the sky.
    \label{fig:quv_maps}}
\end{figure*}

\subsection{Data Acquisition and Pre-processing}
\label{ssec:Data Acquisition and Pre-processing}

The total survey dates included in these results span a period of 4.32 yr; this range excludes downtime for the 90~$\mathrm{GHz}$ telescope installation and the COVID-19 pandemic. 
After data taken during periods of poor weather or systematic instrument malfunction were removed, 45.4\% of the total volume remained. 
At this point, contiguous data were grouped into spans. 
Most spans comprise 22 hr of data, interrupted at midday for changing boresight and \emph{IV} calibration. However, spans may also be shorter if the observations were interrupted. We additionally cut data taken for calibration purposes or that were of lower quality, leaving 28\% of the total volume, equivalent to 86.77 detector$\cdot$years of data.   

The raw data were affected by the finite electro-thermal response of the TES bolometers, which is sufficiently modeled as a single-pole filter with $3$--$4\,\mathrm{ms}$ time constant. The MCE readout electronics also applied an antialiasing filter to the data. Both of these filters were deconvolved from the selected data. The data were then calibrated from binary MCE-units to power using the /emph{IV}-based calibration \citep{appel22}.

\subsection{Demodulation}

As described in more detail in \Linprep, the effect of the VPM is to modulate the amplitude of the polarization signal at a stable frequency near $10\,\mathrm{Hz}$ and its harmonics. This moves the signal in the raw data to sidebands of the modulation frequency components. \emph{Demodulation} is the process by which the modulated signal is recovered. 
For this analysis, the first step was to filter the data with a passband of approximately $9$--$51\,\mathrm{Hz}$, which preserved the encoded sky signal in the modulated data and band limited the noise beyond the instrument signal band. The next step was to use knowledge of the VPM modulation \citep{chuss06}, the source spectra \citep{pysm, petroff20}, the telescope bandpass \citep{dahal22}, and the geometric orientation of the VPM within the telescope \citep{eimer12spie} to solve for the polarization signal incident on the VPM. 
In the process of solving for the signal, the data were further low-pass filtered with a cutoff frequency between 0.5 and 1 $\mathrm{Hz}$---with the value chosen depending on the telescope scan speed. 
The data were then downsampled, and gaps corresponding to cut data were filled by linear interpolation with white noise to limit the introduction of spurious artifacts when treating the data as a complete, uniformly sampled record. 

\subsection{Filtering and Mapmaking}
\label{ssec:map_making}

Before mapping, the demodulated data were filtered to remove systematic errors that were not modeled by the mapmaking procedure. The first filter eliminated pickup from the mount azimuth motor current by removing the first five harmonics of each full $1440^\circ$ (distinguishing between the positive and negative scan velocity) azimuth scan. 
The second filter targeted spurious polarization due to the wind-induced deformation of the thin polypropylene environmental seal at the base of the forebaffle extension. 
Beginning in 2021 September, we removed the environmental seal during good weather, and data taken thereafter did not require this filter. 
A third filter attempted to remove any azimuthally synchronous signal (e.g., ground pickup, electric pickup, or, for circular polarization, Zeeman-splitting of atmospheric oxygen in the Earth's magnetic field). 
For linear (circular) polarization, this filter removed 15 (10) azimuthal harmonics.
These three filters were estimated piecewise over timescales ranging from 2--6 hr depending on the detector set, the era of the survey, and the polarization states.
In other words, although the harmonic functions removed are defined relative to the telescope azimuth scan, they are estimated and held fixed over many scan periods, decreasing their impact on the celestial signal (Section \ref{ssec:transfer-functions}). Finally, from 2017 August to the end of Era~1, the data were notch-filtered at frequencies below the scan frequency due to radio frequency interference from web cameras installed in the telescope. During Era~2, the cameras were kept off during observations. Among these filters, the third filter targeting the azimuthally synchronous pickup removed the most celestial signal on large scales.
A more complete description of the filters and the spurious signals they mitigate can be found in \citetalias{Li23}.
Prior to mapping, the covariance of the filtered demodulated data was estimated by identifying common-mode covariance through singular value decompositions and modeling the remaining noise power as covariant between the detector pairs and Stokes states of each feedhorn \citep[][\Linprep]{dunner13}. In this way, covariance was estimated as a function of frequency (bins), between detectors, and between linear and circular polarization. 
The noise estimation bias due to missing data was iteratively corrected by filling the gaps with mock data that preserved the noise covariance through Gibbs sampling \citep[][\Linprep]{Huffenberger2018}. 
With this noise covariance, linear and circular polarization maps were made via a preconditioned conjugate gradient solver with an adapted version of \texttt{minkasi}\footnote{\url{https://github.com/sievers/minkasi}} \citep{romero20}. 
Once a map solution was achieved, it was subtracted from the demodulated data, the noise covariance was re-estimated, and a new map solution was found. 
This ``template iteration'' was used to make the noise covariance estimation independent from the signal \cite[][\Linprep]{dunner13, aiola20}---a requirement for recovering the signal in the maps, especially on the largest angular scales. \citetalias{Li23} found that five template iterations were sufficient to recover $\geq97.5\% $ of the signal in the filtered demodulated data. 

An initial calibration of the maps in temperature units was based on measurements of the Moon and Jupiter \citep{appel19,xu20,dahal22} and is described in \citet{Li23}. Each detector’s calibration was found to be stable to the subpercent level over the course of the survey, a consequence of the stable atmospheric emission from the CLASS site at 40 $\mathrm{GHz}$ \citep{appel22}. A final calibration correction was performed by comparing the bright regions in the CLASS maps directly to those same regions in WMAP \textit{Ka} and \textit{Q} band maps. This procedure is described in Section \ref{ssec:calib}. The stability of the calibration is further verified by a series of null tests described in Section \ref{ssec:nulls}.

The final linear and circular polarization maps are shown in Figure \ref{fig:quv_maps}. These are represented in HEALPix format with a resolution parameter of $N_{\rm side}=128$ \citep{healpix} ($\sim 27.5'$ pixels over-sampling the telescope beam). We have made maps and associated data products available to NASA LAMBDA for public release.\footnote{\url{https://lambda.gsfc.nasa.gov/product/class/}}

For map-based comparison to previous measurements, we have developed simulations that apply the effect of the CLASS mapping pipeline on other maps.
To generate such a simulation, the $Q$/$U$ polarization maps of another experiment were first smoothed with a beam defined by the ratio of the CLASS beam, discussed below, and the beam of the other experiment, 
thus deconvolving the intrinsic beam from the map and leaving the map smoothed as if observed by CLASS.
Using the CLASS pointing model, the maps were then projected into the time domain as though they were CLASS demodulated data. 
The data selection, filtering, and mapping processes were then applied to form so-called \emph{re-observed} maps. 
For this work, we have used re-observed WMAP \textit{K}-, \textit{Ka}-, and \textit{Q}-band \citep{bennett13} and the \Planck\ 353 $\mathrm{GHz}$ map \citep{planck18III}.
When using a re-observed map, we prepend the name with ``r," e.g., rWMAP \textit{K}-band or simply r\textit{K}.

\subsection{Mapping and Beam Transfer Functions}
\label{ssec:transfer-functions}

To evaluate the impact of time-stream filtering on the maps, \citetalias{Li23} used simulations to estimate the harmonic domain mapping transfer function. This has the form
\begin{equation}
    C_\mathrm{\ell,\mathrm{out}} = F_{\ell\ell'} C_{\ell', \mathrm{in}},  \label{eq:hdtfunc}
\end{equation}
where $C_{\ell', \mathrm{in}}$ refers to the $EE$, $BB$, and $VV$ angular power spectra of the unfiltered sky. The transfer function $F_{\ell\ell'}$ relates these to the angular power spectrum as observed by CLASS, $C_\mathrm{\ell,\mathrm{out}}$, accounting for correlations introduced between multipole moments and between spectra. Simulations showed that when this transfer function is used with a pseudo-$C_\ell$ estimator \citep{polspice}, the recovered estimate of the input spectra is unbiased for $\ell>4$.
The effect of the mapping transfer function is to bias signal power to be low on the largest angular scales. For linear (circular) polarization, the transfer function was found to leave approximately 85\% (93\%), 67\% (85\%), 34\% (47\%), and 6\% (17\%) of the signal in the angular power spectrum at $\ell=30,\,20,\,10,\,\mathrm{and}\, 5$, respectively. See also Figure~14 of \citetalias{Li23}. 

The mapping transfer function shown in \citetalias{Li23} assumed a uniform weight within the CLASS survey region. That form of $F_{\ell\ell'}$ is useful when simulating the impact of the CLASS filtering strategy on the power spectrum of other experiments, as we have done when simulating the noise of other experiments in our calibration (Section \ref{ssec:calib}) and when studying the spatial variation in the synchrotron spectral energy density (SED) (Section \ref{ssec:beta_var}). When correcting for the transfer function bias on CLASS cross-spectra, however, a modified transfer function correction that assumes the nonuniform CLASS weight is used, for example see spectra in Sections \ref{ssec:sync} and \ref{ssec:sed}. The uniform and CLASS survey weighted mapping transfer functions are both included in this data release.

The beam window function for Era~1 was estimated from Moon observations as in \cite{xu20}. However, for this paper we have updated that estimate to include data from Era~2 as well as new analysis steps. The updated analysis is described in \cite{datta23},~and summarized in the Appendix. 
The aggregate beam for the survey combines beam maps from individual detectors weighted by their relative contribution to the survey map. 
The aggregate beam is further symmetrized by combining beams from across the FOV with different elliptical orientations and by including the effects of boresight rotations and scan cross-linking in the survey map. 
The resulting beam has an FWHM of $1.559^\circ\pm0.006^\circ$, a solid angle of $799.4\pm0.5\,\mathrm{\mu sr}$, and an eccentricity of $0.07 \pm 0.01$, where the uncertainties are formal statistical uncertainties from the beam modeling. 
Furthermore, the beam estimates made separately from Era~1 and Era~2 agree well within uncertainties. 
The beam transfer function was estimated as the square of the harmonic transform of the beam (a.k.a. the beam window function). This function indicates the measured angular power spectrum will be reduced to 86\%, 74\%, 49\%, and 9\% of its intrinsic value at $\ell=20,\,30,\,50,\,\mathrm{and}\, 100$, respectively. For more details on the beam and beam window function estimate, see the Appendix. Finally we note that when comparing the Moon-derived beam profile to that of unresolved sources in the survey maps, one must account for the mapping transfer function, which attenuates the larger-scale modes. 

\section{Power spectrum estimation}
\label{sec:spec_est}
Angular power spectra for synchrotron signal analysis and all null tests were estimated by using the pseudo-$C_\ell$ method as implemented in \texttt{PolSpice} \citep{polspice}\footnote{https://www2.iap.fr/users/hivon/software/PolSpice/}. 
This method produces nonbiased $EE$ and $BB$ power spectra accounting for varying survey weight across the sky and the effect of beam convolution. For any given frequency, or \emph{channel}, $i$, the polarization weight $w_i$ was modeled as the scalar weight 
\begin{equation}
w_i = \left(h_\mathrm{QQ}h_\mathrm{UU}-h_\mathrm{QU}^2\right)^{1/2}.
\label{eq:pol_weight}
\end{equation}
Here, $h_{XY}$ are the hits maps, proportional to the inverse per-pixel variance of the map. The CLASS hits maps were described in \citetalias{Li23}. In this work, we have used cross-spectra exclusively. As a result, instrumental noise bias was avoided and systematic errors not shared between the crossed sources are reduced. When estimating the power spectra at a single frequency and single experiment, the cross-spectrum was estimated using a data split. For the CLASS data, the base split is defined in Section \ref{ssec:nulls}. When evaluating cross-spectra for other experiments, split definitions depend upon the experiment. For WMAP channels, covariance weighted coadded even- and odd-numbered years define the split. For the \Planck\ 30 $\mathrm{GHz}$ channel, no splits were considered; only cross-spectra with other experiments were used.

In the following discussion, it is convenient to adopt shorthand expressions for the cross-spectra. 
We use the band letter designation \textit{K}, \textit{Ka} and \textit{Q} to indicate a WMAP channel. The numbers 30, 100, 143, and 353 are used to indicate the \Planck\ channels. We use $C$ to indicate the CLASS 40 $\mathrm{GHz}$ data. Thus, $\mathit{K}\times \mathit{C}$ indicates the cross-spectrum between the full $K$-band maps and the full CLASS maps, and $\mathit{Ka} \times \mathit{Ka}$ denotes the cross-spectrum between the even and odd years of the WMAP \textit{Ka}-band. 

When estimating the CLASS signal cross power spectra, the filtering described in Section \ref{ssec:map_making} causes power to be removed from the map and thus biases the resulting spectrum. The biased spectrum is then corrected by
\begin{equation}
\label{eq:harmonic_tf_correction}
\hat{C}_\mathrm{\ell} = F^{-1}_{\ell\ell'} \hat{C}^\mathrm{polspice}_\mathrm{\ell'},
\end{equation}
where $F_{\ell\ell'}$ is the harmonic domain transfer function described in Equation \ref{eq:hdtfunc}. To correct for bias when estimating the cross-spectrum between CLASS and another experiment, the matrix square root of $F_{\ell\ell'}$ is used to model the required effective combined transfer function; any imaginary portion of the matrix is very small and dropped. 
In cases when portions of the sky were masked, e.g., to block bright sources or the Galactic plane, the initial binary mask was apodized using the NaMaster\footnote{https://github.com/LSSTDESC/NaMaster} \citep{Alonso:2018jzx} \texttt{C2} apodization tool with a $3^\circ$ scale. The map weight used for spectrum estimation was then the product of the weight from Equation \ref{eq:pol_weight} with the apodized mask. Simulations assuming the angular power spectrum followed a power-law model or a nominal fixed synchrotron sky model plus instrument noise were used to confirm the estimators were unbiased. 

Power spectra were evaluated in celestial coordinates, as these are the native coordinates of the CLASS maps. The publicly available WMAP and \Planck\ maps used in this work were first rotated at their full resolution from Galactic coordinates to celestial and smoothed with an antialiasing cosine-like filter \citep{benabed09}
\begin{equation}
    f_\ell= \begin{cases}1 & \text { for } \ell \leq \ell_1, \\ \frac{1}{2}\left[1+\cos \pi \frac{\ell-\ell_1}{\ell_2-\ell_1}\right] & \text { for } \ell_1<\ell \leq \ell_2\\ 0 & \text { for } \ell>\ell_2,\end{cases},
\end{equation} 
where $\ell_1=128$, and $\ell_2=3\times\ell_1$.
The maps were then downgraded to $\textit{N}_\mathrm{side}=128$. In cases when it was necessary to have different channels expressed in a common resolution, the 
map was convolved with the beam ratio of the intrinsic beam and the target beam.
Note for synchrotron-dominated sky regions studied in this paper, the $B$-mode power is a significant fraction of the $E$-mode power. In this case, $E/B$ purification methods are not required and were not applied in the spectra presented here.

Finally, all binned power spectra follow the same binning method; bins are defined between $\ell=[5, 125)$ with fixed width $\Delta \ell = 10$. Binned spectra use the typical $\ell ( \ell + 1)$ weight within each bin. Note, that we adopted this binning to help visualize the results of the null test (see Section \ref{ssec:nulls}) and to help mitigate mode coupling. To be consistent, we adopted these bins for all angular power spectrum results. In all cases, simulations were used to determine the full EE/EB/BB bin--bin covariance matrix. Off-diagonal elements are dominated by the nearest neighboring bins; these elements are $\lesssim 0.1$ for the full region observed by CLASS and can be slightly larger but still $\lesssim 0.15$ when considering the largest foreground masks.

In an effort to quantify the uncertainty in the derived spectra, all estimates were compiled with an ensemble of simulations. 
Each simulation was then processed through the same power spectra estimation method as the data, and the resulting ensemble was used to directly estimate the covariance of the spectra. 
For visualization, the error bars for power spectra are the square root of the diagonal of the covariance matrix. 
For any model fit, however, the full-covariance matrix is used. The details of the simulations used for the covariance estimation vary depending on the quantity being estimated, but they can broadly be broken down into two categories: instrument noise only and sky signal plus instrument noise simulations. 

\begin{figure}
    \centering
    \includegraphics[width=\columnwidth]{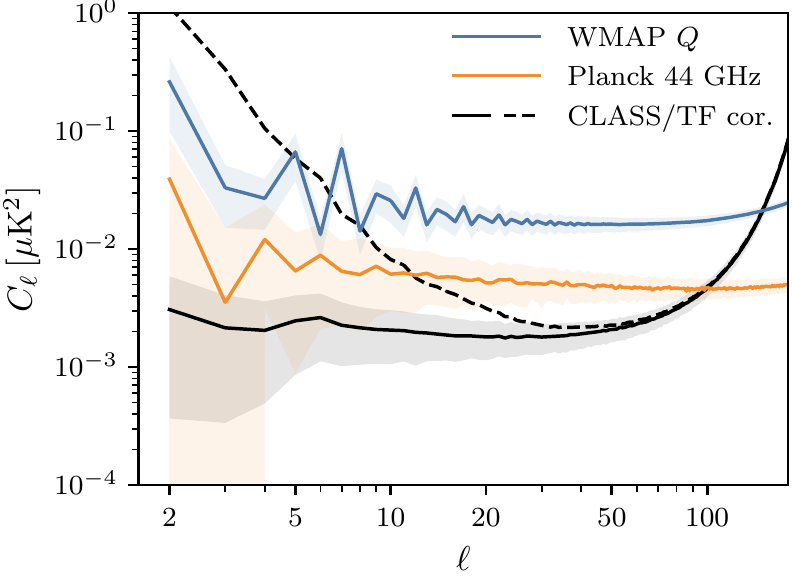}
    \caption{Comparison of the map noise between the full sky regions observed by WMAP and \Planck\ and that of CLASS as measured in the $EE$ cross-spectrum. The CLASS noise spectra are determined via noise-only simulations, for WMAP through sampling the WMAP covariance matrix, and for \Planck\ through the published noise+systematic error simulations. At each frequency, these simulations have been shown to be representative of the noise in the respective total maps. The dashed line illustrates the CLASS noise corrected for the transfer function (TF). The upward curve at higher $\ell$ is due to the CLASS $1.56^\circ$ beam. CLASS has the lowest noise in the range $10 \lesssim \ell \lesssim 100$. 
    \label{fig:noise_comp}}
\end{figure}

The CLASS noise simulations were generated using the Fourier domain noise model inferred during the mapmaking procedure; see \citetalias{Li23}. 
Random time domain realizations constrained to be consistent with the noise spectrum were then mapped. 
Depending on the application, the noise simulation could model the full CLASS data set or half of the CLASS data, i.e., splits.  
For \Planck\, the 300 \texttt{FFP10} end-to-end channel simulations were used \citep{planck15XII}---note these simulations include noise and residual systematic errors. 
For WMAP simulations, large-scale noise realizations consistent with the published low-$\ell$ pixel-pixel noise covariance matrix were combined with high-$\ell$ noise realizations consistent with the polarization hits \citep{larson2011}. 
The WMAP and \Planck\ simulation maps were smoothed with an antialiasing filter and downgraded in resolution to match the CLASS maps. 
The noise estimated from the CLASS noise simulations is shown in comparison with the WMAP \textit{Q}-band and the \Planck\ 44 $\mathrm{GHz}$ full sky simulated noise spectra in Figure~\ref{fig:noise_comp}. After accounting for filtering, the CLASS noise was found to improve upon the comparable space-based measurements in the $10 \lesssim \ell \lesssim 100$ range. 

\section{Internal Consistency Checks}
\label{sec:internal-consistency}

The internal consistency of the survey and stability of the calibration procedure were verified by performing a series of null tests. 
Broadly speaking, a ``null map'' is the difference between two CLASS maps made from nominally independent sets of data. 
The null test procedure checks the statistical consistency between (1) the cross power spectra of two independent CLASS null maps and (2) the cross-spectra of null maps created from simulations of the data. 
Simulations include a sky model and the statistical noise simulations described in Section~\ref{sec:spec_est}. An inconsistency would indicate the presence of systematic errors that may bias the CLASS cross-spectra results.  No such inconsistency was found. 

The filtering outlined in Section \ref{ssec:map_making} was designed to mitigate systematic effects, some of which were initially discovered through early null test investigations. While the current filtering was found to be sufficient to remove systematic errors to the level probed by these consistency checks, the detailed filtering parameters were not precisely tuned to minimize the signal loss. More optimal filtering strategies will be considered in future work. 

\subsection{Null Test Method}
\label{ssec:nulls}

\begin{figure*}
    \centering
    \includegraphics[width=\textwidth]{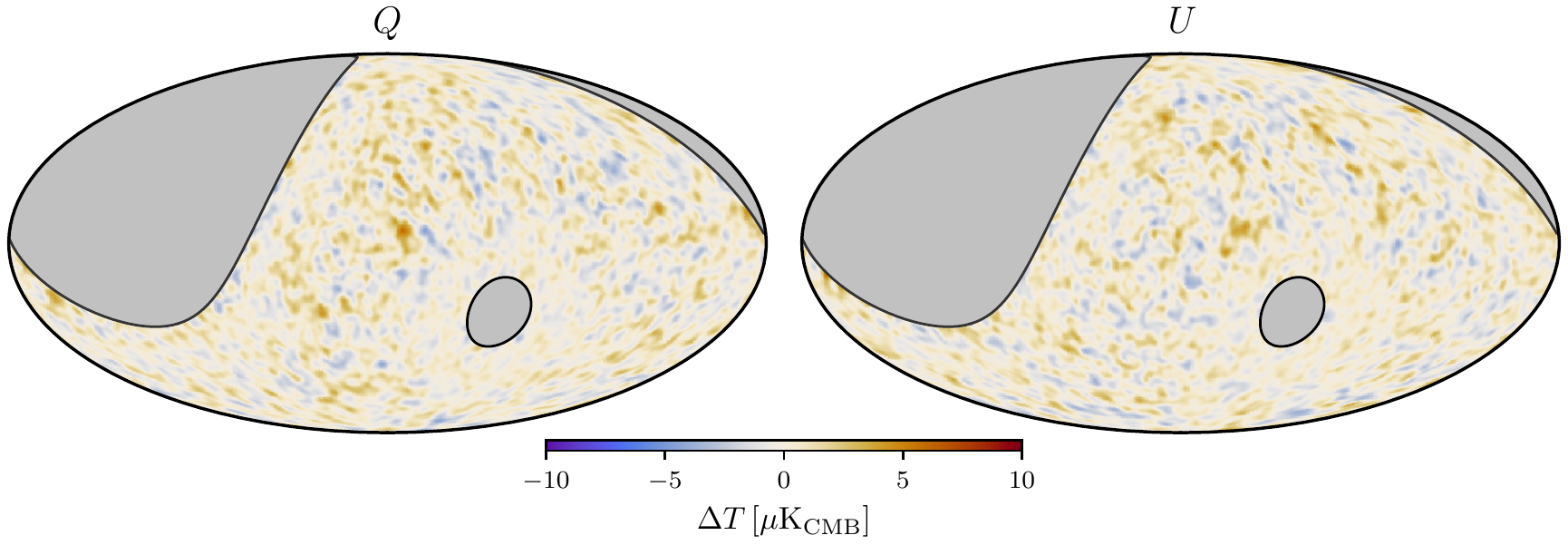}
    \caption{The base split, $m_{1} - m_{2}$, null maps for the $Q$/$U$ data are shown. For clear visualization, the maps have been smoothed with a $3^\circ$ FWHM Gaussian beam. The amplitude of the noise variations visually track the variations in the survey depth; see the hit maps in \Linprep. While not used for the suite of null tests, these null maps demonstrate the overall calibration stability for the base split---the null maps are dominated by noise. 
    \label{fig:basic_null_map}}
\end{figure*}

The time-ordered data spans (Section \ref{ssec:Data Acquisition and Pre-processing}) were first distributed into two base sets, $m_{1}$ and $m_{2}$, which we call the \emph{base split}. 
Each span was placed into one of the two base sets while controlling for the boresight rotation of the telescope, azimuth scanning style, presence of the environmental closeout film, blackening state of the telescope and baffle structures, and observation Era. 
Each base split consists of approximately half of the full mapped data and includes each distinct observational condition in nearly the same proportion as in the full survey map. The null maps of the base split itself are shown in Figure \ref{fig:basic_null_map}.

\begin{figure*}
    \centering
    \includegraphics[width=\textwidth]{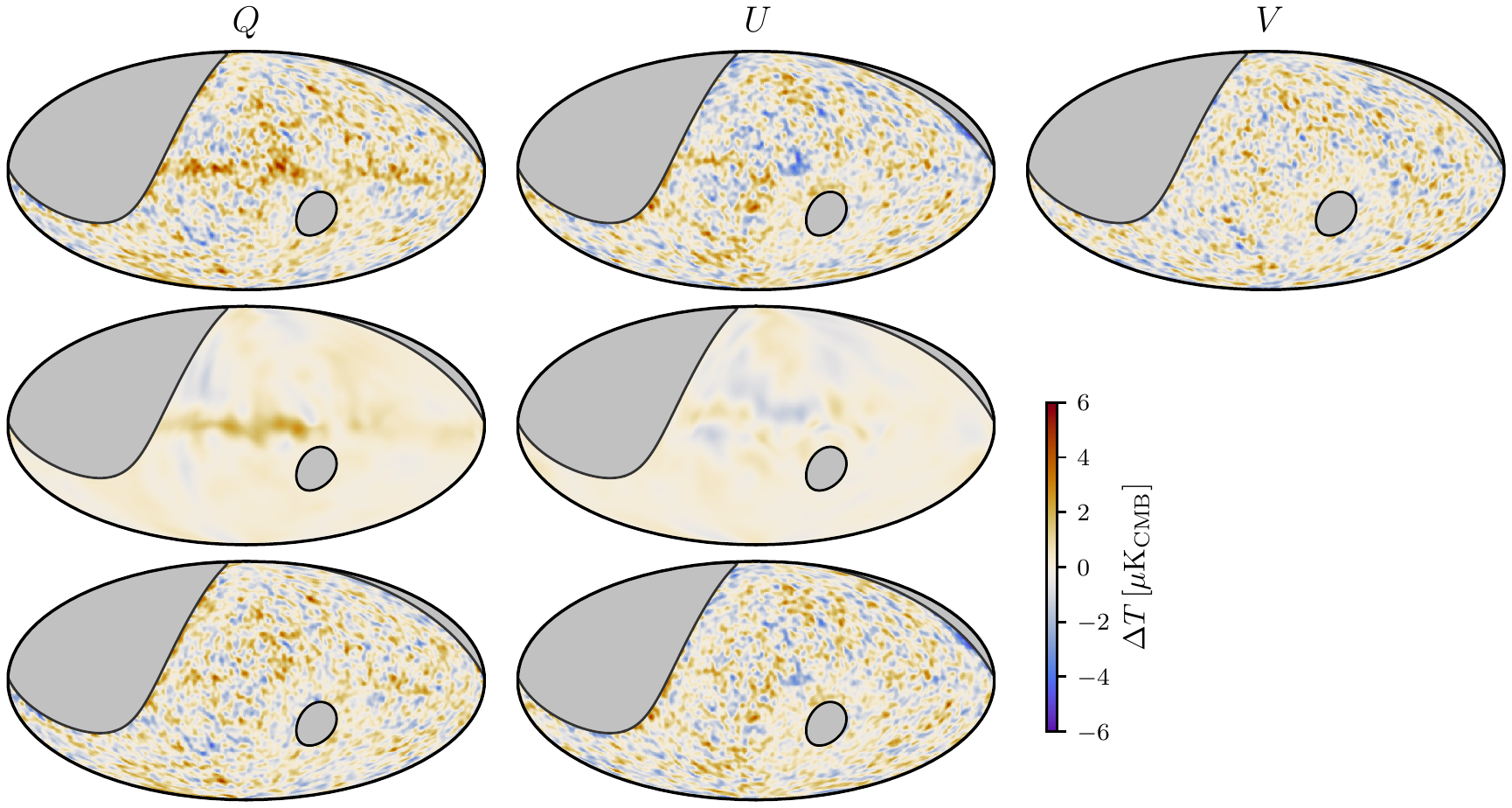}
    \caption{Example null map constructions are shown. The top row shows the Stokes $Q$/$U$/$V$ base null maps $m_{1A} - m_{1B}$ for the 8 hr in/out split---maps were smoothed with a $3^\circ$ FWHM Gaussian beam to make large-scale features more apparent. 
    The middle rows show the equivalent split from a re-observed WMAP \textit{K}-band map scaled to the CLASS bandpass---the features show the expected map difference resulting from imbalanced filtering in the splits. 
    The bottom row is the residual difference between the data and the template, which is statistically consistent with the noise. 
    \label{fig:null_map}}
\end{figure*}

The base splits were further subdivided into subsplits, $A$ and $B$, according to other criteria, e.g., distinct detector sets, defining four maps, $m_{i\rho}$ with $i\in\{1,2\}$ and $\rho \in \{A,B\}$.
The \textit{A}/\textit{B} maps were solved simultaneously similar to the procedure described in \citetalias{Li23}.
This approach ensured that the noise modeling of data from each split was maintained regardless of the detector or temporal division and that the data were combined in the same way as in the total map.
The differences (null maps), $\Delta m_{i} = m_{iA} - m_{iB}$, null the sky signal leaving a biased noise map. 
The full list of splits is enumerated below. 
An example of a typical set of null maps is shown in Figure~\ref{fig:null_map}. 

Differences in filtering and sky coverage between the $A$/$B$ splits will bias the null cross-spectra, and we use simulations to account for these effects. 
For each split, a realization of the CLASS noise model for the split in question was added to a polarization sky signal model projected into the time domain. 
The sky model assumed an atmospheric $V$ dipole \citep{petroff20} and the Stokes $Q$/$U$ maps from WMAP 9-yr \textit{Q}-band smoothed with a $1.5^{\circ}$ Gaussian beam \citep{bennett13}. 
The synthetic time streams were then mapped as described in \citetalias{Li23}. 
In principle, one could then compare the cross power spectra of the null maps $\Delta m_1$ and $\Delta m_2$ to analogous spectra from simulations. 
Such a comparison would fail to compare possible spatial differences between the null maps based on the data versus those from simulations. 
To account for this effect, a null bias template $\Delta m^b_{i}$ was removed from each respective null map to produce a bias-corrected null map:\footnote{While the bias template corrects for the signal imbalance between the split, the mapping transfer function bias is still present in the null map and the simulations.}
\begin{equation}
\label{eq:biasc_null_map}
    \Delta m^c_i = \Delta m_{i} - \Delta m^b_{i}.
\end{equation} 
For the simulated null maps, the $\Delta m^b_{i}$ templates are the average of the simulated null maps $\langle\Delta m_i \rangle$ in which the noise is greatly diminished.  For the null maps made from data, the $\Delta m_i^b$ template is made from a signal-only simulation created by smoothing the synchrotron-dominated WMAP 9 yr \textit{K}-band map \citep{bennett13} with a $1.5^\circ$ Gaussian beam and scaling it by the approximate ratio ($0.17$) in synchrotron intensity between the WMAP \textit{K} band and the CLASS $40\,\mathrm{GHz}$ band. 

Finally, the bias-corrected null spectra from CLASS were compared to the corresponding ensemble of simulations via the $\chi^2$ test
\begin{equation}
    \chi^2 = \left(C_\ell^{1\times2} - \langle C_\ell^{\text{sim}} \rangle \right)^T \text{Cov}_{\ell\ell'}^{-1} \left( C_{\ell'}^{1\times2} - \langle C_{\ell'}^{\text{sim}} \rangle \right),
    \label{eq:chisq}
\end{equation}
where $C_\ell^{1\times2}$ is the binned CLASS null cross-spectra between $\Delta m_1^c$ and $\Delta m_2^c$, $\langle C_\ell^{\text{sim}} \rangle$ is the average of the corresponding simulated binned null power spectra,\footnote{Seventy simulations were used for each data split. 
Using additional simulations did not significantly change the results of the null tests.} $\text{Cov}_{\ell\ell'}$ is the bin--bin covariance matrix determined from the simulations, and summation is implied over repeated indices. 
We compute the probability-to-exceed (PTE) for each null test as the fraction of corresponding simulations having a $\chi^2$ greater than that of the data. 

One important difference between the null tests described here and the angular power spectrum analysis of the CLASS data described in Section \ref{sec:galactic} is the choice of masking. The null tests are evaluated \emph{with no Galactic mask}.  In the angular power spectrum analysis of the data, we mask regions of brightest polarized synchrotron emission (Section \ref{ssec:masks}).  Therefore,  the null tests hold the data and analysis pipeline to the same stringent self-consistency standards in the bright Galactic center as they do in the synchrotron-diffuse regions measured by the angular power spectra.

\subsection{Split Definitions}
\label{ssec:Split definitions}

\begin{figure*}
\includegraphics[width=\linewidth]{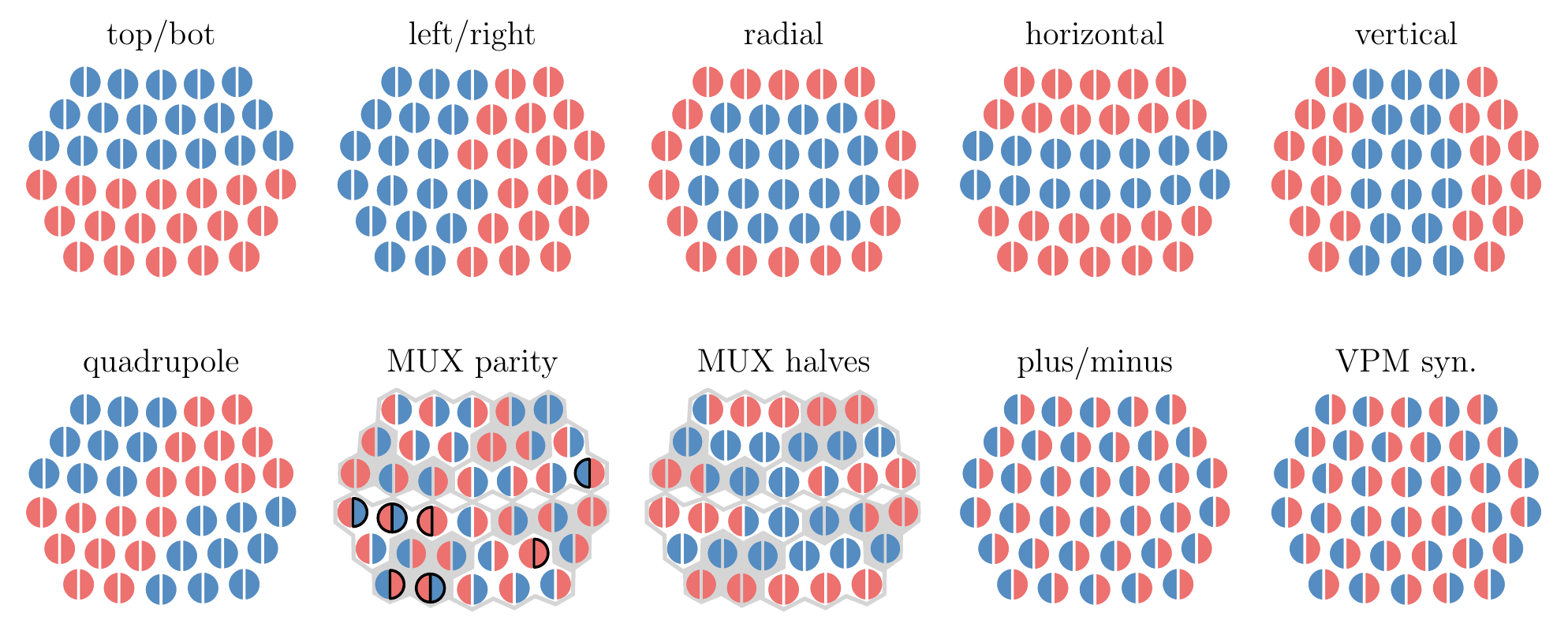}
\caption{\label{fig:null-det-split}
Detector-splits definition of the CLASS 40~$\mathrm{GHz}$ telescope. 
The detector pairs coupled to the same feedhorn are shown as semicircles with the $+45^\circ$/$-45^\circ$ oriented (viewed from the focal plane to the sky) detectors shown on the left/right and are positioned by their pointing offset on the sky. 
In the ``readout rows'' panels, the boundaries of the readout columns are delineated by the gray lines and shades.
During the Era 2 deployment, some of the detector readouts are re-mapped to optimize detector yields; thus, the split based on the odd/even readout row is different for Era 1 and Era 2. The panel above shows the current configuration (Era 2) and has the detectors with different assignments in Era 1 outlined in black.}
\end{figure*}

Here we describe the $A$/$B$ data splits used in the null test suite and the potential systematic errors that motivate the test. Two categories of splits were used. The first category, detector-based splits, divide the $A/B$-split by choosing subsets of half the available detectors; the full collection of detector splits is summarized in Figure \ref{fig:null-det-split}. In addition to testing for sensitivity to possible instrument based systematic effects, some detector splits also test for possible errors in VPM modeling. The second category of splits, temporal splits, divides the $A/B$-split based upon a temporary condition that evolves over time and verifies the total maps are insensitive to that evolving state. 

In the following list, we summarize the $A/B$ splits used in the null tests and describe the systematic errors tested by the specific splits.

\paragraph{Focal plane position}
The splits labeled \texttt{top/bot}, \texttt{left/right}, \texttt{radial}, \texttt{horizontal}, and \texttt{vertical} each break the symmetries of the optical system in different ways and therefore test for beam and VPM based systematic errors that could also follow these subsets. The \texttt{top/bot} split also tests for sensitivity to the atmospheric loading, circular polarization emission from the atmosphere, and residual ground pickup.

The \texttt{quadrupole} split is designed to probe the impact from the wind signal as discussed in \citetalias{Li23}. This split is also sensitive to the modeling of the VPM transfer function. 

\paragraph{Detector addressing} Two splits were defined based on their multiplexing (MUX) address location. The \texttt{odd/even row} split is sensitive to the electric crosstalk between the adjacent detectors in the same column. The \texttt{half rows} split divides the first and second halves of the detectors within each column based on the row index and is another check on possible readout effects. 

\paragraph{Polarization orientation} The \texttt{plus/minus} split divides the detectors according to their polarization orientation relative to the telescope. Within a pair, this split shows opposite signs in the demodulated data for T-to-P signal leakage and VPM synchronous modulated signals. While the total map is less susceptible to these issues, the null test from this split can examine this effect at the extreme limit. 
Single detector Moon observation revealed a faint dipolar residual T-to-P signal.
This effect is shown to be negligible for cosmological analysis, but its impact from leaking bright foreground temperature signal around the Galactic plane into polarization is significant for the \texttt{plus/minus} null test.
Following the prescription in \citetalias{Li23}, templates of this leakage are built by convolving the WMAP \textit{Q}-band temperature maps with the dipole leakage beam inferred from the CLASS Moon-centered maps for the \texttt{plus} and \texttt{minus} splits. The template is subtracted from the CLASS null splits before comparing them to the noise simulations.

The \texttt{VPM syn.} split is the product of the \texttt{left/right} split with the \texttt{plus/minus} split. This split divides the detectors into groups that systematically measure more/less synchronous VPM emission and test for sensitivity to residual emission in the maps.

\paragraph{Mount orientation} The \texttt{bs in/out} split divides the spans into the boresight set $\{-15^\circ, 0^\circ, 15^\circ\}$ and $\{-45^\circ, -30^\circ, 30^\circ, 45^\circ\}$---i.e., the inner versus outer boresight angles. The split tests for map sensitivity between the less tilted versus the more tilted instrument configurations. For example, the cryogenic system could behave differently, or there could be systematic pointing variations. 
In a similar way, the \texttt{bs pos/neg} split separates spans based on the sign of the boresight rotation---boresight 0 is omitted. 
This split also serves as a diagnostic for changes in the optical alignment of the telescopes as any drift in the array center position with respect to the mount boresight will always show up as a separation in either R.A. or decl., depending on whether the drift is in elevation or azimuth.
Due to the projection effect, the leakage from the sky linear polarization to the circular polarization through errors in the VPM transfer function would have the largest contrast when comparing the positive and negative boresight maps. 

\paragraph{Azimuth scanning}
The \texttt{east/west} split divides the data according to the orientation of the telescope boresight with the east half of the split corresponding to the $0^\circ-180^\circ$ azimuth range. This split tests sensitivity to residual ground pickup since the horizon is different between the two directions. For example, most of the nearby mountain features are in the east, but most population centers are in the west. The split also tests sensitivity to observations of the rising versus setting sky, especially in regards to sunrise and sunset. The harmonic domain, azimuth-based filters (Section \ref{ssec:map_making} and \citetalias{Li23}) will generally include data from both splits. For this test, we elect to use the same azimuth filter as defined for the full data and therefore allow the filter definition to include the constraining power from both splits. The approach is duplicated in the simulations to which the split is compared. 
The \texttt{az velocity} split divides the data by the sign of the azimuth scanning direction. This split tests for residual systematic errors related to the azimuth servo motors. Similarly, the \texttt{half sweep} split divides each $720^\circ$ scan of the mount into the first and second $360^\circ$ scans and would be sensitive to the impact of the azimuth servo in a different way. For example, the first (second) $360^\circ$ scan always includes the mount accelerating after (decelerating before) a turn-around. 

\paragraph{Sun and Moon splits}
The \texttt{midnight} split divides each span at the solar midnight. Several environmental factors evolve during the night; for example, wind direction often shifts by $\sim 40^\circ$ toward the north before midnight and the secular warming/cooling cycles of the ground are different when the sun is setting versus rising. 
The \texttt{8h in/out} split divides each span into the middle 8 hr close to the solar midnight and the remaining time close to the twilight. 
This is designed to test the effect of the Sun illumination and the associated environmental change. 
The diurnal cycle of air temperature, wind speed, and PWV follows well with local time and peaks in the afternoon at around 14:00, 16:00, and 18:00 (UT1-4), respectively.
Therefore this split approximately divides the data into times when the temperature, wind, and PWV are low and high.
It is also sensitive to the clouds, as the cloud cover at the site is observed to decrease during the night and rise again in the morning. 
Finally, the \texttt{Moon} split divides each span depending on whether the Moon elevation is below/above the horizon. This is sensitive to the Moon's illumination through the far sidelobes.

\paragraph{Half survey} The final split, \texttt{survey}, divides the data before/after 2019 May 21, which is approximately the half point of the survey. This probes a potential secular change in the instrument or sky during the 6 yr survey.

\subsection{Null Test Result Summary}

Using the two bias-corrected null maps derived from Equation \ref{eq:biasc_null_map}, the cross power spectra were computed for each of the splits described above. The resulting null spectra (and summarized $\chi^2$ PTEs in \refpte) from the detector-based splits are collected in Figure \ref{fig:nulltest_chi_p1}, and the temporal splits are in Figure~\ref{fig:nulltest_chi_p2}. In each spectrum, the points shown are the ratio of the null spectra from the maps to the diagonal of the noise simulation-based covariance matrix. 
The reduced $\chi^2$ value for each spectrum is consistent with unity and is shown in the lower-right corner of each subplot; this is computed using the full-covariance matrix as in Equation \ref{eq:chisq}. The $\chi^2$ PTEs for all spectra are collected in \refpte. The split names indicated in the Figures follow the definitions outlined in the text. As the full survey maps combine all of these data, the final results benefit from reducing any residual systematic by restoring the symmetry that individual tests are designed to break. In fact, the symmetries, e.g., polarization angle, focal plane side, etc., and degeneracies, e.g., boresight angle, scan direction, etc., are \emph{designed} in the instrument and survey specifically to reduce possible systematic errors in the combined data set. 

The lowest PTE value, 0.01, occurs in the \texttt{midnight} split of the $VE$ cross-spectrum. Careful study of these null maps and spectrum do not indicate a clear problem. The low PTE value is driven by the prevalence of slightly negative band powers in the null spectrum, seen in Figure \ref{fig:nulltest_chi_p2}. It is also possible there is some very small signal or unknown systematic effect remaining in this particular split; it could also just be noise. The number of PTE values near 1 indicate that the noise might be slightly overestimated in some cases. All uncertainties, however, retain this noise model. 

The presented null tests are not necessarily independent of one another, and therefore the collected PTEs are not sampled from a uniform distribution. Rather than a strict comparison to a uniform distribution, these results should be viewed as a demonstration of the overall robust rejection of systematic errors even when the data are split in ways that maximize the impact of the potential systematic errors probed by each split.



\begin{figure}
    \label{fig:null-pte}
    \centering
    \includegraphics[trim=5.4cm 10cm 5.4cm 4.58cm, clip, width=\linewidth]{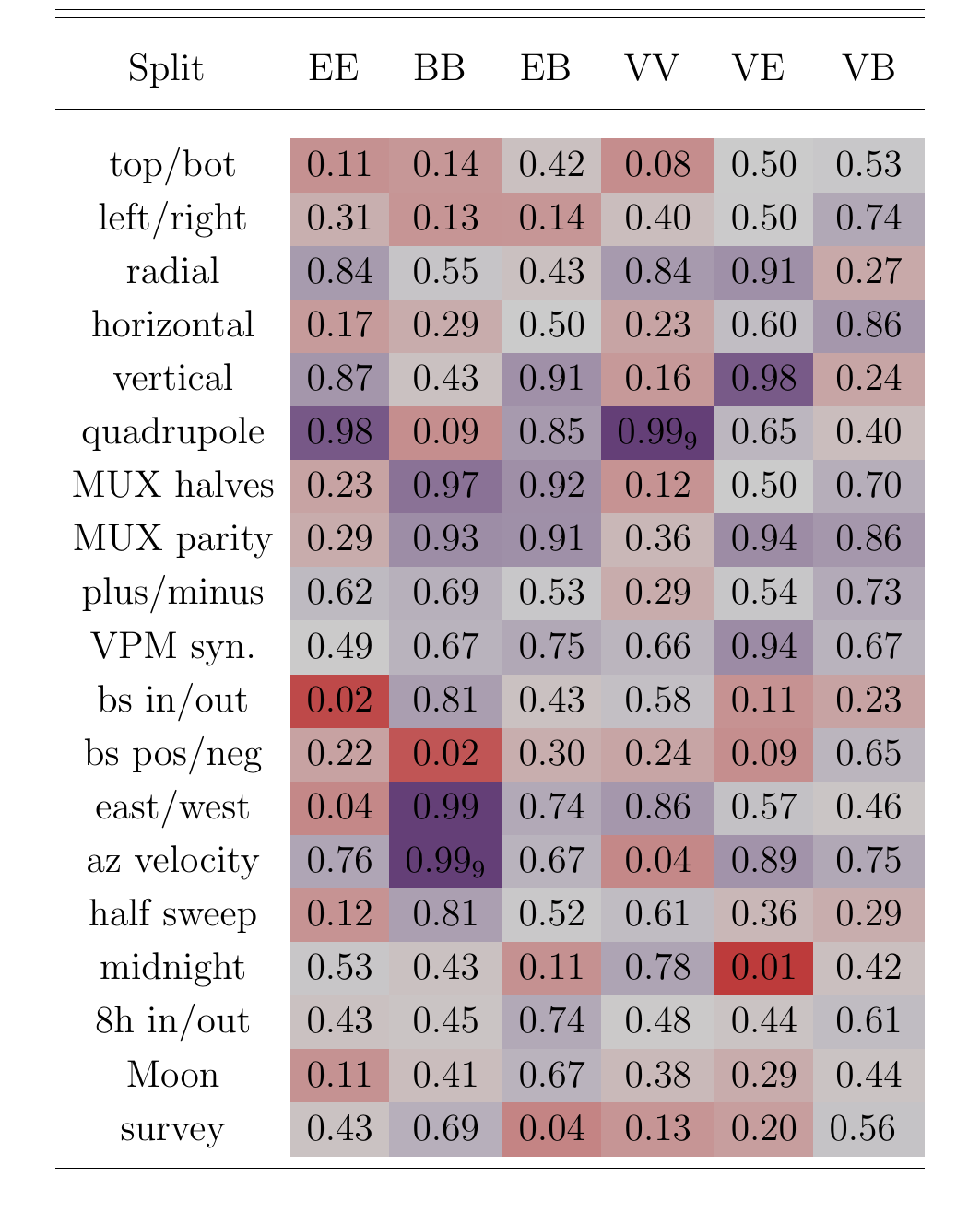}
    \caption{
The PTE values are tabulated for every null test split and each of the six polarization (cross) spectra.  
The cell color indicates the PTE value---red (purple) for PTE values less (greater) than 0.5, meaning that the simulation shows less (more) scatter than the data.
The saturation of the color scales with the deviation of the PTE from $0.5$.
A subscript third significant digit is included in cells otherwise rounding to 1 or 0.
    }
    \label{fig:pte-table}
\end{figure}

\begin{figure*}
    \includegraphics[width=\textwidth]{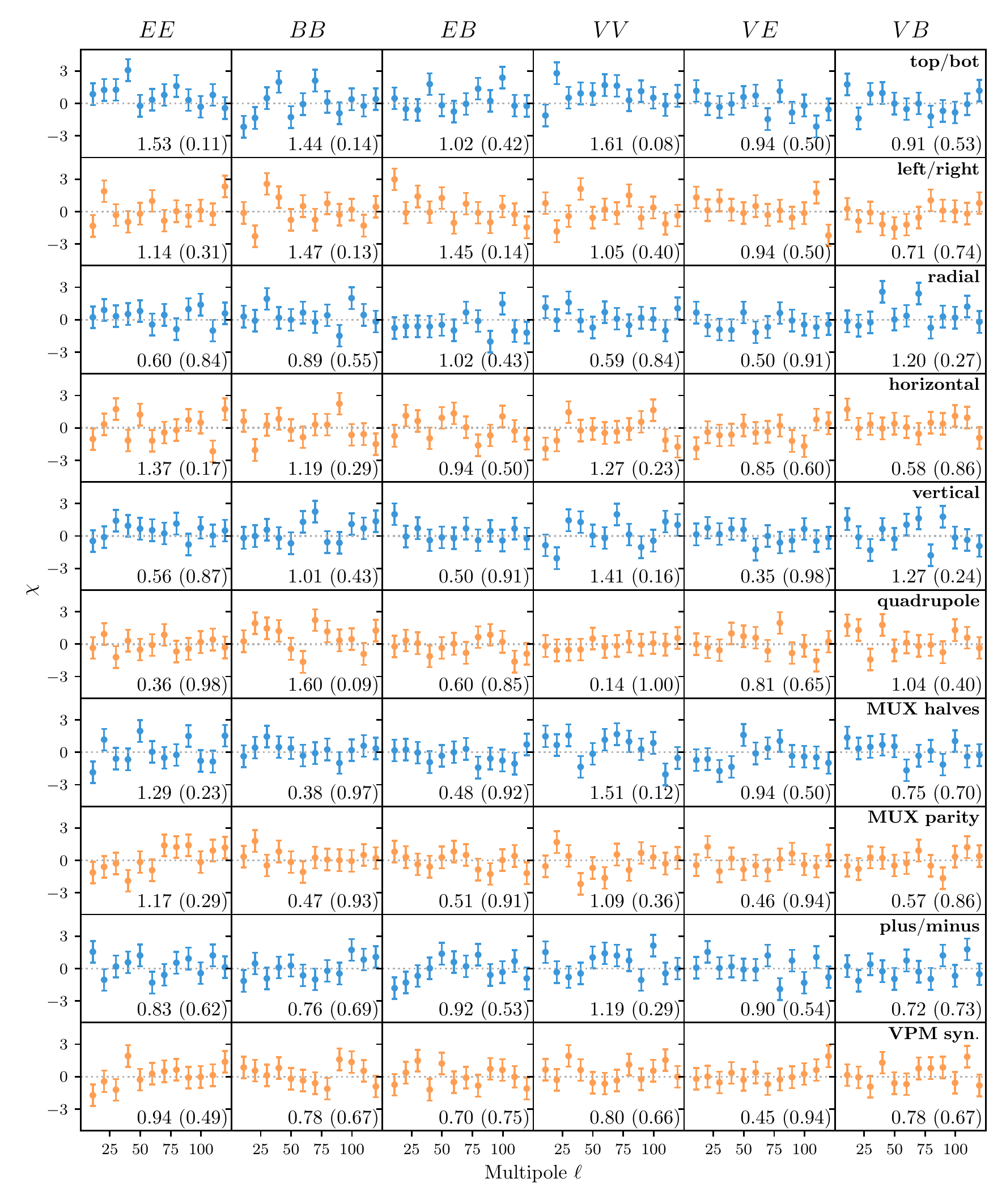}
    \caption{\label{fig:nulltest_chi_p1}
    Null test spectra for detector splits.
    Each row corresponds to the six null spectra of a null test split.
    The null spectra (colored points) are shown as the $\chi$ value defined by the ratio of the null spectra band power and the diagonal of the simulation covariance matrix.
    For visualization purposes, the error bars of all of the data points are set to unity.
    In the null test, the $\chi^2$ statistics are based on the full-covariance matrix (Equation \ref{eq:chisq}); the reduced-$\chi^2$ and PTE (in parentheses) values are displayed in the lower-right corner of each panel. The corresponding PTE values are collected in \refpte.}
\end{figure*}

\begin{figure*}
 \includegraphics[width=\textwidth]{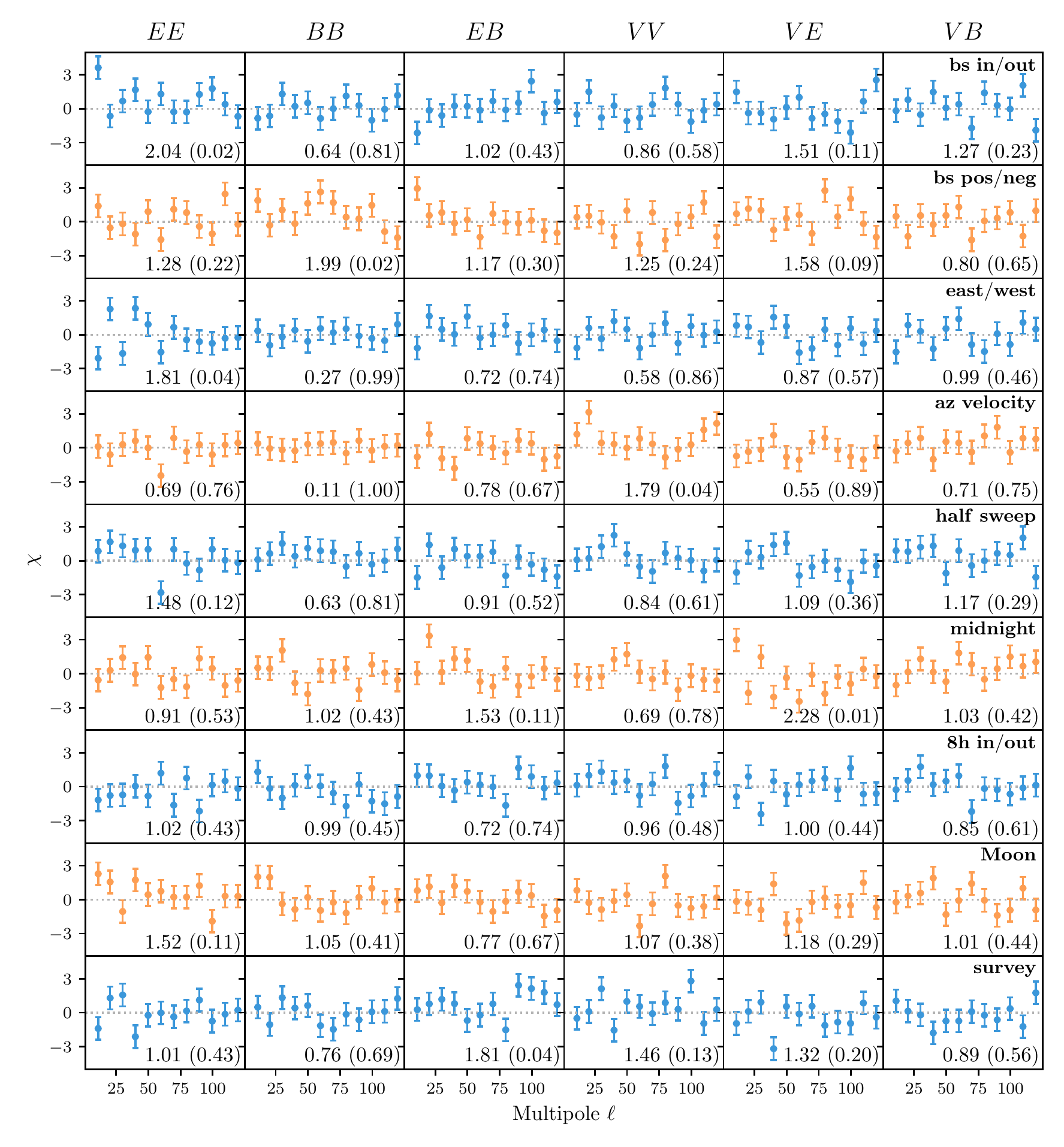}
    \caption{\label{fig:nulltest_chi_p2} Same as Figure \ref{fig:nulltest_chi_p1}, but for the temporal splits.}
\end{figure*}

\section{Comparison and Calibration to Previous Measurements}
\label{sec:compare}
\subsection{Masks}
\label{ssec:masks}
For our analysis, we have defined a series of sky masks. The sky masks considered were defined to mask sequentially larger regions with the brightest polarized synchrotron (\texttt{sX}) or dust (\texttt{dX}) emission, where \texttt{X} is an integer $0$--$9$. For example, \texttt{s0} masks the smallest region with the most intense synchrotron polarization. Additionally, all masks exclude data outside of the CLASS survey boundary at declinations $\delta=-76^\circ$ and $\delta=30^\circ$ and a bright-source mask.

The \texttt{sX} and \texttt{dX} masks exclude regions of polarized intensity based on the \Planck\ \texttt{Commander} synchrotron and dust component maps, respectively \citep{planck18IV}. Each mask was defined following a five-step process:

\begin{enumerate}
    \item Each \texttt{Commander} map, e.g., the synchrotron polarized intensity map, was smoothed with a Gaussian beam with FWHM selected from $0.5^\circ$--$12^\circ$ in 10 logarithmically spaced intervals.
    \item A threshold, $t_X$, was chosen at each smoothing scale. Pixels exceeding the threshold were masked to form a series of 10 masks for each component map. 
    \item The resulting binary mask was smoothed again with a $5^\circ$ Gaussian beam to remove sharp features and then reset to a binary state with a threshold at 0.25.
    \item The binary masks, at the native \texttt{Commander} resolution of $N_{\rm side}=2048$, were downgraded to $N_{\rm side}=128$, masking any lower-resolution pixel if it contained any flagged $N_{\rm side}=2048$ pixels. 
    \item Any pixel beyond the CLASS survey region or within the bright-source mask, described below, was also masked.
\end{enumerate}
The thresholds, $t_X$, used in step two, are defined as 
\begin{equation}
t_X \propto \sum_{\ell}(2\ell+1)\ell^{-1}b^2_{X, \ell},
\end{equation}
where $b_{X, \ell}$ is the smoothing window function for each respective Gaussian beam used in step one.
This construction corresponds to the pixel variance for a Gaussian random field with a power-law spectrum.
The only free parameter is the overall scaling, which was fine-tuned to visually reject the Galactic features at each scale.

\begin{figure}
    \includegraphics[width=\linewidth]{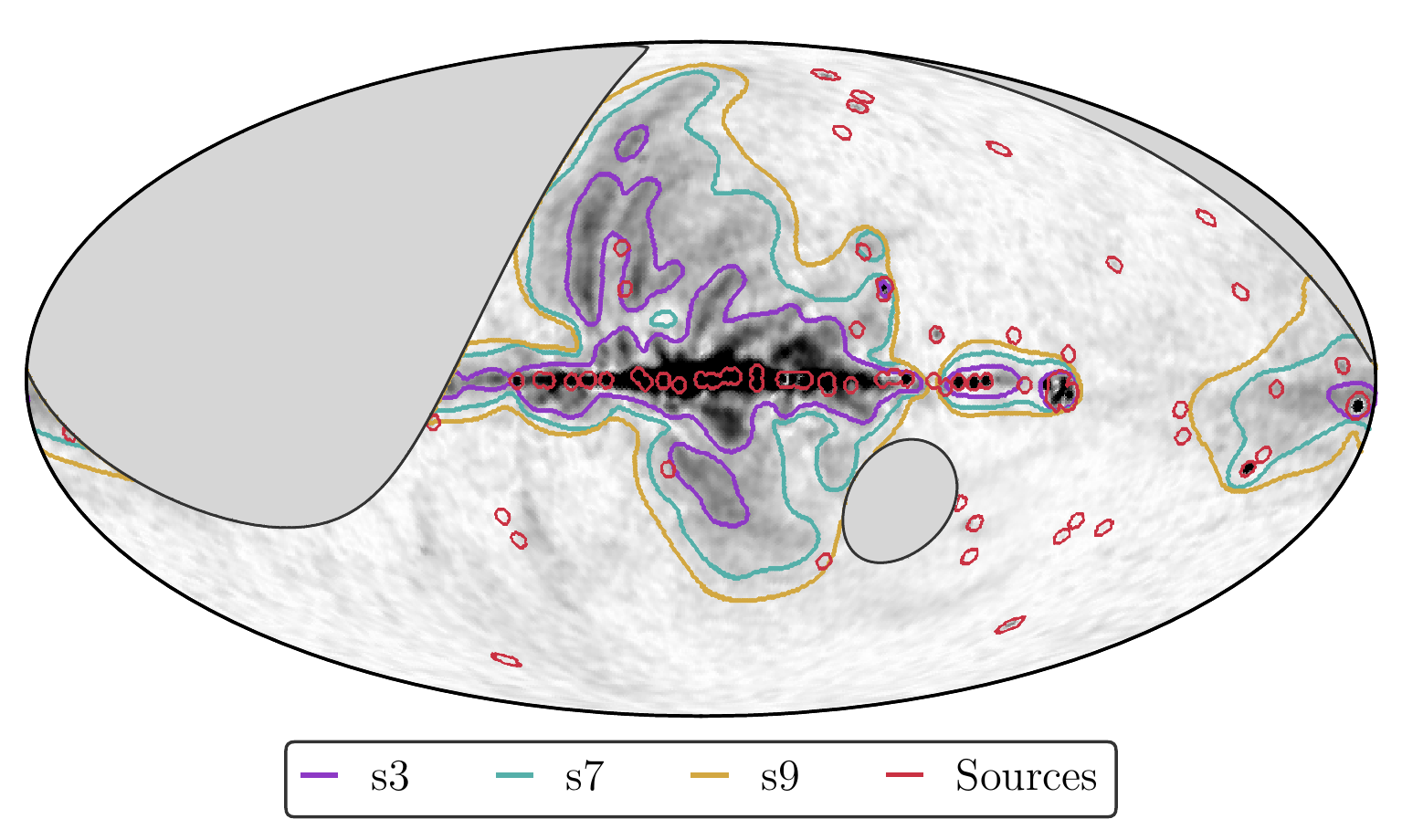}
    \caption{The CLASS 40 $\mathrm{GHz}$ analysis masks \texttt{s3} \fcolorbox{black}{s3_c}{\rule{0pt}{0pt}\rule{0pt}{0pt}}, \texttt{s7} \fcolorbox{black}{s7_c}{\rule{0pt}{0pt}\rule{0pt}{0pt}}, and \texttt{s9} \fcolorbox{black}{s9_c}{\rule{0pt}{0pt}\rule{0pt}{0pt}} are shown.
    The masks share the CLASS decl.~ limitation (gray with black border) and bright-source \mbox{mask \fcolorbox{black}{ps_c}{\rule{0pt}{0pt}\rule{0pt}{0pt}}}. They differ in terms of the masked Galactic synchrotron emission region.
    The background grayscale image shows the polarization intensity of
the \Planck\ component-separated synchrotron map using the \texttt{Commander} algorithm \citep{planck18IV}.
     \label{fig:masks}}
\end{figure}

The Galactic masks for \texttt{s3}, \texttt{s7}, and \texttt{s9} boundaries are shown in purple, teal, and yellow, respectively, in Figure~\ref{fig:masks}.
Because different \texttt{Commander} maps were used in constructing the maps for each component, the contour index is not directly relatable between foreground components.
For instance, \texttt{s1} and \texttt{d1} do not share common properties such as masked sky fraction.

As seen in Figure \ref{fig:masks},  the morphologies of the \texttt{sX} masks follow the Galactic plane with excursions to mid-to-high latitudes in the regions of the polar spurs. However, the synchrotron intensity distribution has a marked low intensity region in the approximate $l\sim(180^\circ,300^\circ)$ range in Galactic latitude. The \texttt{dX} masks tend to be more confined to the plane with less interruption in longitude direction. 

Unless otherwise stated, all masks also account for bright sources based on the \Planck\ 30 and 44 $\mathrm{GHz}$ polarized compact source catalog that fall within the CLASS survey region \citep{planck15XXVI}. 
In cases where sources have angular separation less than $1^\circ$, only the brightest source of the collection was kept. Of the remaining catalog, the brightest 70 sources were masked. The source mask has a $3^\circ$ diameter exclusion region centered at each of the selected points (twice the CLASS beamwidth), and custom regions are tailored to surround a few extended sources. 
Finally, on account of its exceptionally high intensity, a $5.4^\circ$ diameter region is excluded around Tau~A. The source mask is shown in red in Figure \ref{fig:masks}. Nearly all of the masked sources reside within the Galactic plane and would already be covered by the foreground mask. The individual sources within the plane are still useful when considering the bright synchrotron near the plane, but not associated with particular compact regions. All masks were included in the public release hosted on the NASA LAMBDA site.

\subsection{Color correction}
\label{ssec:cc}
\begin{figure}
    \centering
    \includegraphics[width=\linewidth]{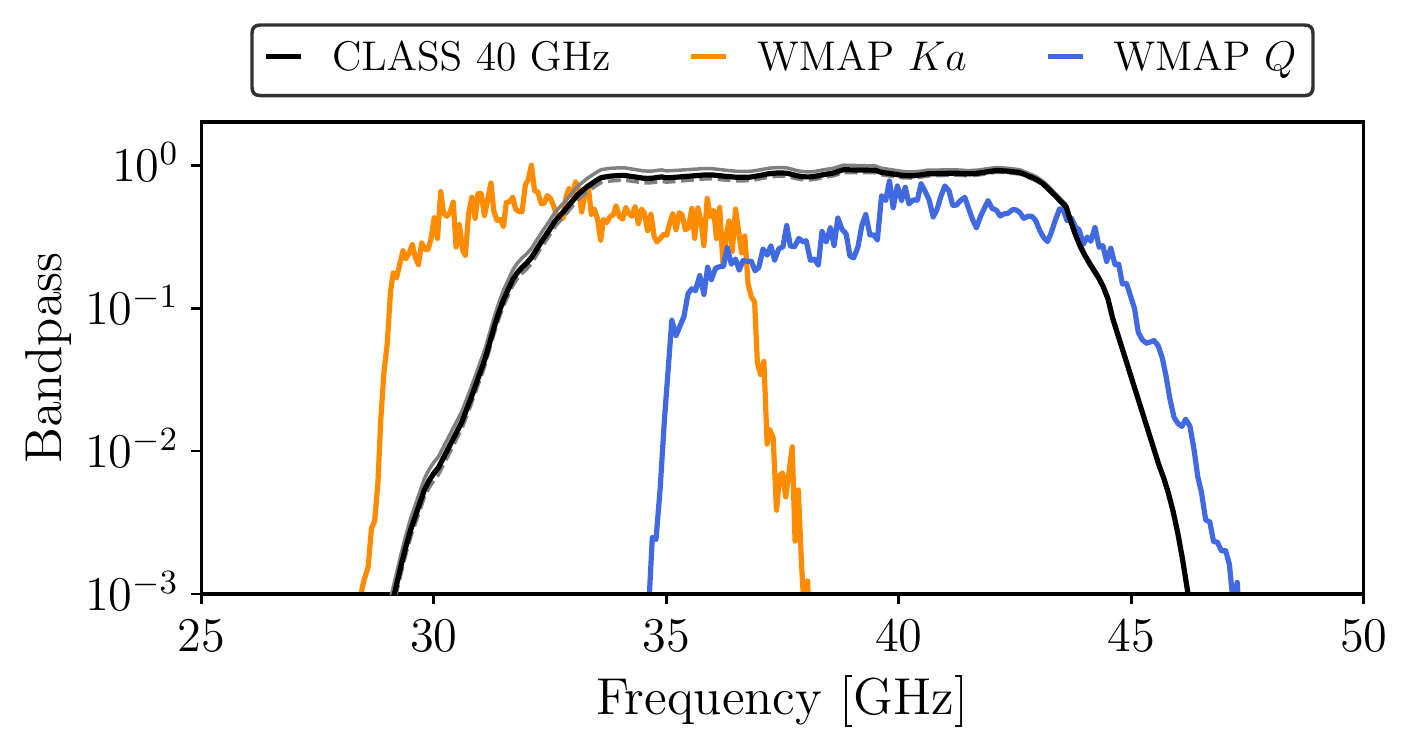}
    \caption{The era-weighted effective bandpass model of CLASS $40\GHz$ is shown. For comparison, the detector average of the WMAP \textit{Ka}-band and WMAP \textit{Q}-band measurements \citep{jarosik03} is also included. The bandpass model for CLASS $40\GHz$ era1 (era2) is shown by solid (dashed) gray line.}
    \label{fig:bp}
\end{figure}

To collect as much signal from the sky as possible, incoherent detectors used for surveys observing near the CMB foreground minimum are typically designed to have fractional bandwidths in the range of 0.2--0.3. Such a broadband detector integrates the incoming signal over the SED of all sources along a line of sight---different SED shapes can produce different detector responses. One method to account for this effect, called color correction, is described in Appendix E of \cite{bennett13}. To facilitate the comparison of CLASS maps with other experiments for astrophysical sources with nonCMB spectra, we provide the color correction factors based on the bandpass presented in \cite{dahal22}.
Following \cite{bennett13}, the measured source temperature in thermodynamic units is
\begin{equation}
\label{eq:cc}
    T_\mathrm{CMB}=\frac{\Delta T_\mathrm{CMB}}{\Delta T_\mathrm{A}}\sum_{i=1}^{3} \omega_i T_\mathrm{A}(\nu_i),
\end{equation}
where $\Delta T_\mathrm{CMB}/\Delta T_\mathrm{A}=1.038$ is the antenna-to-thermodynamic unit conversion factor at 38~$\mathrm{GHz}$ \citep{dahal22}--the approximate band center of the CLASS 40 $\mathrm{GHz}$ channel; $T_\mathrm{A}(\nu_i)$ are the source brightness temperatures evaluated at fixed frequencies, and $\omega_i$ are the weights that encode the shape of the bandpass function. 
The weighting here is the Gaussian quadrature approximation to the actual bandpass integral and is accurate for sources with a smooth spectral shape over the bandpass. 
The frequencies $\nu_i$ are somewhat arbitrary and were chosen to even out the weights $w_i$. Table \ref{tab: colorcorrection} summarizes these quantities for the CLASS 40 $\mathrm{GHz}$ maps.
Since the CLASS effective bandpass has changed in Era~2 due to the deployment of an optical low-pass filter \citep{dahal22}, we use the average bandpass function from Era 1 and Era 2 weighted by their relative statistical weights in the final maps; see Figure \ref{fig:bp}.

\begin{deluxetable}{rrrrrr}
\tablecaption{40~$\mathrm{GHz}$ Interpolation Data for Color Correction Using {Equation \ref{eq:cc}\tablenotemark{a}}.\label{tab: colorcorrection}}
\tablehead{
\colhead{$\nu_1$\tablenotemark{b}} & \colhead{$\nu_2$\tablenotemark{b}} & \colhead{$\nu_3$\tablenotemark{b}} & \colhead{$\omega_1$} & \colhead{$\omega_2$} & \colhead{$\omega_3$}}
\startdata
33.9 & 38.0  & 42.1 & 0.3060 & 0.3448 & 0.3491
\enddata
\tablenotetext{a}{As explained in the text, these values result from a detailed consideration of the detector bandpass, but they should not be interpreted as bandpass values themselves.}
\tablenotetext{b}{Frequencies are given in gigahertz. The choice of frequencies is arbitrary; therefore, the value reported here should not be taken as the central frequencies or the bandwidth of the detector.}
\end{deluxetable}

\subsection{Calibration}
\label{ssec:calib}

The initial calibration of the maps, as described in \citet{Li23}, was based on measurements of the Moon and Jupiter \citep{appel19,xu20,dahal22}. 
Following this, a final adjustment to the absolute polarization calibration was made by comparing the CLASS data to the bright diffuse synchrotron signal in re-observed WMAP \textit{Ka} and \textit{Q} bands. Thus, the CLASS calibration depends upon the low calibration uncertainty from the nearby frequency channels of WMAP and removes any dependence on the initial Moon and Jupiter calibration.

\begin{figure*}
    \centering
    \includegraphics[width=\linewidth]{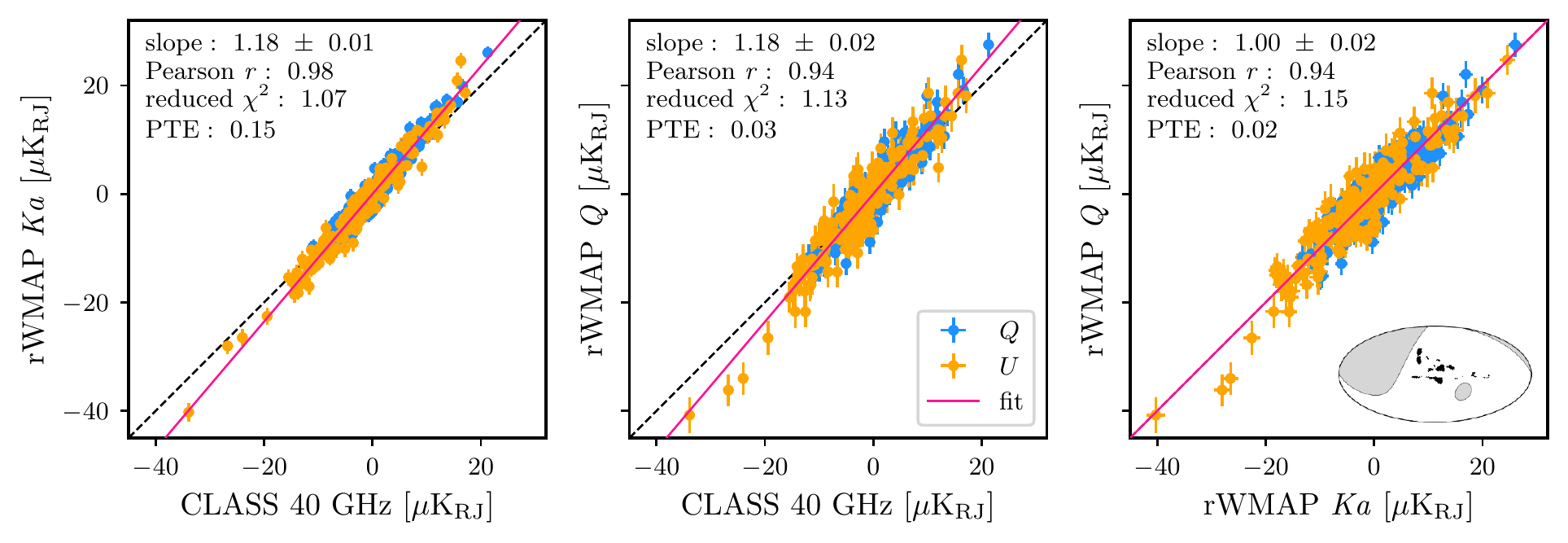}
    \caption{Correlation of Stokes $Q$ (blue) and $U$ (orange) between different bands. 
    All data are converted to antenna temperature at $\nu_\mathrm{eff}=37.56\GHz$.
    \textit{Left}: rWMAP \textit{Ka} to CLASS $40\GHz$. 
    \textit{Middle}: rWMAP \textit{Q} to CLASS $40\GHz$. 
    \textit{Right}: rWMAP \textit{Q} to rWMAP \textit{Ka}.
    The pink solid line is the best fit for the data, and the black dashed line has a slope equal to 1 for reference.
    The top left of each panel shows the best-fit slope, the Pearson correlation coefficient $r$, the reduced $\chi^2$, and the PTE.
    The number of degrees of freedom of the fitting is 417.
    The slightly low PTE values result from the intrinsic scattering of the synchrotron signal within the selected region---bootstrapped uncertainties are $\sim 10\%$ larger than the uncertainties shown here, which rounds up to the same values.
    The inset within the right panel shows the region selected by \texttt{s1-d1} mask (black) at $N_{\rm side}=32$, where the CLASS decl.~ limit is in gray.
    }
    \label{fig:syncalib}
\end{figure*}

The CLASS $40\GHz$, and rWMAP \textit{Ka}/\textit{Q}-band maps were smoothed to a common resolution of $2^\circ$ and downgraded to $N_{\rm side}=32$ ($1.8^\circ$ pixels).
The bright source mask was applied prior to smoothing to diminish the impact from bright-sources, and only $N_{\rm side}=32$ pixels derived from more than 5 of the possible 16 unmasked $N_{\rm side}=128$ subpixels were kept for further usage.
To select regions with bright diffuse synchrotron signal but avoid regions with the largest impact from thermal dust emission, we kept pixels inside \texttt{s1} but outside \texttt{d1}, and we denote this mask as \texttt{s1-d1}; see the inset in Figure \ref{fig:syncalib}.

The uncertainty per $N_{\rm side}=32$  map pixel for each dataset is the standard deviation across 200 different combinations of the corresponding noise simulations and the CMB simulations.
For rWMAP, full-sky noise simulations were generated at $N_{\rm side}=16$ by sampling according to the full noise covariance matrix.\footnote{\texttt{wmap\_band\_quninv\_r4\_9yr\_Ka\_v5.fits}\\\texttt{wmap\_band\_quninv\_r4\_9yr\_Q\_v5.fits}}
White noise realizations were created at $N_{\rm side}=128$ by sampling according to the per-pixel $Q$/$U$ covariance matrices.
The simulations were converted to spherical harmonic coefficients $a_{\ell m}$, and composite spectra were made by concatenating the full-covariance $a_{\ell m}$’s at $\ell\leq48$ with the white noise simulations at $\ell>48$. 
The combined spectra were then converted back to $N_{\rm side}=128$, smoothed to $2^\circ$ FWHM, and downgraded to $N_{\rm side}=32$.
For CLASS, details of the noise simulations can be found in \citetalias{Li23}. These were masked, smoothed, and downgraded in the same way as the CLASS data.
Gaussian random CMB simulations were generated at $N_{\rm side}=128$ according to the \Planck\ best-fit $\Lambda$CDM CMB power spectra\footnote{\texttt{COM\_PowerSpect\_CMB-base-plikHM-TTTEEE-lowl-lowE-\\lensing-minimum-theory\_R3.01.txt}} and then masked and smoothed to $2^\circ$ FWHM, and downgraded to $N_{\rm side}=32$.
To create the effect of reobservation, we applied the CLASS harmonic domain mapping transfer function (Equation \ref{eq:hdtfunc}) to all simulations except the CLASS noise simulations (already filtered).

The calibration factor was obtained via a two-step process. 
First, the spectral index $\beta_s$ was estimated within the region selected by \texttt{s1-d1} as:
\begin{equation}
    \hat\beta_s=\underset{\beta_s}{\mathrm{argmin}}\sum_p\frac{\left[\tilde d_Y(p,\nu,\beta_s)-\tilde d_X(p,\nu,\beta_s)\right]^2}{\tilde\sigma_Y^2(p,\nu,\beta_s)+\tilde\sigma_X^2(p,\nu,\beta_s)},
    \label{eqn:bestbeta}
    \end{equation}
where $d_X$, $\sigma_X$ are the amplitude and uncertainty of Stokes $Q$/$U$ measured in band $X$, and the sum is over all pixels $p$ in \texttt{s1-d1}. 
The notation 
$\tilde x(\nu,\beta_s)$ means that the quantity $x$ has been converted into antenna-temperature units and color corrected to frequency $\nu$ assuming a power-law spectrum with index $\beta_s$.
In this step, we used the rWMAP \textit{Ka} and rWMAP \textit{Q} bands data color corrected to the same frequency $\nu=38\GHz$ using parameters from Table 20 of \citet{bennett13}. 
The best-fit value is $\hat\beta_s=-3.23$.

In the second step, we began by computing the effective frequency for CLASS $40\GHz$ ($\nu_\mathrm{eff}$) where the color correction factor is unity for the best-fit $\hat\beta_s$.
The $\nu_\mathrm{eff}$ can be solved using Equation \ref{eq:cc}, by setting 
$\sum_{i=1}^3\omega_i(\nu_i/\nu_\mathrm{eff})^{\hat{\beta}_s}=1$.
The effective frequency is $\nu_\mathrm{eff}=37.56\GHz$.

We then converted CLASS $40\GHz$, rWMAP \textit{Ka}, and \textit{Q} data into antenna-temperature units and then color corrected the rWMAP \textit{Ka}- and \textit{Q}-band data to $\nu_\mathrm{eff}$. Provided the signal in the selected sky region scales like the assumed power law, all three maps should have the same signal up to an overall calibration factor, $\eta$. To find this factor, the model $y=\eta x$ was fit with $y$ as either rWMAP \textit{Ka}- or \textit{Q}-band and $x$ as CLASS $40\GHz$, taking into consideration the uncertainties on both axes.  As a check of the first step, we also fit rWMAP \textit{Q} against rWMAP \textit{Ka} with the expectation that the slope be consistent with unity.
The correlations between pixels and between the Stokes $Q$/$U$ signal within a pixel were not considered.
The slope, the Pearson correlation coefficient $r$, the reduced $\chi^2$ of the fit, and the PTE values are shown in Figure \ref{fig:syncalib}. In all of the fits, a strong linear relationship is present with $r\ge0.94$, and the reduced $\chi^2$ is close to unity.
As expected, the slope between WMAP \textit{Ka} and \textit{Q} bands is consistent with unity. The calibration factors (slopes) derived from the fits between rWMAP \textit{Ka}/\textit{Q} bands and CLASS $40\GHz$ are $\eta_\mathit{Ka} = 1.18 \pm 0.01$ and $\eta_\mathit{Q} = 1.18\pm0.02$, where the smaller uncertainty from the rWMAP \textit{Ka} band fit is a result of its higher synchrotron signal-to-noise ratio (S/N). The fact that the fitted slopes using the CLASS data match each other well and that from the rWMAP fit is consistent with 1 means that the frequency spectrum of the data we selected can be well explained by a single power law with index $\beta_s=-3.23$.

We express the absolute calibration factor as $\eta=\eta_0\pm\sigma_\eta^\mathrm{(stat)}\pm\sigma_\eta^\mathrm{(sys)}$.
The calibration value $\eta_0=1.18$ is the inverse-variance weighted average of $\eta_{Ka}$ and $\eta_Q$.
The statistical uncertainty $\sigma_\eta^\mathrm{stat}=0.01$ is obtained by propagating the uncertainties on each of the two slopes to the weighted average.
We note that this treatment neglected the correlation caused by the fact that CLASS $40~\mathrm{GHz}$ data were used in both, though the correlation would be minimal as the CLASS $40~\mathrm{GHz}$ data has the smallest uncertainty.
The PTE values of the fitting are slightly low because the intrinsic scattering of the synchrotron signal starts to matter at this relatively high S/N region. 
The uncertainty estimated by bootstrapping is about 10\% larger than the propagated uncertainty, which rounds up to the same value of $0.01$.
The systematic uncertainty $\sigma_\eta^\mathrm{sys}=0.05$ results from the $0.5~\mathrm{GHz}$ uncertainty on the band center of CLASS 40 $\mathrm{GHz}$ \citep{dahal22}, which is the difference in the calibration value when converting rWMAP \textit{Ka} and \textit{Q} bands to antenna temperature at $\nu_\mathrm{eff}\pm0.5~\mathrm{GHz}$.
As a final check, we tried all combinations of \texttt{sX-dY} with \texttt{X,Y} ranging from \texttt{0} to \texttt{9}, and for all cases with $r>0.9$ and PTE of slopes within $(0.01, 0.99)$ (9 out of 100 in total), the calibration values are consistent with $\eta_0=1.18$.

The final calibration factor is $\eta=1.18\pm0.01\mathrm{(stat)}\pm0.05\mathrm{(sys)}$. At this time, it is unclear why this polarization calibration differs from the Moon and Jupiter temperature-based calibrations. 
A portion of this final correction includes accounting for the stable, $\sim3\%$, atmospheric opacity that is not included in the baseline temperature calibration. The uncertainty of the initial Moon and Jupiter based calibration is $\sim2\%$. However the remaining discrepancy is still $\sim14\pm5\%$ and cannot be fully accounted for by the known calibration uncertainties and the atmospheric opacity correction. 

To check that the \texttt{d1} mask provides robust protection against calibration bias from dust contamination, we performed the same procedure using the smaller \texttt{d0} mask and subtracted a dust template formed by scaling rPlanck $353\GHz$--$38\,\mathrm{GHz}$ following a modified blackbody spectrum with $\beta_d=1.53$ and $T_\mathrm{dust}=19.6 \mathrm{K}$. The derived calibration differed by only $0.001$ indicating dust negligibly impacts our calibration.

\begin{figure}
    \centering
    \includegraphics{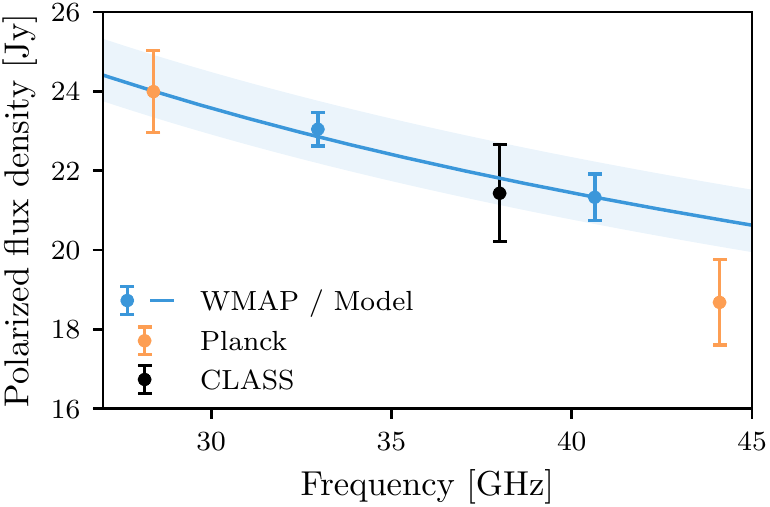}
    \caption{The measured Tau~A polarization flux density of ${21.4\pm1.2\,\mathrm{Jy}}$ at an effective frequency of 38~$\mathrm{GHz}$ is consistent with previous measurements. The uncertainty on the CLASS Tau~A measurement is driven by the uncertainty on the map absolute calibration factor. The polarization flux densities measured by WMAP \citep{weiland11} and \Planck\ \citep{planck15XXVI} are referenced to the year 2019 by correcting for the expected secular decrease in Tau~A flux density\citep{weiland11}. The blue line shows the best-fit WMAP Tau~A polarization flux density model (referenced to 2019) as a function of frequency. The shaded region is the 1$\sigma$ contour of the model’s prediction including spectral and time evolution uncertainty.
    \label{fig:tauA_flux}}
\end{figure}

As a check on the survey-based calibration, we confirmed that the measured polarized flux density of Tau~A, using aperture photometry, is consistent with previous measurements; see Figure \ref{fig:tauA_flux}. We measured the Tau~A flux density by integrating the polarization $P=\sqrt{Q^2+U^2}$ in antenna temperature within a disk of radius 1.85$^\circ$ centered at R.A. = 83.633$^\circ$ and decl.~ = 22.014$^\circ$. The CLASS beam profile suppresses the Tau~A $P$ signal to 0.6\% of the peak value at a radius of 1.85$^\circ$. At this radius, the Tau~A signal amplitude matches the background noise level of the CLASS $P$ map. An estimate of the nearby diffuse Galactic emission was subtracted from the $Q$ and $U$ disk amplitudes to account for the Galactic signal that the large CLASS beam integrates together with Tau~A (see Figure~\ref{fig:plane_comp}). This baseline correction increased the measured flux density by $\sim8$\%. An additional $\sim4$\% flux density was added to account for the beam solid angle outside the disk radius (see the Appendix). The integrated Tau~A antenna temperature (in units of K-sr) was divided by $\Gamma \Omega= c^2/(2 k \nu_\mathrm{eff}^2) $ to obtain the Tau~A flux density in units of janskys (Jy). We subtracted a $0.2$ Jy noise bias from the CLASS Tau~A flux measurement.  The bias was estimated by applying the same algorithm to compute the flux density across the 200 noise only CLASS map simulations.
Finally, we note that the survey maps were produced assuming a VPM transfer function based on the flux density of the sky having a spectral index $\alpha = -1$ as opposed to the known Tau~A index $\alpha=-0.35$ \citep{weiland11}. To correct for this difference, based on simulations testing the impact of this effect, the map-based CLASS Tau~A polarization flux density measurement was divided by 1.04. 
We applied the same aperture photometry method to re-observed WMAP and \Planck\ maps and found Tau~A flux density values consistent with the published values. 
Simulations showed that the CLASS mapping transfer function (Equation~\ref{eq:hdtfunc}) diminishes flux density measurements of unresolved sources (like Tau~A) by $\le1\%$. For simplicity, a correction of this bias was not applied to this Tau~A measurement.

\begin{figure*}
    \centering
    \includegraphics[]{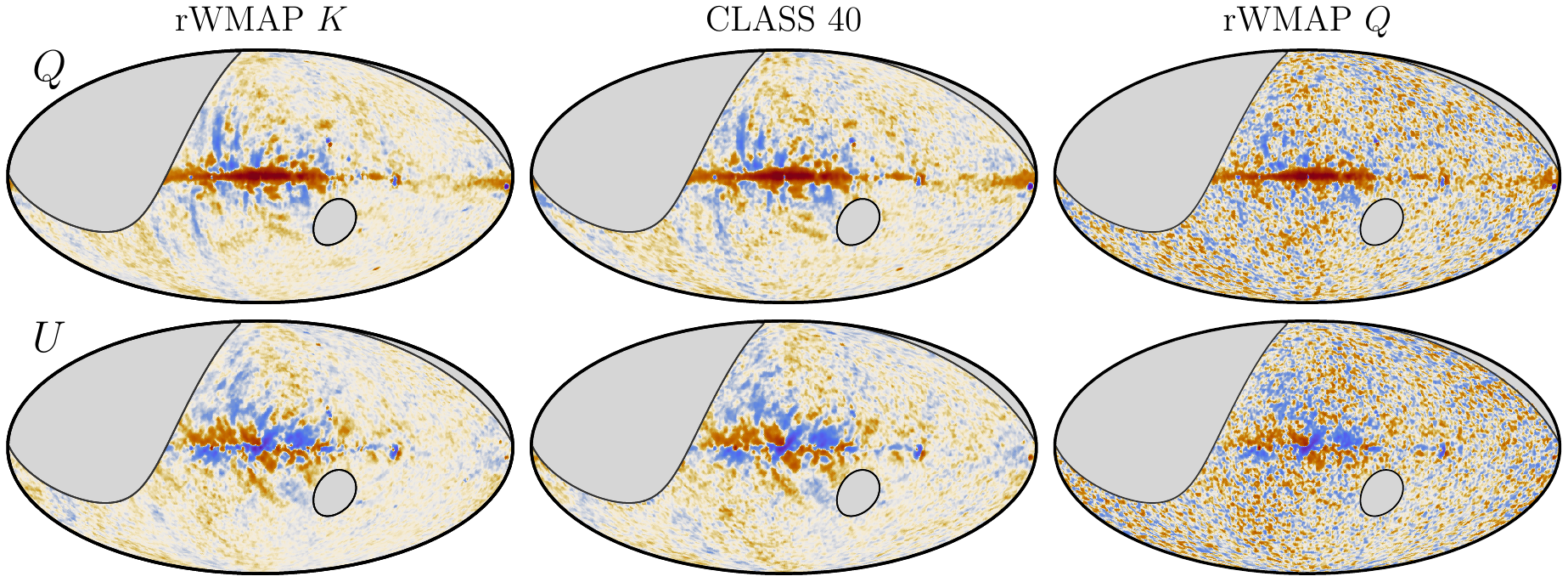}
    \caption{A comparison of the CLASS 40 $\mathrm{GHz}$ survey maps (middle column) to the rWMAP \textit{K}-band (left column) and rWMAP \textit{Q}-band (right column) channels. Stokes $Q$ ($U$) maps are in the top (bottom) row. All bands are smoothed with a 1.5$^\circ$ Gaussian beam for visualization purposes. This causes the signal features to be smoothed to $2.25^\circ$ in the CLASS maps. The noise between the maps can still be visually compared. The synchrotron foreground is much brighter at the lower-frequency (23 $\mathrm{GHz}$) \textit{K}-band, so the maps were scaled (by $0.2$) to roughly match the CLASS amplitude. The agreement between the high signal rWMAP \textit{K}-band map and the CLASS map is striking---as is the sensitivity improvement relative to rWMAP \textit{Q}-band. The CLASS color scale is the same as Figure \ref{fig:quv_maps}.}
    \label{fig:map_comp}
\end{figure*}

\subsection{Comparison with Previous Measurements}
\label{ssec:prev_m}

The CLASS 40 $\mathrm{GHz}$ survey complements the collection of available surveys targeting the synchrotron foreground component of the polarized millimeter sky---with the WMAP \textit{Q}-band \citep[$41\,\mathrm{GHz}$,][]{bennett13} being the closest in frequency and sky footprint to the maps presented in this work. To the extent that the emission is dominated by synchrotron, the SED is well described by a power law in antenna-temperature units with index $\beta \approx -3.1$; though this value is known to vary over the sky \citep{weiland22}, a point we consider in Section \ref{ssec:beta_var}. Taking advantage of this spectral ``lever arm,'' we will also use WMAP \textit{K}-band \citep[$23\,\mathrm{GHz}$,][]{bennett13} data with their high-S/N polarized synchrotron emission to qualitatively validate the CLASS maps in this Section.

Figure \ref{fig:map_comp} shows full-sky comparisons between CLASS polarization maps and re-observed WMAP \textit{K}- and \textit{Q}-band maps. The rWMAP \textit{K}-band maps have been scaled by a factor of 0.2 to approximately reproduce the expected signal level in the CLASS maps. 
No scaling has been applied to rWMAP \textit{Q}-band. 
Note that for the WMAP data, the reobservation process already smooths the map with a $1.5^\circ$ beam---no additional smoothing was performed to those maps. For visualization purposes, the CLASS maps have been smoothed with a $1.5^\circ$ beam. Smoothing the CLASS map has the effect making the noise comparable between the three maps, but the signal in the CLASS maps is smoothed to a slightly lower resolution, $2.25^\circ$---this effect is too small to be visible in the figures. 
There are two main qualitative conclusions to draw from Figure \ref{fig:map_comp}. First, the CLASS maps have significantly less noise than the rWMAP \textit{Q}-band maps. 
This is consistent with angular noise spectra of CLASS being the lowest in the range $10 < \ell < 100$ in \mbox{Figure \ref{fig:noise_comp}}. 
The second is that the synchrotron emission qualitatively matches the rWMAP \textit{K}-band signal on the largest angular scales. 

\begin{figure*}
    \centering
    \includegraphics[]{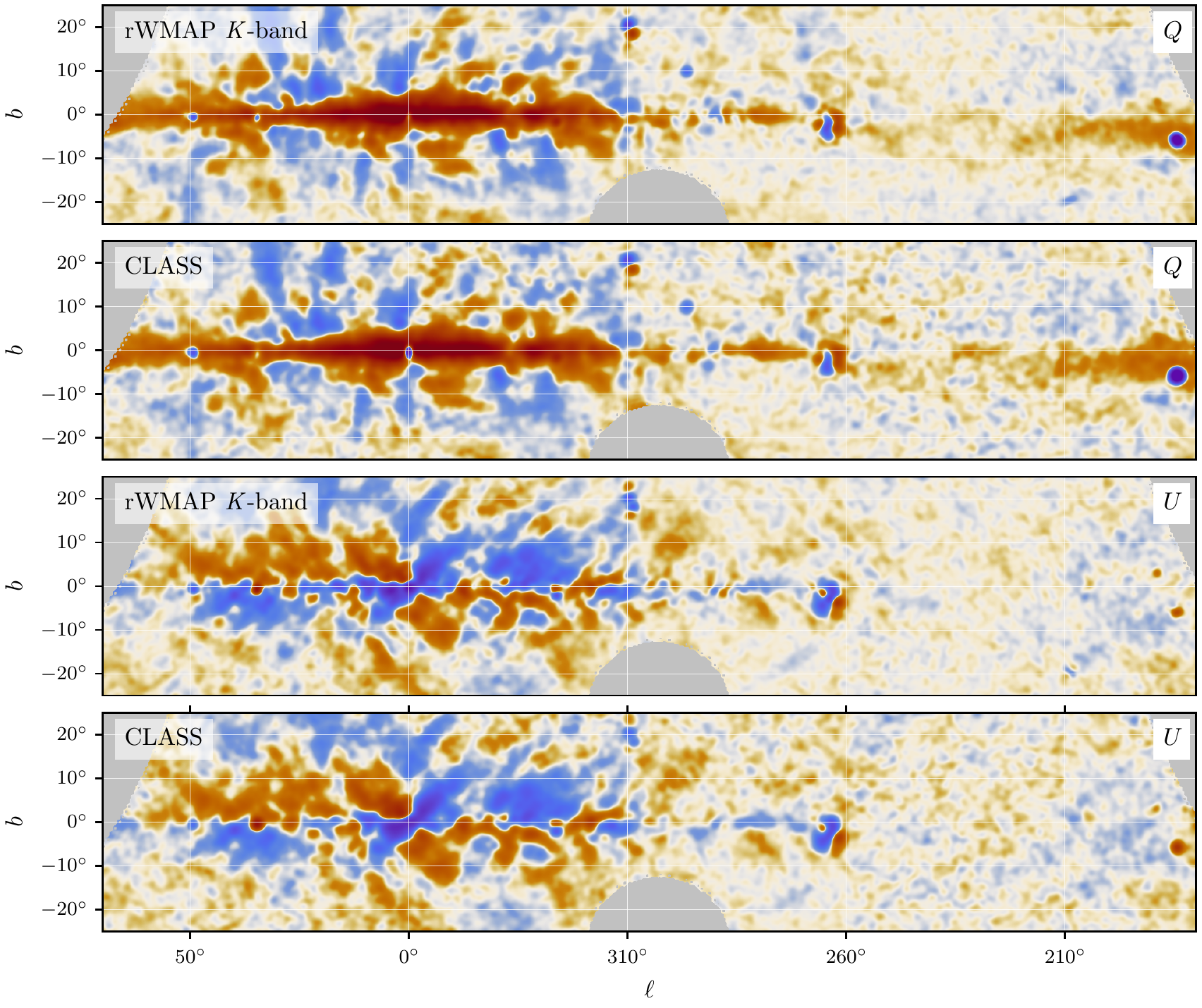}
    \caption{A comparison of the CLASS 40 $\mathrm{GHz}$ survey maps to the re-observed WMAP \textit{K}-band along the Galactic plane. Stokes $Q$ ($U$) maps are in the top (bottom) two panels. The maps have been smoothed with a 1.5$^\circ$ Gaussian beam for visualization purposes. This causes the signal features to be smoothed to $2.25^\circ$ in the CLASS maps. The noise between the maps can still be visually compared. 
    The r\textit{K}-band map has been scaled by a factor of 0.21 to enable direct visual comparison with the same color scale. 
    Given the diversity of astrophysical processes contained within this map and the substantial difference in frequencies at which the signals are measured, the differences near compact sources and in some diffuse regions can be expected. Overall, the CLASS result shows remarkable similarity to rWMAP, despite being from a ground-based observation.} 
    \label{fig:plane_comp}
\end{figure*}

\begin{figure*}
    \centering
    \includegraphics[width=\textwidth]{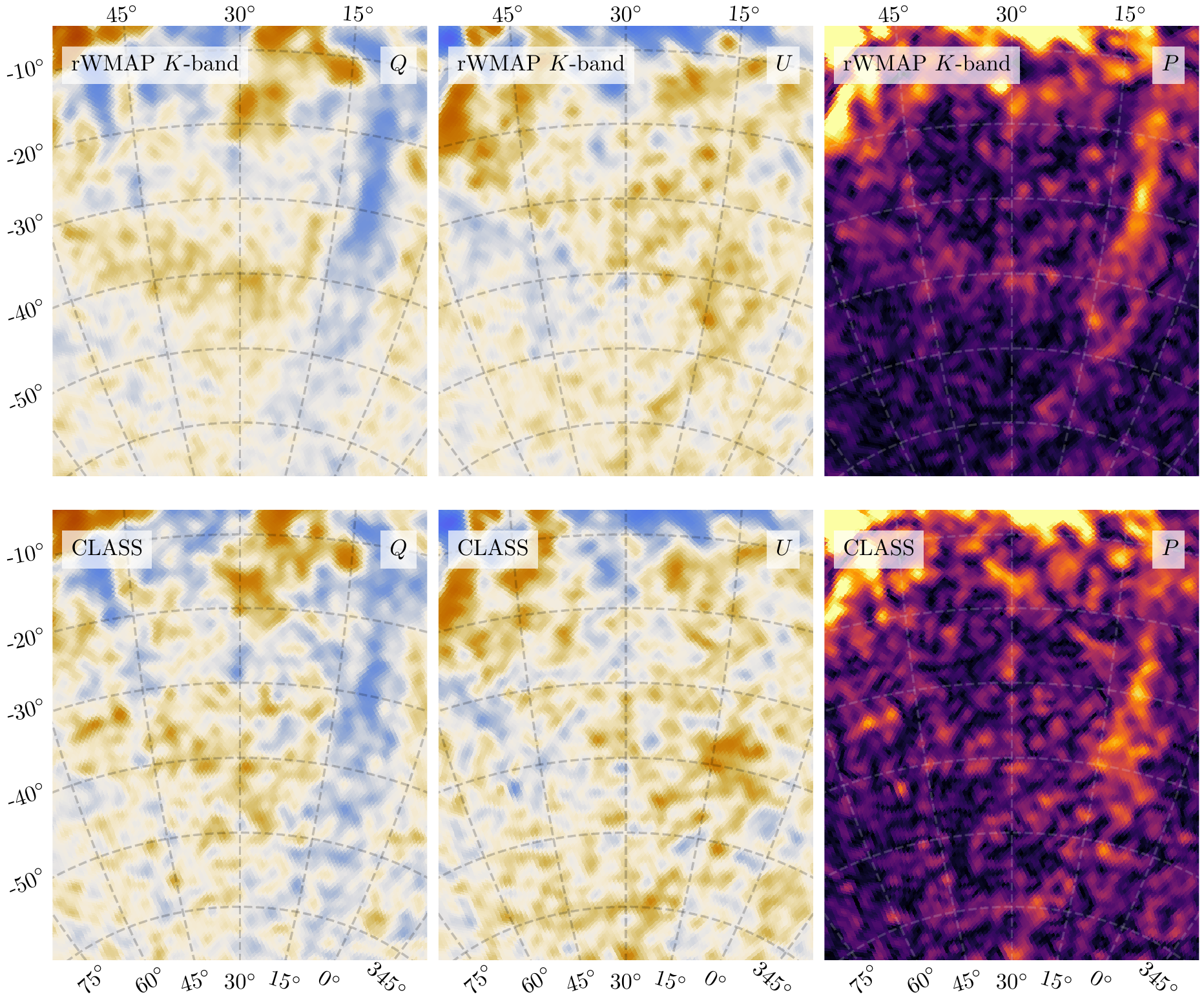}
    \caption{A comparison of the CLASS (bottom row) and rWMAP \textit{K}-band (top row) maps in an area of relatively diffuse emission centered on $(\ell, b) = (22^\circ, -50^\circ)$. The columns, from left to right, show the Stokes $Q$, $U$, and polarization intensity. CLASS has replicated the rWMAP space measurement and with improved sensitivity using a ground-band observatory.  This comparison includes lower surface brightness ($5-10\,\mathrm{\mu K}$ at 40 $\mathrm{GHz}$) features extending $40^\circ$ across the map. }
    \label{fig:southern_comp}
\end{figure*}

Figures \ref{fig:plane_comp} and \ref{fig:southern_comp} provide a zoomed-in comparison between the CLASS polarization maps to the rWMAP \textit{K}-band data - no further image processing is performed. 
Figure \ref{fig:plane_comp} gives a view of the Galactic plane in the range $-25^\circ<b<25^\circ$. 
Overall there is excellent visual agreement between the two data sets. 
Some prominent Galactic features, such as Tau~A at $(\ell,b)\approx(185^\circ,-5^\circ)$ and the Galactic Center, appear discrepant due to frequency spectra that differ significantly from the overall spectra implicit in the applied scaling.
Otherwise, the overwhelming majority of features, including familiar sources such as the Gum Nebula at $(\ell,b)\approx(260^\circ,0^\circ)$ and the Centaurus A radio galaxy at $(\ell,b)\approx(310^\circ,20^\circ)$, match in morphological detail. Figure \ref{fig:southern_comp} is a more detailed look at a subset of diffuse polarized features in the maps. The amplitude of the features is approximately $5-10\,\mathrm{\mu K}$. 
At this brightness contrast, the apparent discrepancies between the maps on the smallest scales are due to the rWMAP \textit{K}-band data having lower noise after being rescaled by the 0.2 factor to match the CLASS synchrotron brightness. On larger scales, we see agreement between CLASS and rWMAP for extended lower-surface brightness features extending $40^\circ$ across the maps.

\section{Galactic synchrotron analysis}
\label{sec:galactic}

Polarized diffuse Galactic emission predominantly arises from two physical mechanisms: synchrotron emission from relativistic electrons spiraling about Galactic magnetic field lines, and thermal emission from dust grains that are preferentially aligned within that field.  
Polarized free--free emission is expected to be negligible and is ignored in this work. Spinning dust, a hypothesized explanation for anomalous microwave emission (AME), is also expected to be negligible  \citep{draine16} in line with current upper limits \citep{beyondplanckXV}. Polarized emission from magnetic dust is also expected to be very faint \citep{hoang16}. Thus, we do not consider AME and magnetic dust in this work.
Emission from these components significantly surpasses 
the polarized CMB signal on large angular scales, and thus accurate removal of the Galactic emission is essential to studies of the CMB polarization.
In general, polarized foreground emission from synchrotron dominates over that of thermal dust at frequencies $\nu \lesssim 60$~$\mathrm{GHz}$, although the exact mixture is dependent on sky position and angular scale \citep{planck18IV, weiland22}.
As the primary component for CLASS 40~$\mathrm{GHz}$,  we examine behavior of synchrotron emission in both harmonic and pixel space.

 In units of antenna-temperature, synchrotron emission is often approximated as a power-law over a limited range of frequencies, with 
 $T\propto \nu^{\bs}$, where $\bs$ is the spectral index.
Previous studies using data from, e.g., WMAP, \Planck\, S-PASS, and QUIJOTE have firmly established spatial variations of $\bs$ with a mean value near $-3.1$  \citep[e.g.,][]{planck18IV, krachmalnicoff18, quijote23, quijote23VIII, fuskeland14, weiland22}, though the degree to which $\bs$ varies has not yet been well characterized over the full sky.

In this Section, we consider how the CLASS measurement contributes to furthering the characterization of the synchrotron foreground component.

\subsection{Synchrotron Power Spectra}
\label{ssec:sync}
The distribution of synchrotron radiation is highly anisotropic---emission is dominated by structures within the Galactic plane---and non-Gaussian, and therefore suffers information loss when compressed to a power spectrum representation. 
Nevertheless, the power spectrum can be computed and formally interpreted as a representative measurement of how the angular power varies within the region of interest, which, for CMB studies, generally excludes the brightest emission from the plane and other Galactic features that are the most anisotropic and non-Gaussian. 

When evaluating properties of the synchrotron signal, this analysis used only the pseudo-$C_\mathrm{\ell}$ estimator \citep[\texttt{PolSpice},][]{polspice} to estimate the cross power spectrum over the $\ell$-range (5, 125). 
While less optimal than a quadratic estimator \citep[e.g., xQML,][]{xQML} for $\ell < 20$, there are fundamental difficulties in combining the two estimators in a single spectrum.
For example, a low-resolution mask appropriate for xQML evaluation that is exactly upgraded to the full map resolution has sharp edges that cause excess mode mixing in the \texttt{PolSpice} spectrum estimation. 
Normally this pseudo-$C_\ell$ issue is ameliorated by apodizing the mask edges---fundamentally decreasing the power relative to the low-resolution map version. 
The estimator could adjust for the mask distinction only if the underlying field were isotropic and purely Gaussian. 
Rather than compromising the consistency of the measured power between the estimators, we simplify this analysis by using only the single \texttt{PolSpice} estimator at the expense of less optimal uncertainties for the first two band powers.

\begin{figure*}
 \label{fig:sync_spec}
    \centering
    \includegraphics[width=\textwidth]{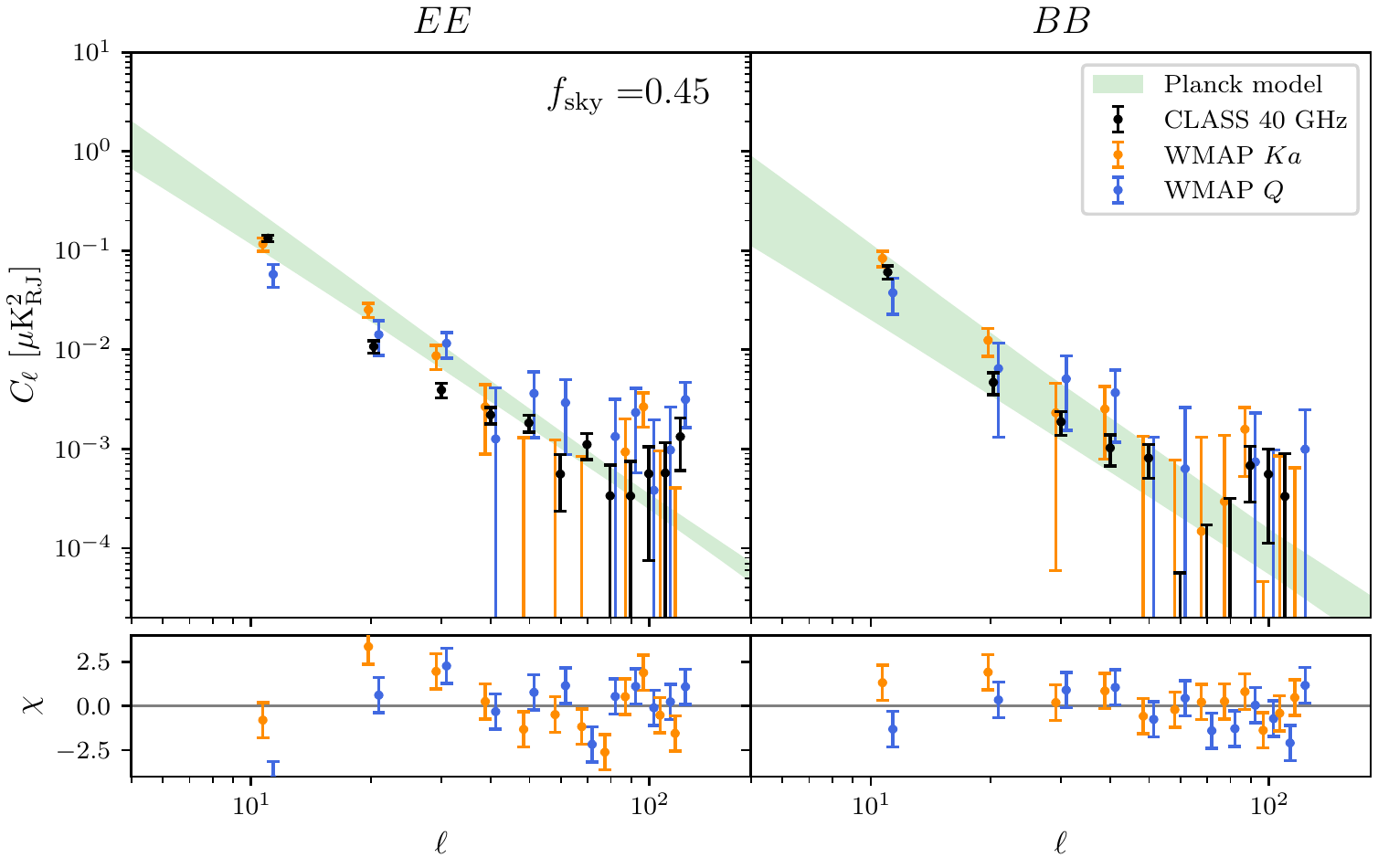}
    \caption{\emph{Top panels:} the angular cross power spectra of the diffuse synchrotron component of the CLASS 40 $\mathrm{GHz}$ data are shown in units of $\mathrm{\mu K}^2$ antenna temperature outside of the \texttt{s9} mask. For comparison, the synchrotron component of the WMAP \textit{Q}-band and WMAP \textit{Ka}-band  \citep{bennett13} cross-spectra between the even and odd years are also shown. Each spectrum has been color corrected to 38 $\mathrm{GHz}$ assuming a fixed synchrotron spectral index of $\bs = -3.1$. Error bars are the binned noise of each measurement and do not include sample variance. The shaded swaths are bounds of simulations of synchrotron $EE$ and $BB$ models from the \Planck\ \texttt{Commander} component analysis \citep{planck18IV}, which has been referenced to $38\,\mathrm{GHz}$ also assuming $\bs=-3.1$. The \Planck\ best-fit CMB $EE$ angular power spectrum \citep{planck18VI} was subtracted from the $EE$ data sets as plotted to isolate the synchrotron components. \emph{Bottom panels:} the uncertainty-normalized difference between the CLASS 40 and WMAP-\textit{Ka}/\textit{Q} band. }
\end{figure*}

The statistical properties of the CLASS 40 $\mathrm{GHz}$ maps were summarized by estimating the angular power spectrum with the \texttt{s9} mask applied, as described in Section \ref{ssec:masks}. 
Following the cross-spectrum estimation approach described in Section \ref{sec:spec_est}, the $EE$ and $BB$ power spectra are shown in Figure \ref{fig:sync_spec}. For comparison, the cross-spectrum for the WMAP \textit{Q}- and \textit{Ka}-bands within the same sky region is also shown. The spectra are in antenna-temperature units assuming a constant spectral index $\bs = -3.1$ and color corrected to a common frequency of 38 $\mathrm{GHz}$ using the factors in Table \ref{tab: colorcorrection} for CLASS and the analogous table in \cite{bennett13}, Appendix E, for WMAP.
To isolate the synchrotron component, the \Planck\ best-fit CMB angular power spectrum \citep{planck18VI} was binned and directly subtracted from the $EE$ spectra prior to color correction. 
The resulting uncertainty is estimated using the respective noise simulations described in Section \ref{sec:spec_est}. 
For CLASS, noise simulations for the base data split were randomly paired to form an ensemble of 10,000 CLASS noise cross-spectra simulations---individual split simulations were allowed to be reused, but no pair was duplicated. The WMAP \textit{Q}/\textit{Ka}-band noise cross-spectra was estimated by generating 500 WMAP cross-spectra simulations with independent noise realization for the even/odd year split as described in Section \ref{sec:spec_est}. The uncertainties on the spectrum are the square root of the diagonal of the inferred binned covariance matrix. 
The measured CLASS spectra are in reasonable agreement with the WMAP results within the same sky region. 
Differences between the spectra might result from the bandpass differences between the two channels---recall the WMAP \textit{Q}-band approximately includes 38--46 $\mathrm{GHz}$ \citep[8 $\mathrm{GHz}$ bandwidth,][]{page03optics}, the WMAP \textit{Ka}-band is 28--37 $\mathrm{GHz}$ \citep[6.9 $\mathrm{GHz}$ bandwidth,][]{page03optics}, while the CLASS 40 $\mathrm{GHz}$ channel uses a somewhat larger 32.8--43.6 $\mathrm{GHz}$ (10.9 $\mathrm{GHz}$ bandwidth) range \citep{dahal22}. See also Figure \ref{fig:bp}. Given the anisotropic nature of the foreground signal, the differences between the map weights of each experiment could manifest as somewhat different resulting spectra as well. 

The measured spectra follow the expected overall trend found in the \texttt{Commander} synchrotron model \citep{planck18IV} which is shown as the shaded band in Figure \ref{fig:sync_spec}. 
These spectra indicate that diffuse polarized synchrotron emission still dominates the polarized CMB signal outside the \texttt{s9} mask. The new CLASS measurement is seen to improve upon the measured spectra from the nearest frequency space-based measurement. Since the \texttt{s9} mask was designed to block a large fraction of the anisotropic synchrotron power that was included in the region described by the \Planck\ model, it is expected to find the CLASS and WMAP spectra to fall somewhat lower than the model.

Visually, there is a suggestion that the angular power spectrum is flattening, especially for $\ell \ge 50$ in the WMAP \textit{Q}-band spectra. While this is consistent with the WMAP \textit{Ka} and CLASS measurements, we do not attempt to constrain a flattened component using the low-S/N measurements in this region. 
As a check on this trend, we extended the WMAP cross-spectra to higher multipoles, but the signal is too faint for a clear indication of a flattened region. 
Since point sources can contribute a flat spectral component (e.g., \citealt{wright09, krachmalnicoff18}),
we also tried masking the remaining \emph{Planck} identified 30 and 44 GHz polarized point sources. This had a negligible impact on the spectra.
Future measurements with higher sensitivity will help improve our understanding of this portion of the spectra.

\subsection{Synchrotron SED}
\label{ssec:sed}

Within the frequency range from a few gigahertz up to ~100 GHz, the SED of the diffuse synchrotron radiation has been measured to follow a power law with index $\beta_s \approx -3$ with small differences depending upon the sky region and angular scale measured, see for example \cite{krachmalnicoff18, 2022JCAP...04..003M, planck18IV, choi&page15}. In this Section we explore the inclusion of the CLASS data to those publicly available data dominated by synchrotron, but not significantly impacted by polarization decorrelation and rotation from Faraday rotation effects \citep{Vidal15}. To this end, we considered data from WMAP \textit{K}-, \textit{Ka}-, and \textit{Q}-bands, \Planck\ 30 $\mathrm{GHz}$, and the CLASS 40 $\mathrm{GHz}$ channel. 

In the sky regions where polarized synchrotron radiation is faint, the harmonic domain description of the data can be used to quantify the brightness for a given channel.
Following the conventions from \cite{krachmalnicoff18} and \cite{quijote23}, we modeled the diffuse synchrotron emission as
\begin{equation}
\label{eq:ang_pl}
    C^{XY}_\ell = A^{XY}_{40} \left( \frac{\ell}{40} \right)^{\alpha_{XY}}, 
\end{equation}
where $\{X,Y\}$ indicated the data sets being crossed. Provided the index $\alpha$ is consistent between different channels, the amplitude, $A_{40}$, measures the synchrotron SED at the effective frequency $\nu_\mathrm{eff} = \sqrt{\nu_X \nu_Y}$. The frequencies $\nu_{X(Y)}$ are the effective frequency for each individual channel accounting for the finite bandwidth of the detectors and the shape of the input spectrum. 

The SEDs for each of the $E$ and $B$ components are assumed to follow a simple model
\begin{equation}
\label{eq:sed_pl}
    D^{\nu_1 \times \nu_2} = A_0 \left(\frac{\nu_1\nu_2}{\nu_0^2} \right)^{\beta_s}, 
\end{equation}
where we have adopted the pivot frequency $\nu_0 = 30$ $\mathrm{GHz}$ as it is near the geometric mean of the range of effective frequencies considered.

Measurement of the SED followed two methods. In the first method, the power law in Equation \ref{eq:ang_pl} was fit to each cross-spectra $C^{XY}$ for each $EE$ and $BB$ pair. As explained in Section \ref{sec:spec_est}, power spectra covariance are estimated using simulations. For spectra not including CLASS data, each map simulation comprised the noise sampled as described in Section \ref{sec:spec_est}, the CMB realizations following the \Planck\ best-fit \lcdm model \citep{planck18VI}, and the Gaussian realizations of a fiducial synchrotron model.\footnote{The exact parameters of this model would affect the fitting result through its contribution to the sample variance. We have iterated this procedure with the best-fit results from the data and find it converged within five steps.} 
Power spectra were estimated for these map simulations in the same way as the data; therefore, their covariance naturally incorporates the statistical variance from the map noise, the signal sample variance, and the inter-bin covariance from the estimator due to mode coupling.

For CLASS data, the noise simulations provided by the pipeline \citepalias{Li23} represent the map noise after filtering. 
To correct for the mapping transfer function, the simulated cross power spectra were computed separately for the signal, noise, and cross terms---taking advantage of the associative property of the two-point statistic. 
The harmonic transfer function correction was only applied to terms where the CLASS noise simulations were involved: there is no need for signal term corrections as the filtering was not included in the signal simulations. 
For cross-spectrum terms that included map pairs that both required transfer function correction, the spectrum was corrected with $F^{-1}_\mathrm{\ell\ell'}$, and for terms with only one CLASS noise map, the correction was done with $F^{-1/2}_\mathrm{\ell\ell'}$.

\begin{figure*}
    \centering
    \includegraphics[width=\textwidth]{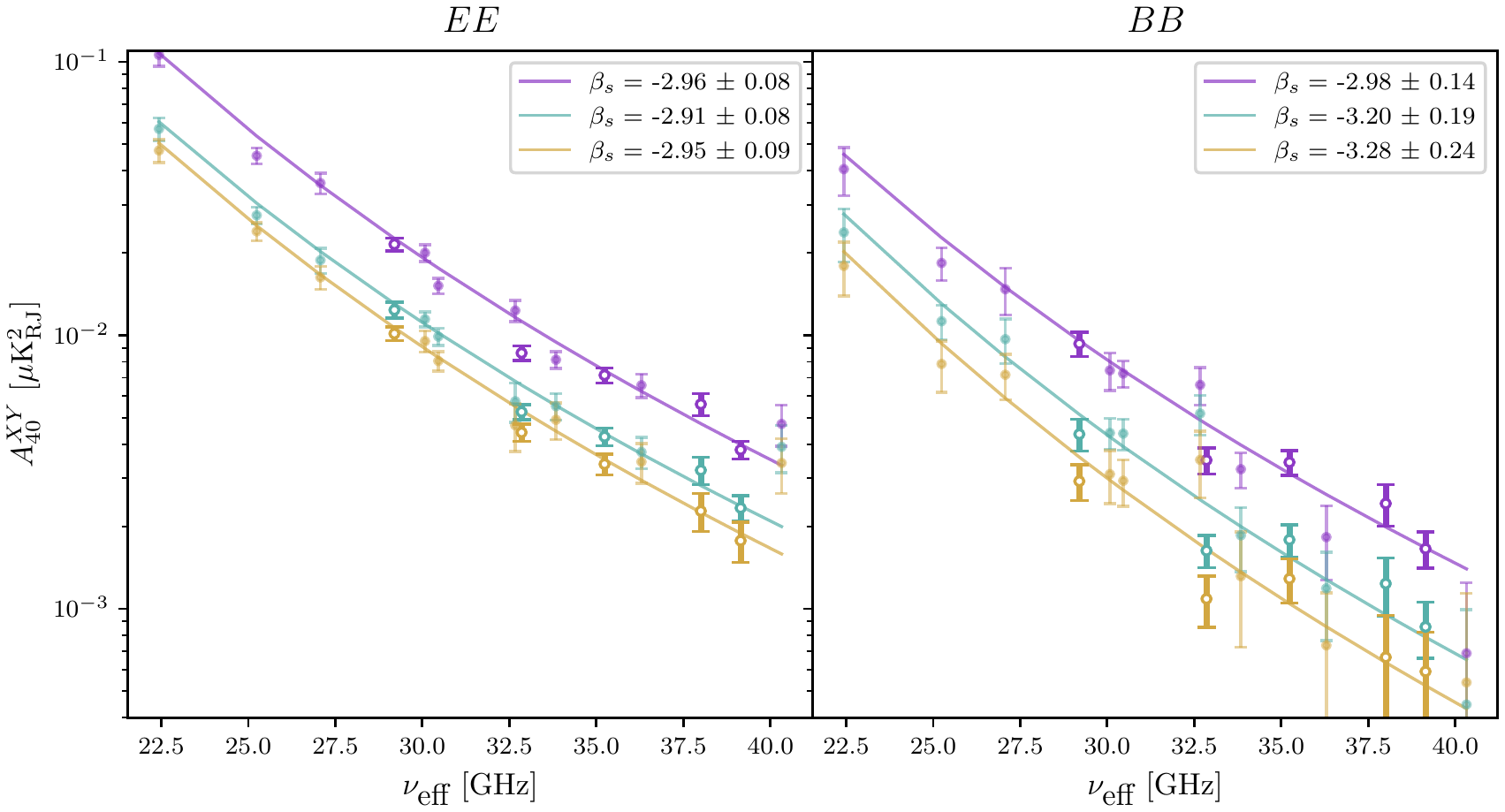}
    \caption{Measurement of the synchrotron SED for the $EE$ ($BB$) modes is shown on the left (right). Colors indicate the mask applied when evaluating the signal with \texttt{s3} \cbox{s3_c}, \texttt{s7} \cbox{s7_c}, and \texttt{s9} \cbox{s9_c} masks. The points and error bars are measured by fitting a power-law model to each individual angular cross-spectrum.  Open circle, heavier lined points indicate cross-spectra that include the CLASS data. The channel, best-fit amplitude, and uncertainties are collected in Table \ref{tab:spec_fits}. The curve, and legend values, are from the joint fit of all spectra together assuming the amplitudes follow the power-law SED in Equation \ref{eq:sed_pl}.
    \label{fig:sed}}
\end{figure*}

The resulting $A_{40}$ amplitudes and uncertainties for the \texttt{s3}, \texttt{s7}, and \texttt{s9} masks are collected in Table \ref{tab:spec_fits} and shown in Figure \ref{fig:sed}. The cross-spectra including CLASS data are indicated with the open circled, relatively larger, and bolder colored points in the plot. The power-law index from Equation \ref{eq:ang_pl} stays roughly constant for all cross-spectra and masks. The index for the \texttt{s9} mask, for example, is $\alpha^{EE} = -2.42 \pm 0.14$ and $\alpha^{BB} = -2.84 \pm 0.25$ where we report the mean value and median error taken over the 14 cross power spectra. 

\begin{deluxetable*}{c|c|ccc|ccc}
\tablecaption{Synchrotron power spectra amplitude values, $A_{40}$ values for cross-spectra synchrotron SED.  \label{tab:spec_fits}}
\tablehead{
\multicolumn{2}{c}{ } & \multicolumn{3}{|c|}{$EE$ ($10^2 \times \mathrm{\mu K}^2$)} & \multicolumn{3}{|c}{$BB$ ($10^2 \times \mathrm{\mu K}^2$)} \\
 \hline
\colhead{Channel\tablenotemark{a}}& \colhead{$\nu_\mathrm {eff}$ ($\mathrm{GHz}$)} &  \colhead{\texttt{s3}} & \colhead{\texttt{s7}} & \colhead{\texttt{s9}} & \colhead{\texttt{s3}} & \colhead{\texttt{s7}} & \colhead{\texttt{s9}}
}
\startdata
\textit{K}$\times$\textit{K} & 22.4 & 10.60 $\pm$ 0.97 & 5.69 $\pm$ 0.55 & 4.74 $\pm$ 0.46 & 4.05 $\pm$ 0.81 & 2.38 $\pm$ 0.52 & 1.79 $\pm$ 0.40 \\
\textit{K}$\times$\textit{30} & 25.2 & 4.54 $\pm$ 0.31 & 2.75 $\pm$ 0.19 & 2.40 $\pm$ 0.18 & 1.84 $\pm$ 0.25 & 1.13 $\pm$ 0.16 & 0.79 $\pm$ 0.17 \\
\textit{K}$\times$\textit{Ka} & 27.1 & 3.60 $\pm$ 0.31 & 1.88 $\pm$ 0.20 & 1.63 $\pm$ 0.16 & 1.47 $\pm$ 0.29 & 0.97 $\pm$ 0.18 & 0.72 $\pm$ 0.14 \\
\textit{K}$\times$\textit{C} & 29.2 & 2.15 $\pm$ 0.12 & 1.24 $\pm$ 0.08 & 1.01 $\pm$ 0.06 & 0.93 $\pm$ 0.09 & 0.44 $\pm$ 0.06 & 0.29 $\pm$ 0.04 \\
\textit{K}$\times$\textit{Q} & 30.1 & 2.00 $\pm$ 0.14 & 1.15 $\pm$ 0.07 & 0.95 $\pm$ 0.09 & 0.74 $\pm$ 0.12 & 0.44 $\pm$ 0.06 & 0.31 $\pm$ 0.07 \\
\textit{30}$\times$\textit{Ka} & 30.5 & 1.52 $\pm$ 0.10 & 0.99 $\pm$ 0.07 & 0.81 $\pm$ 0.07 & 0.72 $\pm$ 0.08 & 0.44 $\pm$ 0.06 & 0.29 $\pm$ 0.06 \\
\textit{Ka}$\times$\textit{Ka} & 32.7 & 1.23 $\pm$ 0.11 & 0.57 $\pm$ 0.09 & 0.47 $\pm$ 0.09 & 0.66 $\pm$ 0.10 & 0.52 $\pm$ 0.09 & 0.35 $\pm$ 0.10 \\
\textit{30}$\times$\textit{C} & 32.9 & 0.86 $\pm$ 0.05 & 0.52 $\pm$ 0.03 & 0.44 $\pm$ 0.03 & 0.35 $\pm$ 0.04 & 0.16 $\pm$ 0.02 & 0.11 $\pm$ 0.02 \\
\textit{30}$\times$\textit{Q} & 33.8 & 0.81 $\pm$ 0.06 & 0.55 $\pm$ 0.06 & 0.49 $\pm$ 0.08 & 0.32 $\pm$ 0.05 & 0.19 $\pm$ 0.05 & 0.13 $\pm$ 0.06 \\
\textit{Ka}$\times$\textit{C} & 35.2 & 0.71 $\pm$ 0.05 & 0.43 $\pm$ 0.03 & 0.34 $\pm$ 0.03 & 0.34 $\pm$ 0.04 & 0.18 $\pm$ 0.02 & 0.13 $\pm$ 0.02 \\
\textit{Ka}$\times$\textit{Q} & 36.3 & 0.66 $\pm$ 0.07 & 0.37 $\pm$ 0.05 & 0.34 $\pm$ 0.06 & 0.18 $\pm$ 0.06 & 0.12 $\pm$ 0.04 & 0.07 $\pm$ 0.04 \\
\textit{C}$\times$\textit{C} & 38.0 & 0.56 $\pm$ 0.05 & 0.32 $\pm$ 0.04 & 0.23 $\pm$ 0.04 & 0.24 $\pm$ 0.04 & 0.12 $\pm$ 0.03 & 0.07 $\pm$ 0.03 \\
\textit{C}$\times$\textit{Q} & 39.1 & 0.38 $\pm$ 0.03 & 0.23 $\pm$ 0.03 & 0.18 $\pm$ 0.03 & 0.17 $\pm$ 0.03 & 0.09 $\pm$ 0.02 & 0.06 $\pm$ 0.02 \\
\textit{Q}$\times$\textit{Q} & 40.3 & 0.47 $\pm$ 0.08 & 0.39 $\pm$ 0.08 & 0.34 $\pm$ 0.08 & 0.07 $\pm$ 0.06 & 0.04 $\pm$ 0.05 & 0.05 $\pm$ 0.06 \\
\enddata
\tablenotetext{a}{The cross-spectrum channel is indicated using an abbreviation of the channels involved. The abbreviations $K$, $Ka$, $Q$ are used for the WMAP bands, the \Planck\ 30 GHz channel is denoted 30, and the CLASS 40 $\mathrm{GHz}$ channel is denoted $C$.}
\end{deluxetable*}

The second method was to fit jointly Equation \ref{eq:sed_pl} to the full set of cross power spectra. 
In this case, the power law from Equation \ref{eq:ang_pl} is still fit to the angular power spectrum of each frequency, but now the amplitudes were constrained to follow Equation \ref{eq:sed_pl}. 
The resulting best-fit power laws for the \texttt{s3}, \texttt{s7}, and \texttt{s9} masks are shown in Figure \ref{fig:sed} and show good agreement with the single-frequency measurements. 
For this fit, the $EE$ and $BB$ spectra were fit separately. 
Furthermore, the $168\times168$ covariance matrix (12 bins $\times$ 14 cross-spectra) was assumed to be block diagonal with only the internal cross-spectra $12\times12$ sub-block considered to be nonzero. The best-fit power-law index, $\beta_s$, is shown in the Figure legend. The ratio of the best-fit $BB/EE$ amplitudes is $0.42 \pm 0.02$, $0.39 \pm 0.02$, and $0.33 \pm 0.02$
for the \texttt{s3}, \texttt{s7}, and \texttt{s9} masks. Since the shapes of the $EE$ and $BB$ SED profiles differ, this amplitude ratio is frequency dependent. It still summarizes the relative contributing power in this frequency range critical for CMB component separation efforts. 

In this study, we have not investigated the possible $\beta_s$ variation with angular scale as has been hinted at by other investigations \citep{choi&page15, krachmalnicoff18}, but that topic will be interesting to pursue with future work. Here we are focusing on the overall consistency of the CLASS data with previous satellite measurements and the improved characterization of the large-scale synchrotron properties near 40 $\mathrm{GHz}$. 

\subsection{Pixel-based Spectral Index Estimation}
\label{ssec:beta_var}
Beyond improving knowledge of the Galactic processes involved in generating the polarized synchrotron emission, improved measurements of the spatial variation of the spectral index may be essential for high-precision component separation---especially for the detection of the primordial $B$-modes. 
For example, \citet{2019PhRvD..99d3529E} found that foreground cleaning with a single spectral index could result in spurious detection of the $r$ of the order of 0.01.

In this Section, we present a polarized synchrotron spectral index ($\beta_s$) map made by analyzing the CLASS 40 $\mathrm{GHz}$ data together with the rWMAP $K$-band data.
The $K$-band data were chosen due to their high-S/N measurement of synchrotron emission and lack of significant Faraday rotation and depolarization effects. Using the \emph{re-observed} WMAP data mitigates the bias on $\beta_s$ that would result from comparing the raw (unbiased) WMAP data to the filtered CLASS data. We investigate potential residual bias from the reobservation processing below.

The bright-source mask, described in Section \ref{ssec:masks}, is applied at $N_{\rm side}=128$ to avoid bias on $\beta_s$ from the different physical conditions within those regions.
To mitigate the impact of polarized thermal dust emission in the CLASS ${40\,\mathrm{GHz}}$ maps, we scaled and subtracted the rPlanck 353 $\mathrm{GHz}$ data assuming a modified blackbody spectrum with a uniform $\beta_d=1.53$ and $T_d=19.6~\mathrm{K}$ across the sky \citep{planck18IV}. Later in this Section, we test the impact of this dust model assumption. Following the process used in Section \ref{ssec:calib}, the masked and dust subtracted maps were smoothed and downgraded to $N_{\rm side}=32$ ($1.8^\circ$ pixels).

\begin{deluxetable}{lccccc}
\tablecaption{Frequencies, color correction, and thermodynamic-to-antenna unit conversion ($\Delta T_\mathrm{A}/\Delta T_\mathrm{CMB}$) factors. \label{tab:cmb2ant}}
\tablehead{
\colhead{} & \colhead{\hspace{.5cm}$\nu_K^{s}$}\hspace{.5cm} & \colhead{\hspace{.5cm}$\nu_\mathrm {40}^{s}$}\hspace{.5cm} & \colhead{\hspace{.5cm}$\nu_\mathrm{353}^{d}$~~~~~~}}
\startdata
Reference Freq. ($\mathrm{GHz}$) & 22.8 & 38 & 353\\
Color correction & 0.9496 & 0.9663 & 0.9036 \\
$\Delta T_\mathrm{A}/\Delta T_\mathrm{CMB}$ & 0.9867 & 0.9634 & 0.0773
\enddata
\tablecomments{From left to right: frequency for synchrotron at WMAP \textit{K}-band $\nu_K^s$, synchrotron at CLASS 40 $\mathrm{GHz}$ $\nu_\mathrm{40}^s$ and thermal dust at \Planck\ 353 $\mathrm{GHz}$ $\nu_{353}^d$. 
We used Table 20 in \citet{bennett13} to compute the factors for WMAP, Table \ref{tab: colorcorrection} for CLASS and 
the public code \texttt{fastcc} \citep{2022RNAAS...6..252P} to compute the color correction factor for \Planck.
When computing the color correction factors, $\beta_s=-3.1$ is assumed for synchrotron, and $\beta_d=1.53$, $T_d=19.6~\mathrm{K}$ for thermal dust.
}
\end{deluxetable}

The maps of $\beta_s$ were computed in the Galactic coordinate system at resolution $N_{\rm side}=8$ ($7.3^\circ$ pixels) following the linear correlation method \citep{fuskeland14,fuskeland21,weiland22}.
Within each $N_{\rm side}=8$ ``superpixel'', an $x$--$y$ scatter plot was formed using the unmasked $N_{\rm side}=32$ Stokes $Q$ and $U$ subpixel data from rWMAP \textit{K}-band ($\bm d_x$) and CLASS ${40\,\mathrm{GHz}}$ ($\bm d_y$). The data were fit using a line with slope $k$ and zero intercept. 
The estimation of the slope in each $N_{\rm side}=8$ superpixel was obtained by total least-squares fitting as
\begin{equation}
    \hat k=\underset{k}{\mathrm{argmin}}\left[\bm d(k)^T\mathsf\Sigma^{-1}\bm d(k)\right],\label{eq:beta_tls}
\end{equation}
where $\bm d(k)\equiv\bm d_y-k\bm d_x$, and $\mathsf\Sigma\equiv k^2\mathsf N_x+\mathsf N_y$ is the covariance matrix.
The $\mathsf N_x$ and $\mathsf N_y$ are the $Q$/$U$ covariance matrices obtained from rWMAP \textit{K}-band noise and CMB~+~CLASS $\mathrm{40~GHz}$ noise simulations, as described in Section \ref{ssec:calib}.
The uncertainty on the slope $\Delta k$ was obtained by taking half of the difference between the 16th and 84th percentiles of the likelihood function $\mathcal L\propto\exp\left[-\frac12\bm d(k)^T\mathsf\Sigma^{-1}\bm d(k)\right]$.

We converted the fitted slope $\hat k$ to $\hat\beta_s$ as
\begin{equation}
    \hat\beta_s=\frac{\log{\hat k}}{\log(\nu_\mathrm{40}^s/\nu_K^s)},\label{eq:betas}
\end{equation}
where $\nu_\mathrm{40}^s$ and $\nu_K^s$ can be found in Table \ref{tab:cmb2ant}.
As $\bm d(k)$ depends on the color correction, and thus $\beta_s$, the values were computed iteratively, with initial value \mbox{$\beta_s=-3.1$}. Convergence to within 0.01\% was achieved in five iterations.

\begin{figure}
    \includegraphics[width=\linewidth]{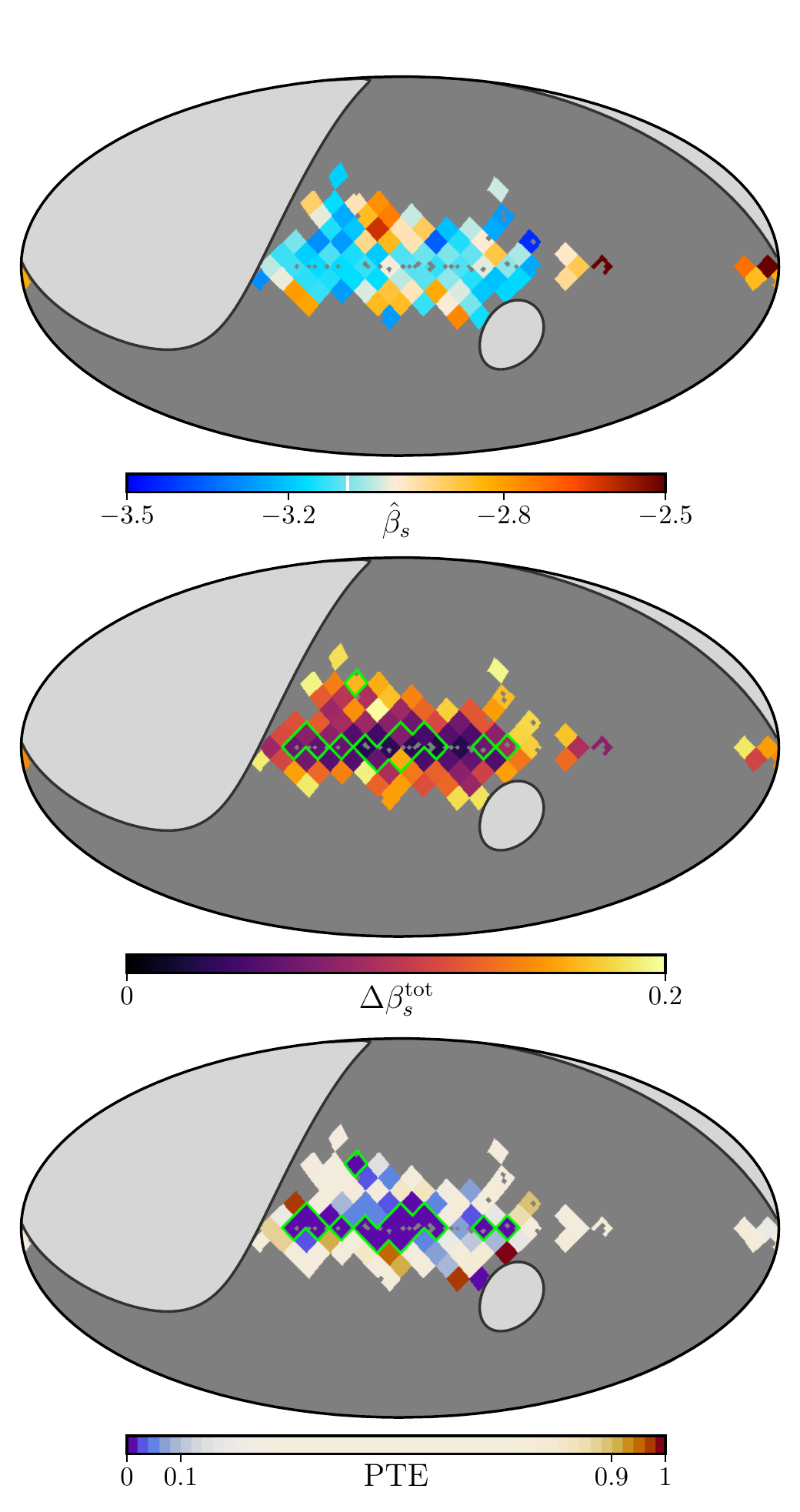}
    \caption{
    \textit{Top image}: the $\hat\beta_s$ map at $N_{\rm side}=8$ resolution using rWMAP \textit{K}-band and CLASS $40~\mathrm{GHz}$ data. 
    The vertical white line in the color bar marks the median $\hat\beta_s=-3.09$.
    \textit{Middle image}: the total uncertainty on the spectral index $\Delta\beta_s^\mathrm{tot}$.
    \textit{Bottom image}: the PTE values.
    Regions with $\Delta\beta_s^\mathrm{tot}>0.2$ and those masked by the decl.~ limit (light gray) and point-source mask are excluded.
    In the middle and bottom panels, pixels with $\mathrm{PTE}<0.01$ are highlighted by green borders.
    }
    \label{fig:beta_synch_maps}
\end{figure}

Figure \ref{fig:beta_synch_maps} shows the $\beta_s$ map, its total uncertainty $\Delta\beta_s^\mathrm{tot}$ map, and a map of the PTE associated with the linear fit.
The distribution of spectral indices has median $\hat \beta_s = -3.09^{+0.23}_{-0.11}$ (16th/84th percentiles)---consistent with previous studies that computed $\beta_s$ along the plane \citep{wmap3kogut07,quiet15, weiland22}.

The total uncertainty is defined as 
\begin{equation}
\Delta\beta_s^\mathrm{tot}\equiv\sqrt{\left(\Delta\beta_s^\mathrm{stat}\right)^2+\left(\Delta\beta_s^\mathrm{sys}\right)^2},\label{eq:Db_tot}
\end{equation}
where $\Delta\beta_s^\mathrm{stat}$ ($\Delta\beta_s^\mathrm{sys}$) is the statistical (systematic) uncertainty.
Regions with $\Delta\beta_s^\mathrm{tot}>0.2$ and those masked by the decl.~ limit (light gray) and the bright-source mask are excluded.
In the $\Delta\beta_s^\mathrm{tot}$ and $\mathrm{PTE}$ maps, pixels with $\mathrm{PTE}<0.01$ are highlighted by green borders.
As shown in the bottom panels of Figure \ref{fig:beta_synch_maps}, regions with high synchrotron S/N tend to have these low PTE values, which is at least in part because the intrinsic scatter of the synchrotron signal is not considered in the covariance matrix used in Equation \ref{eq:beta_tls}.
In these pixels, it could be possible to resolve finer-scale spectral index variations. For this investigation, however, we capture this scatter in the statistical uncertainty using bootstrapping.
We determined the bootstrapping uncertainty, $\Delta\beta_s^\mathrm{BS}$, by fitting $\hat k$ for 5000 different data resamplings using only the diagonal components of the covariance matrix (resampled in the same way).
The $\Delta\beta_s^\mathrm{BS}$ was estimated as the standard deviation of the spectral indexes inferred from different samples.

\begin{figure}
    \includegraphics[width=\linewidth]{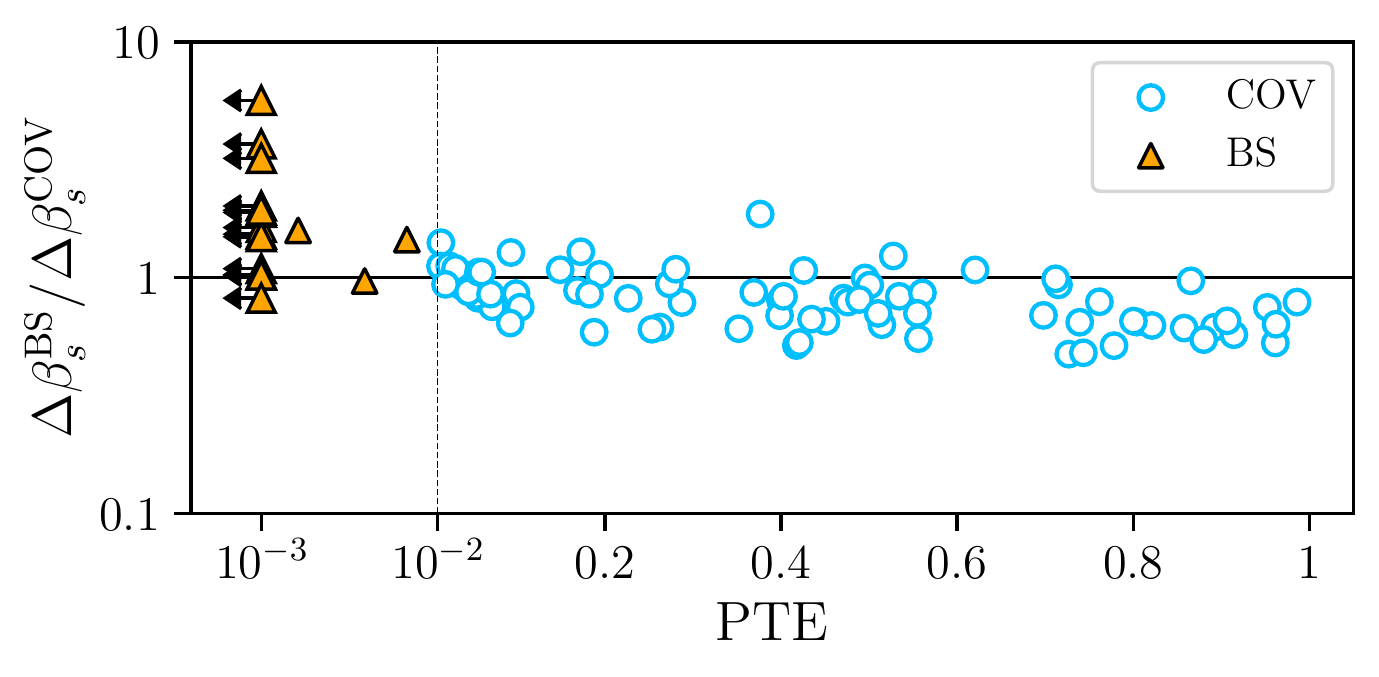}
    \caption{The ratio of the statistical uncertainty on $\beta_s$ using the bootstrapping method ($\Delta\beta_s^\mathrm{BS}$) to that obtained by propagating the $\Delta\hat k$ estimated with the full-covariance matrices ($\Delta\beta_s^\mathrm{COV}$), as a function of the $\mathrm{PTE}$ values.
    The $x$-axis goes to log-scale at $\mathrm{PTE}<0.01$.
    The vertical dashed line ($\mathrm{PTE}=0.01$) and the horizontal line ($\Delta\beta_s^\mathrm{BS}/\Delta\beta_s^\mathrm{COV}=1$) are included for reference.
    All data points with $\mathrm{PTE}<10^{-3}$ are shifted to $\mathrm{PTE}=10^{-3}$ and marked with horizontal left arrows to show that they have smaller $\mathrm{PTE}$ values.
    For pixels with $\mathrm{PTE}<0.01$ (orange triangles), we adopt the $\Delta\beta_s^\mathrm{BS}$ as the statistical uncertainty.
    For the remaining pixels (blue circles), the $\Delta\beta_s^\mathrm{COV}$ is chosen as the statistical uncertainty.
    }
    \label{fig:Db_ratio}
\end{figure}

For all other pixels, the statistical uncertainty was obtained by propagating the $\Delta k$ estimated with the full-covariance matrices as $\Delta\beta_s^\mathrm{COV}=\Delta k/[\hat k\log(\nu_\mathrm{40}^s/\nu_K^s)]$. A comparison of the two statistical uncertainty estimates for all mapped pixels is shown in Figure \ref{fig:Db_ratio}.
The noise correlation between subpixels is most significant in the relatively low-S/N region.
Neglecting the off-diagonal components in the covariance matrix tends to give smaller estimations of the uncertainty, which is reflected by the ratio in Figure \ref{fig:Db_ratio} with larger PTE values.

We estimated the systematic uncertainty, $\Delta\beta_s^\mathrm{sys}$, by considering the impact of variations in the synchrotron morphological and spectral properties. 
As a basis for comparison, we adopted the \texttt{PySM} \citep{pysm} synchrotron amplitude model at WMAP \textit{K}-band.  First, to test for a dependence on the morphology of the synchrotron signal, we generated an ensemble of perturbations of this template by adding to it 100 \textit{K}-band noise simulations. Each instance was then scaled to CLASS ${40\,\mathrm{GHz}}$ according to the \texttt{PySM} \texttt{s1} spectral index variation model (unrelated to \texttt{s1} masks in Section \ref{ssec:masks})\footnote{In this Section, we exclusively use the \texttt{PySM} \texttt{s1} synchrotron spectral index variation model at $N_{\rm side}=8$---upgraded to higher $N_{\rm side}$ when necessary.}. The ensembles of \textit{K}-band and scaled maps were then re-observed with the CLASS pipeline and combined to estimate $\beta_s$ at $N_{\rm side}=8$ following the procedure described above. 
The differences between the recovered $\beta_s$ map and the \texttt{PySM} model were negligible, and therefore the uncertainty on the synchrotron morphology does not contribute to the systematic uncertainty. 
Second, to check for a dependence of the assumed spectral index variation model, we fixed the \textit{K}-band template and perturbed the \texttt{s1} model. 
Gaussian random values with standard deviation equal to 0.1 were added to each pixel in the \texttt{s1} template. 
The single \textit{K}-band model was then scaled to CLASS ${40\,\mathrm{GHz}}$ using 100 different perturbed \texttt{s1} models. As before, the \textit{K}-band map and scaled maps were re-observed with the CLASS pipeline and combined to estimate $\beta_s$.
In this case, the variations in the recovered spectral index had a median deviation $0.033^{+0.035}_{-0.015}$ (16th/84th percentiles), which we identify as the systematic uncertainty on the $\beta_s$ estimation, $\Delta\beta_s^\mathrm{sys}$. 

As a check on a possible mean offset between the recovered $\beta_s$ and the input model due to the mapping transfer function, we scaled the \textit{K}-band template to the CLASS ${40\,\mathrm{GHz}}$ band using the estimated values shown in Figure \ref{fig:beta_synch_maps} with missing values filled by $-3.1$. After both maps were re-observed, we fit for $\beta_s$ using the same covariance matrices as used for the main result.
The difference between the fitted $\beta_s$ and the input model is consistent with the $\Delta\beta_s^\mathrm{sys}$ estimate. Therefore, the bias in $\beta_s$ introduced by the mapping transfer function is captured within this systematic uncertainty estimate.

The impact from a multiplicative bias can be bounded by the CLASS calibration uncertainty. A shift in calibration by $\pm 5\%$ is equivalent to shifting the $\beta_s$ measurement in all $N_{\rm side}=8$ pixels of Figure \ref{fig:beta_synch_maps} by a common $\pm 0.1$.
The variation in the spectral index about its median value is essentially independent of a 5\% change in the normalization of the ${40\,\mathrm{GHz}}$ data. The impact of this uncertainty is not included in the $\Delta\beta_s^\mathrm{tot}$ estimate, but it should be kept in mind when interpreting the map of $\beta_s$.

To verify $\hat k$ in Equation~\ref{eq:beta_tls} is unbiased and $\Delta k$ is near optimal, we performed an additional series of simulations. 
We again used the \texttt{PySM} model at WMAP \textit{K}-band and a scaled map at CLASS $40\,\mathrm{GHz}$ using the \texttt{s1} index model. An ensemble of \textit{K}-band maps was made by adding 100 \textit{K}-band noise simulations after which each instance was re-observed with the CLASS pipeline. Similarly, 100 CLASS noise simulations and CMB simulations are added to the synchrotron model at the CLASS $40\,\mathrm{GHz}$ band and re-observed with the CLASS pipeline. 
The pairs of simulated maps were then used to generate $\beta_s$ maps following the procedure used for the main result.
The per-pixel difference between the ensemble mean of $\beta_s$ and the \texttt{s1} model, normalized by the ensemble standard deviation on the mean, follows a standard normal distribution with values mostly within $\pm3$ across the map, indicating $\hat k$ is unbiased.
The fractional difference between the ensemble average of $\Delta k$ to the ensemble standard deviation of the $\hat k$ is largely within $\pm20\%$, indicating $\Delta k$ is near optimal. We note, using an analogous set of simulations with only CMB or only noise being added to the CLASS $40\,\mathrm{GHz}$ band synchrotron model, that we found the CMB contributed $10\%$--$25\%$ of the error relative to the CLASS noise alone.

\begin{figure}
    \centering
    \includegraphics[width=\linewidth]{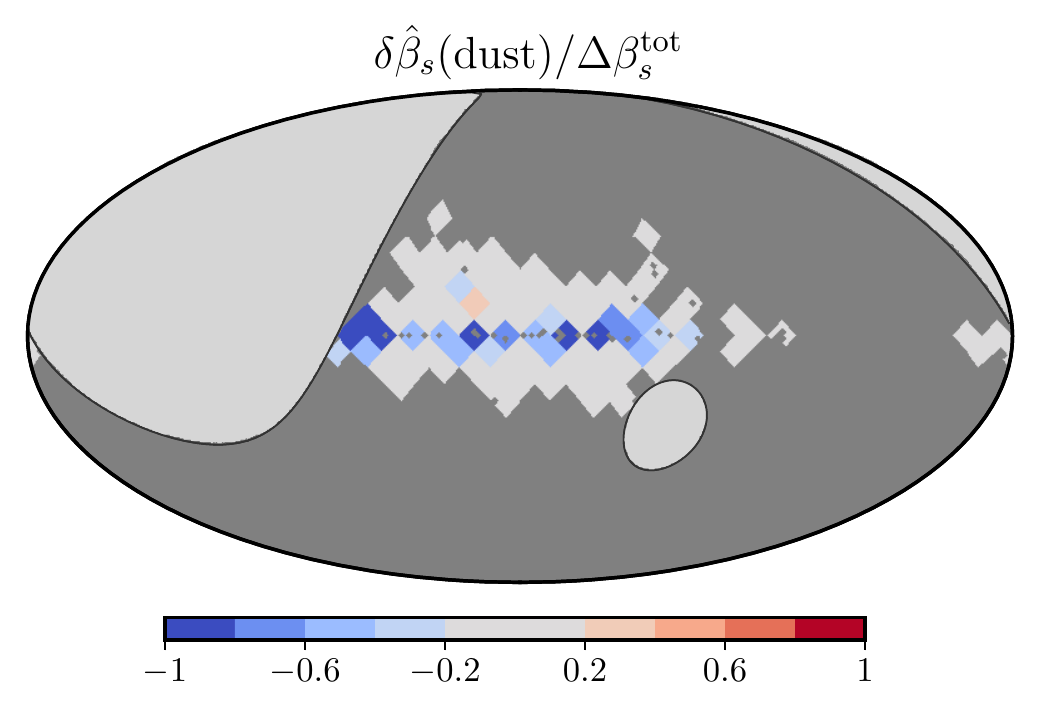}
    \caption{The difference of the $\hat\beta_s$ fitted with different dust frequency scaling models ($\delta\hat\beta_s(\mathrm{dust})$) scaled by $\Delta\beta_s^\mathrm{tot}$. 
    The $\hat\beta_s$ fitted using the $\beta_d$ and $T_d$ templates with spatial variation is subtracted from that fitted with uniform $\beta_d=1.53$ and $T_d=19.6~\mathrm{K}$.
    Regions with $\Delta\beta_s^\mathrm{tot}>0.2$ and those masked by the declination limit (light gray) and point-source mask are excluded.
    }
    \label{fig:beta_s_dustmodel}
\end{figure}

Next, we tested how these results depend upon the assumed dust model. Rather than using a constant spectral index for the polarized dust emission, we used the $\beta_d$ and $T_d$ templates with spatial variation derived with the \texttt{Commander} component separation algorithm using the \Planck\ 2015 data \citep{planck15X}. 
The difference of the $\hat\beta_s$ fitted with different dust frequency scaling models ($\delta\hat\beta_s(\mathrm{dust})$) scaled by the $\Delta\beta_s^\mathrm{tot}$ is shown in Figure \ref{fig:beta_s_dustmodel}.
The $\delta\hat\beta_s(\mathrm{dust})$ has a median  of $-0.005^{+0.01}_{-0.02}$ (16th/84th percentiles), and we found that $|\delta\hat\beta_s(\mathrm{dust})|<0.2\times\Delta\beta_s^\mathrm{tot}$ is achieved in most directions, meaning that the uncertainty in the dust frequency scaling model does not contribute much to the uncertainty on $\hat\beta_s$ in most regions.
The $\delta\hat\beta_s(\mathrm{dust})$ is significant in regions closer to the Galactic plane, as expected from the complexity of astrophysical processes contributing to the dust signal within that region.
Better characterization of the thermal dust frequency scaling is required for more precise $\beta_s$ estimation in these regions.

\section{Cosmology}
\label{sec:cosmo}
The CLASS ${40\,\mathrm{GHz}}$ channel was designed specifically to target Galactic synchrotron emission with sufficient sensitivity to enable high-precision cleaning of regions on the sky with relatively low Galactic foreground contamination. Using external data sets, however, these foreground-dominated maps have shown sufficient sensitivity to probe CMB power. 

\subsection{CMB $EE$ Spectrum}
On large scales ($\ell < 30$), the $EE$ angular power spectrum contains key information for understanding the reionization history of the Universe \citep{zaldarriaga97}. The difficulties in measuring this signal are underscored by the fact that the parameter directly tied to the $EE$ amplitude on these scales, the optical depth to reionization, $\tau$, remains the least well-measured \lcdm parameter \citep{pagano19}. 
To demonstrate the progress made with these new maps collected by a ground-based platform, we compute cross-spectra between CLASS and external data to verify the large angular scale CMB power has been preserved in the CLASS ${40\,\mathrm{GHz}}$ maps. 
We use a pseudo-$C_\ell$ estimator \citep{polspice} for this purpose, while recognizing it is nonoptimal for the largest-scale bins.
A more optimal estimator, such as xQML \citep{xQML}, will be used for future work---especially as the remaining higher-sensitivity CLASS channels are incorporated in our analysis framework. 

\begin{figure}
    \includegraphics[width=\linewidth]{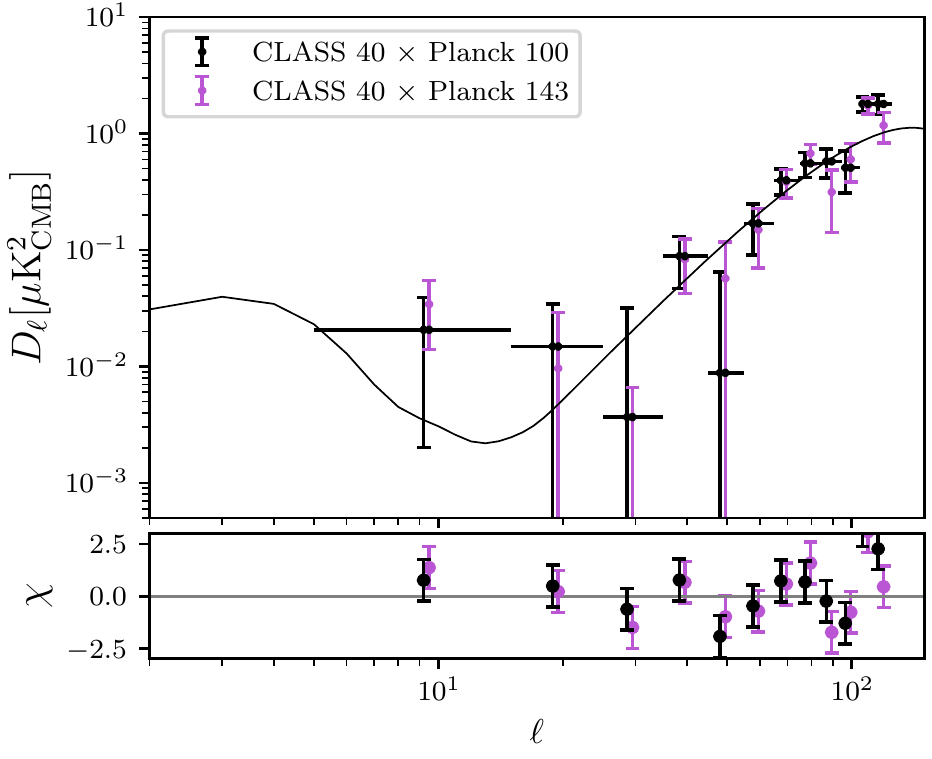}
    \caption{\emph{Top:} CMB $EE$ spectrum computed over the range  $ 5 < \ell < 125$ from foreground-reduced CLASS 40 $\mathrm{GHz}$ $\times$ \Planck\ PR3 100 $\mathrm{GHz}$ or \Planck\ PR3 143 $\mathrm{GHz}$.  The spectrum is plotted in $D_\ell = \ell (\ell +1) C_\ell /2 \pi$. Uncertainties are estimated using simulations and are dominated by CLASS noise. As the synchrotron foreground channel, it is expected that the CLASS 40 $\mathrm{GHz}$ noise would dominate here---it is the CLASS 90 $\mathrm{GHz}$ channel that is optimized for sensitivity to the CMB. Horizontal bars indicate the bins. As described in the text, template subtraction has been used to remove an estimate of the foreground emission within the \texttt{s9} mask.  The solid line is the \Planck\ best-fit \lcdm spectrum \citep{planck18VI}. \emph{Bottom:} The uncertainty-normalized difference between each spectra and the binned \lcdm expectation is shown. 
    \label{fig:cmb_ee}}
\end{figure}

The measurement of the CMB $EE$ spectrum using CLASS ${40\,\mathrm{GHz}}$ $\times$ \Planck\ ${100\,\mathrm{GHz}}$ and ${40\,\mathrm{GHz}}$ $\times$ \Planck\ ${143\,\mathrm{GHz}}$ is shown in Figure~\ref{fig:cmb_ee}---the \Planck\ PR3 maps \citep{planck18III} were used to avoid the extra computational complication needed to estimate and correct for the mapping transfer functions associated with the later PR4 release \citep{planckLVII}. Evaluation of the PR4 mapping transfer function within our sky region could be used in future work. The method to estimate these spectra is described below.

To target the CMB, synchrotron and dust templates are subtracted from each channel. The rWMAP \textit{K}-band (WMAP \textit{Ka}-band) was used as the synchrotron template for CLASS (\Planck\ channels). The \Planck\ 353 $\mathrm{GHz}$ channel was used as a dust template in the \Planck\ 100 and $143\,\mathrm{GHz}$ maps---within the mask region considered and the level of sensitivity achieved, dust is expected to be a negligible component of the CLASS $40\,\mathrm{GHz}$ maps. The foreground-cleaned maps are then defined as
\begin{equation}
\label{eq:fg_cleaning}
    \hat{m}_{x}=\frac{{m}_{x}-c_1 {m}_\mathrm{sync}-c_2 {m}_{353}}{1-c_1-c_2},
\end{equation}
where $m_{x}$ is the $Q$ or $U$ map of the channel to be cleaned, $m_\mathrm{sync}$ is rWMAP $K$-band (WMAP $Ka$-band) for cleaning 
synchrotron from CLASS (\Planck), and $m_{353}$ is the \Planck\ 353 $\mathrm{GHz}$ channel used for dust removal. 
The final template coefficients are collected in Table \ref{tab:template_coeffs}. The dust coefficients are directly adopted from \cite{pagano19}. For the synchrotron coefficients, we assume the very simple model of $\beta_s=-3.1$, and scale the rWMAP $K$-band (WMAP $Ka$-band) to the CLASS (\Planck) channel. 
A more careful analysis minimizing residual foregrounds and considering more complicated foreground models will be needed when the remaining higher-sensitivity CLASS channels are included in future analysis. For our purpose of demonstrating recovery of CMB power without attempting to place constraints on particular cosmological parameters, the current simple approach is sufficient. 

\begin{deluxetable}{lcc}
\tablecaption{Polarized foreground removal template coefficients. \label{tab:template_coeffs}}
\tablehead{
\colhead{Channel} & \colhead{\hspace{0.75cm}$c_1$ (Synchrotron)} & \colhead{\hspace{0.75cm}$c_2$ (Dust)}}  
\startdata
CLASS 40 & 0.207 & 0 \\
\Planck\ 100 & 0.039 & 0.0186 \\
\Planck\ 143 & 0.0165 & 0.0394 \\
\enddata
\tablecomments{Template scaling coefficients apply to maps in thermodynamic temperature.  The synchrotron template is WMAP \textit{K}-band for CLASS 40 $\mathrm{GHz}$ and WMAP \textit{Ka}-band for \Planck\ 100 $\mathrm{GHz}$ or 143 $\mathrm{GHz}$.  The dust template is the \Planck\ PR3 353 $\mathrm{GHz}$ map.
}
\end{deluxetable}

The maps are weighted using an estimate for the inverse-variance maps with the foreground-reduction procedure applied. These weight maps are estimated using simulations. The CLASS noise was modeled by coadding random pairs of the 200 split noise simulations---repeated use of a single split was allowed, but no pairing was duplicated. The \texttt{FFP10} noise simulations were used for the \Planck\ channels \citep{planck15XII}; each simulation was antialias filtered, rotated to celestial coordinates, and downsampled to $N_{\mathrm{side}}=128$. The WMAP noise simulations were performed following the same procedure described in Section \ref{sec:spec_est}. Finally, since the re-observed $K$-band was used, the $K$-band noise simulations were re-observed as well. Since $Ka$-band was used as the synchrotron cleaning template for both \Planck\ channels, all data and simulations used for the \Planck\ channels were smoothed to match the $Ka$-band resolution. The weight map for the cleaned CLASS map was estimated by applying the cleaning procedure in Equation \ref{eq:fg_cleaning} to the combined CLASS noise and r$K$-band noise simulations. The pixel covariance was then directly estimated from an ensemble of 5000 noise simulations (allowing repeat instances, but no repeat of any particular combination), from which the weight map was computed following Equation \ref{eq:pol_weight}. An analogous procedure was used to estimate the weight in the \Planck\ channels. Finally, the CLASS weights were multiplied by the apodized \texttt{s9} mask, and the \Planck\ channels were multiplied by the respective \texttt{Plik} frequency masks combined with the total \Planck\ point-source mask \citep{planck18V}.  

Uncertainties were estimated using the same suite of simulations described above, except now including a re-observed CMB realization as part of the CLASS channel and the same CMB instance (but not re-observed) for each simulation in the \Planck\ channel. Cross-spectra were estimated assuming the map weights described above.
The square root of the diagonal of the resulting estimated covariance matrix is shown as the uncertainties in the 
$40 \times 100$ and $40 \times 143$ cross-spectra in Figure ~\ref{fig:cmb_ee}.
Both cross-spectra are found to be consistent with each other and with the best-fit \lcdm model \citep{planck18VI}. Efforts for the inclusion of the more sensitive higher-frequency CLASS channels are well underway.

\subsection{Circular Polarization Limits}

\begin{figure}
    \includegraphics[width=\linewidth]{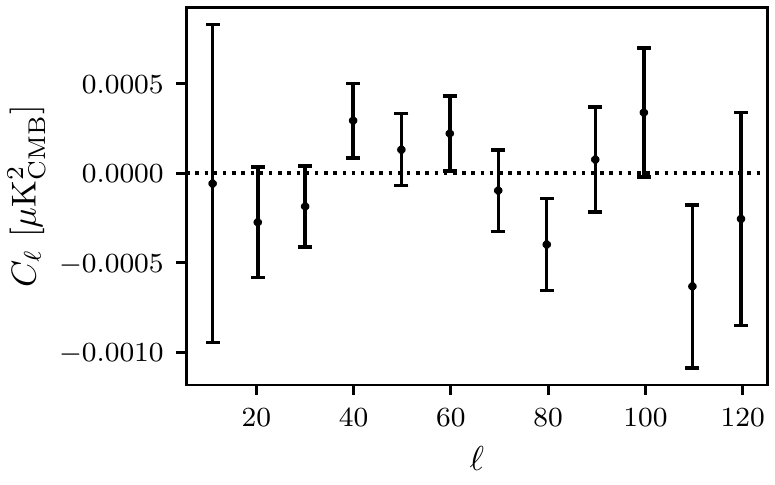}
    \caption{The CLASS ${40\,\mathrm{GHz}}$ $VV$ spectrum is shown in $\mathrm{\mu K^2}$ thermodynamic units. The spectrum is evaluated over the full survey area accounting for the mapping and beam window functions. The result is consistent with no circular polarization signal in our maps. 
    \label{fig:vv_spec}}
\end{figure}

\begin{figure}
    \includegraphics[width=\linewidth]{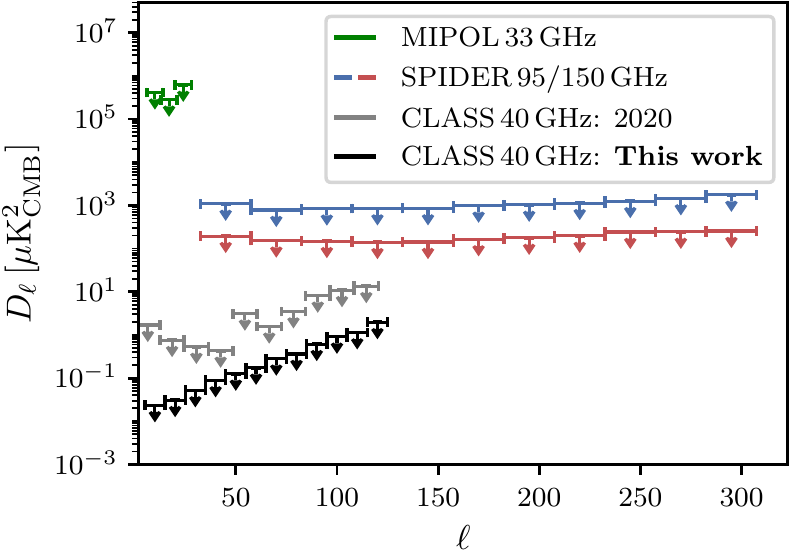}
    \caption{Upper limits ($95\%$ confidence level) on cosmological circular polarization spectrum. The spectrum is plotted in $D_\ell = \ell (\ell +1) C_\ell /2 \pi$. Former measurements from MIPOL \citep{mainini13} at $33\,\mathrm{GHz}$ and \textsc{Spider} \citep{nagy17} at $95/150\,\mathrm{GHz}$ are shown in comparison with the CLASS measurement at $40\,\mathrm{GHz}$. 
    The constraints from \cite{padilla20} used only the Era~1 nighttime data, and the maps were inverse-variance weighted. 
    \label{fig:vv_lim}}
\end{figure}

A possible circular polarization background has been predicted by multiple theories, e.g., Lorentz violating physics \citep{PhysRevD.79.063524, 2023JCAP...03..018C} and Faraday conversion 
 \citep{PhysRevD.92.123506}, among others. The unique capacity for CLASS to measure circular polarization allows us to search for such a signal. 
In the current work, we improve on the results already demonstrated \citep{padilla20}. The new mapping procedure, \citetalias{Li23}, and null test framework, Section \ref{ssec:nulls}, have been applied to the extended data described here. 
The $V$ maps shown in Section \ref{sec:maps} appear to be free from any significant signal above the noise. 
Similar to the transfer function correction described in Section \ref{sec:spec_est}, the $V$ spectra are corrected to create a nonbiased estimate of the signal power. 
Note that the filtering applied to the $V$ channel removed less power than the filtering applied to the linear polarization maps, and therefore the $VV$ spectra require less of a correction. The $VV$ cross power spectra, evaluated on the base data split and using the respective $w_{VV}$ pixel weights over the full survey area is shown in Figure \ref{fig:vv_spec}. 
The band power signal covariance is estimated using 10,000 noise-only CLASS simulations---where the mapping transfer function is corrected in each simulation as well.
The binned $VV$ spectrum has $\chi^2 = 10.6$ with 12 degrees of freedom and is therefore consistent with noise. 
The resulting $95\%$ quantiles of the band power distribution were then interpreted as upper limits in Figure \ref{fig:vv_lim}, e.g., the first bin places the upper limit $D_\ell < 0.023$ $\mathrm{\mu K^2_{CMB}}$. 
After accounting for binning differences, the new spectrum reduces the upper limit on large-scale circular polarization by over 2 orders of magnitude over results from others \citep{nagy17, mainini13}, and approximately an order of magnitude over our previous CLASS result \citep{padilla20}. The improvement in this work comes from the increase in the data volume and the optimal mapmaking that suppresses the correlated noise at large angular scales \citepalias{Li23}.

\section{Conclusions}
\label{sec:conc}

In this paper, we have presented the initial results collected from the 40 $\mathrm{GHz}$ channel of CLASS---the first of four frequency channels. 
Data from the survey operating from 2016 August 31 to 2022 May 19 have been reduced to maps and have been demonstrated to have improved noise performance over previous measurements at nearby frequencies in the critical $10 < \ell < 100$ range. The low-$\ell$ performance is set by the current filtering strategy used in the mapmaking process to reduce large angular scale systematic errors, and the resultant maps have been shown to pass a broad array of null tests. The maps and associated data products have been made available to NASA LAMBDA for public release.

The systematic effects motivating the filtering described in \citetalias{Li23} are responsible for the decreased sensitivity as one moves to the lowest $\ell$ values. As explained in \citetalias{Li23}, the cause of some of these systematic effects has already been fixed. The remaining systematic effects are the object of ongoing instrument improvements from which we anticipate further improvements on the largest scales. 

The increase in sensitivity over previous measurements surveying large fractions of the sky near 40 $\mathrm{GHz}$ is significant---map sensitivity is up to a factor of 2.1 (1.9) times higher than WMAP \textit{Q} (\Planck{} 44). Furthermore, we have found agreement between re-observed higher S/N measurements of synchrotron emission from lower frequencies once scaled to be comparable to the CLASS 40 $\mathrm{GHz}$ maps; this includes bright regions along the Galactic plane and lower brightness regions well removed from the plane. CLASS has replicated and improved upon measurements of the same sky signal as compared to space missions.

As a synchrotron-dominated channel, we have characterized the power-law behavior of the angular power spectrum and found general agreement with previous measurements. 
With the inclusion of external data sets, we have measured the synchrotron SED off the Galactic plane by comparing the amplitude of the signal in the harmonic domain. In the most restrictive synchrotron mask, \texttt{s9}, we find the diffuse synchrotron is characterized by a power law with $\beta_s=-2.95 \pm 0.09$ for $EE$ and $\beta_s=-3.28 \pm 0.24$ for $BB$. The ratio of the amplitude $BB/EE$ for this region is $0.33\pm0.02$. Furthermore, we have obtained a new $\beta_s$ map at $N_{\rm side}=8$ resolution, and measured the variations of the SED of the total polarized signal in an expanded region along the Galactic plane relative to previous efforts. 

The unique capability of the CLASS array to measure circular polarization continues to allow for the search for a cosmic circular polarization signal. We have presented new upper limits on the existence of such a signal, improving on the CLASS-set state of the art by an order of magnitude---$D_\ell < 0.023$ $\mathrm{\mu K^2_{CMB}}$ at 95\% confidence for the lowest $\ell$ bin. 

We have shown, through cross-spectra between foreground-reduced versions of the CLASS and \Planck\ 100 and 143 $\mathrm{GHz}$ maps, that the $EE$ power of the CMB has been measured even in the CLASS 40 $\mathrm{GHz}$ foreground channel. This result foreshadows CMB-focused analyses when the other CLASS frequency channel data are included---an effort well underway. The low-$\ell$ noise performance of the 40 $\mathrm{GHz}$ survey already matches the optimistic forecast of upcoming major observatories \citep[e.g.,][]{simons19whitepaper}.  Through our self-consistency tests, we have made significant progress in understanding the nature of the systematic errors that must be overcome to make progress at the lowest $\ell$. Therefore, the CLASS project is at the forefront of pushing ground-based polarization measurements to the lowest $\ell$'s to provide access to the reionization optical depth and eventually primordial $B$-modes through the reionization signal.

\section{Acknowledgments}

We acknowledge the National Science Foundation Division of Astronomical Sciences for their support of CLASS under grant Nos. 0959349, 1429236, 1636634, 1654494, 2034400, and 2109311. We thank Johns Hopkins University President R. Daniels and the Deans of the Kreiger School of Arts and Sciences for their steadfast support of CLASS. We further acknowledge the very generous support of Jim and Heather Murren (JHU A\&S ’88), Matthew Polk (JHU A\&S Physics BS ’71), David Nicholson, and Michael Bloomberg (JHU Engineering ’64). The CLASS project employs detector technology developed in collaboration between JHU and Goddard Space Flight Center under several NASA grants.  Detector development work at JHU was funded by NASA cooperative agreement 80NSSC19M0005. We acknowledge the use of the Legacy Archive for Microwave Background Data Analysis (LAMBDA), part of the High Energy Astrophysics Science Archive Center (HEASARC). HEASARC/LAMBDA is a service of the Astrophysics Science Division at the NASA Goddard Space Flight Center. CLASS is located in the Parque Astron\'omico Atacama in northern Chile under the auspices of the Agencia Nacional de Investigaci\'on y Desarrollo (ANID). 

We acknowledge scientific and engineering contributions from Max Abitbol, Fletcher Boone, David Carcamo, 
Manwei Chan, 
Benjamín Edwing Fernández Ríos, Joey Golec, Dominik Gothe, 
Francisco Espinoza, 
Ted Grunberg, 
Mark Halpern, 
Saianeesh Haridas, 
Kyle Helson, Gene Hilton, 
Connor Henley, 
Lindsay Lowry, 
Jeffrey~John McMahon, 
Nick Mehrle, 
Carolina~Morales Perez, 
Ivan~L. Padilla, Gonzalo Palma, Lucas Parker, 
Sasha Novack, 
Bastian Pradenas, Isu Ravi, 
Carl~D. Reintsema, 
Gary Rhoades, Daniel Swartz, Bingjie Wang, Qinan Wang, Tiffany Wei, Zi\'ang Yan, and Zhuo Zhang. For essential logistical support, we thank Jill Hanson, William Deysher, Joseph Zolenas, LaVera Jackson, Miguel Angel D\'iaz, Mar\'ia Jos\'e Amaral, and Chantal Boisvert. We acknowledge productive collaboration with Dean Carpenter and the JHU Physical Sciences Machine Shop team.

S.D. is supported by an appointment to the NASA Postdoctoral Program at the NASA Goddard Space Flight Center, administered by Oak Ridge Associated Universities under contract with NASA. K.H. was supported by NASA under award No. 80GSFC21M0002. R.R. acknowledges support from ANID BASAL project FB210003. Z.X. is supported by the Gordon and Betty Moore Foundation through grant GBMF5215 to the Massachusetts Institute of Technology. R.D. thanks ANID for grants BASAL CATA FB210003 and FONDEF ID21I10236.

This paper uses data products derived from observations obtained with \Planck\ (http://www.esa.int/Planck), an ESA science mission with instruments and contributions directly funded by ESA Member States, NASA, and Canada.

Finally we thank the anonymous reviewer, whose feedback has allowed us to improve this work.


\software{
numpy \citep{numpy20}, 
scipy \citep{scipy}, 
matplotlib \citep{matplotlib},
astropy \citep{astropy}, 
HEALPix \citep{healpix},
camb \citep{camb},
fastcc \citep{2022RNAAS...6..252P},
pysm \citep{pysm},
PolSpice \citep{polspice},
NaMaster \citep{Alonso:2018jzx}
}




\bibliographystyle{aasjournal_apj}
\bibliography{cosmology, class_pub, cmb, foreground, software, non_cmb, hardware, references}

\clearpage
\appendix

\section{Cosmology Beam and Window Function}
\label{app:beam}
In this Appendix, we construct the effective cosmology beam for the 40 $\mathrm{GHz}$ telescope as defined in \citet{xu20}, the radial beam profile, and the window function derived from it. 

\subsection{Observations}

Dedicated observations of the Moon were used to construct the effective cosmology beam. These observations were performed by scanning the telescope back and forth in azimuth $\pm 14^{\circ}$ on the sky at a constant elevation while the Moon rose or set through the array. Separate beams were constructed for Era 1 and Era 2 and then combined to form the final aggregate cosmology beam. The Era 1 (Era 2) beam utilized 309 (91) dedicated observations. The beams were constructed in four steps: each sequential step is described in the following subsections.

\subsection{Analysis: Position and Amplitude}

During the first step, for each observation, the TOD for each detector is passed through a 5 Hz low-pass filter to eliminate the signal from the VPM, downsampled by a factor of 10 from $\sim200$ to $\sim20$ Hz, and rotated from sky coordinates into the \textit{receiver coordinate system}. The receiver coordinates are a spherical coordinate system centered on zero latitude and longitude and oriented such that lines of latitude correspond to horizontal and lines of longitude correspond to vertical in the instrument frame. The data are then fit to an elliptical Gaussian to determine amplitude and position. Position is determined by the position of the Moon at the time corresponding to the fitted peak of the beam. The fitted beam parameters are recorded, and a 10$^{\circ}$ radius map centered on the beam peak is output. 
After all of the detector data are processed, a minimization is performed comparing the fitted position of each detector with its expected position with respect to array center to locate the position of the array center. 
The offsets of each detector's fitted position with respect to the array center from its expected position are then recorded. 

\subsection{Analysis: Individual Detector Averaging}

During the second step, the 10$^{\circ}$ radius maps for all observations are read for each detector. 
Maps offset by more than 0.1$^{\circ}$ from the detector's expected position with respect to array center are rejected. 
A linear baseline is removed from each pass over the map using data outside a radius of 6$^{\circ}$ from the center of the map, and an estimation of the RMS noise is calculated from the standard deviation of all data outside this radius. Simulations show that the baseline removal induces a bias on the final window function that is $\leq 0.2\%$ for $\ell<200$. 
For simplicity, we choose to not correct for this small bias in this analysis. 
The maps are then rotated by the boresight angle at the time of each observation. Since there are seven boresight angles encompassing 90$^{\circ}$ of rotation, this rotation tends to symmetrize the beam when all observations are averaged. The maps are scaled to a fiducial angular diameter of the Moon and to a constant brightness temperature using a model based on the 37 $\mathrm{GHz}$ Chang'E satellite data \citep{2012Icar..219..194Z} slightly modified to yield a flat result versus Moon phase over all of the Era 1 observations.
\begin{figure*}[ht]
    \centering
    \includegraphics[width=.8\linewidth]{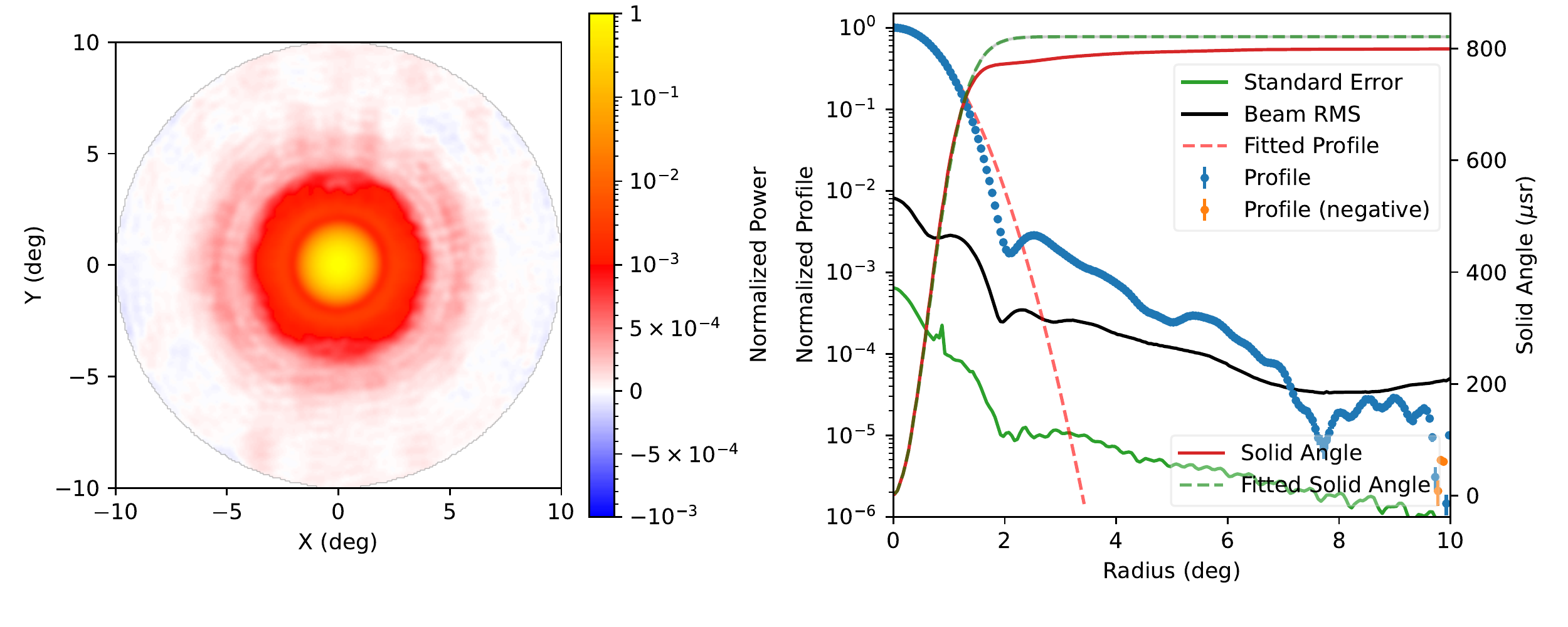}
    \includegraphics[width=.8\linewidth]{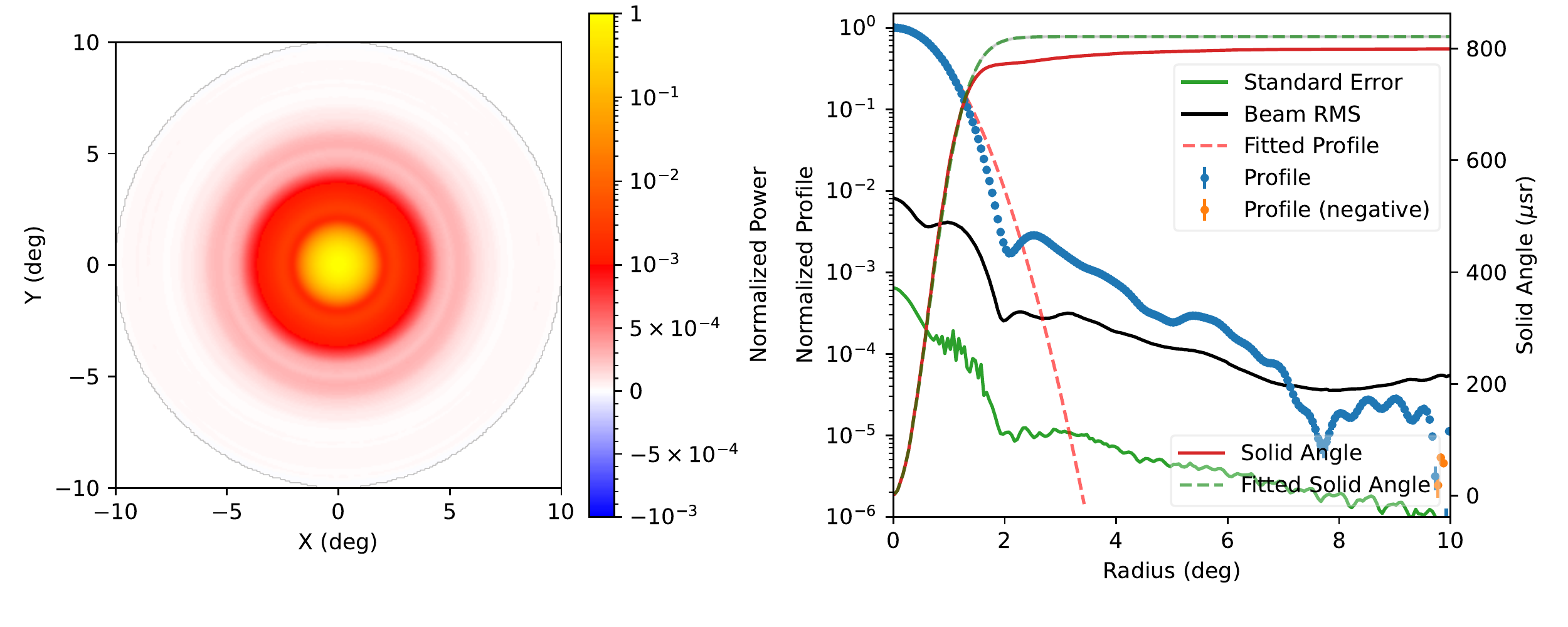}
    \caption{Top panels: combined era cosmology beam and profile \emph{without} cross-linking. Here, the profile labeled ``Fitted Profile'' is the azimuthal average of the fitted elliptical Gaussian beam. The profile labeled ``Fitted Solid Angle'' is its integral. The profile labeled ``Solid Angle'' is the integral of the beam map. The one labeled ``Beam RMS'' is the square root of the azimuthal average of the beam variance map, and the one labeled ``Standard Error'' is the standard deviation of the 100 sample radial beam profiles. Bottom panels: combined era cosmology beam and profile \emph{with} cross-linking.
    \label{fig:cos_beam}}
\end{figure*}

Once all of the observations are processed, the median unscaled RMS and median scaled peak amplitude are calculated. Any map that does not satisfy either of the following:
\begin{align}
\mathrm{RMS} &< 3 \ \mathrm{median(RMS)} \\
0.5 \ \mathrm{median(Amp)} &< \mathrm{Amp}  < 2 \ \mathrm{median(Amp)}
\end{align}
is rejected. The surviving maps are then subjected to random selection with replacement to yield 100 combinations of all of the surviving maps. 
Each combination is binned onto a regular grid at 0.05$^{\circ}$ resolution and averaged weighted by the S/N defined as the peak amplitude divided by the RMS multiplied by boresight angle dependent weights derived from the accumulated integration time for each boresight angle used in the CMB mapping. A variance map is also created using the variance of the weighted mean in each bin/pixel. The averaged amplitude map is fitted to an elliptical Gaussian out to a radius of 1$^{\circ}$ from beam center and then the averaged map along with the variance map and fitted beam parameters are saved in a file.

\subsection{Analysis: Individual Era Aggregate Beam and Window Function}

During the third step, for each of the 100 sample combinations, the files saved during the second step are read for all of the detectors. Each detector's amplitude and variance map is scaled by the inverse of the detector's relative efficiency. A variance is calculated from the scaled variance map using the mean of the data outside a radius of 6$^{\circ}$ from beam center. Then all of the detector's amplitude maps are averaged weighted by the inverse of the calculated variance multiplied by the boresight angle dependent weights summed over all seven boresight angles to yield a sample aggregate cosmology beam map. An additional weighted variance map is also created using the sum of the variance of each detector's beam with respect to the average and each detector's individual variance map.

Since the amplitude map is convolved with the Moonm, and a single cosmology beam and window function are needed independent of decl.~in the CMB maps, an additional two steps are performed. First, the amplitude map is deconvolved by dividing the two-dimensional Fourier transform of the amplitude map by the two-dimensional Fourier transform of a 0.25$^{\circ}$ radius disk limited to 25\% of the amplitude of the resulting Jinc function before its first zero. It was found that trying to extend this lower did not make much difference and started to cause the appearance of background ripple at the cutoff spatial frequency. All data outside of this radius are set to zero in the transform, and then the data are transformed back to derive an estimate of the deconvolved aggregate cosmology beam. Second, an azimuth angle is assigned to each of the 100 sample aggregate maps by incrementing the azimuth angle from 0$^{\circ}$ to 180$^{\circ}$ in 100 steps. Each aggregate map is then rotated by plus and minus the parallactic angle at the assigned azimuth and further averaged, as is the corresponding variance map, thus taking into account the east--west cross-linking at that azimuth.

After subtracting a constant baseline derived from the average of the data from a radius of 9.75$^{\circ}$--10$^{\circ}$, each sample aggregate amplitude map is fitted to an elliptical Gaussian as above. However, since the beam tends to be flatter than a Gaussian near beam center, an additional fit is done using data out to only 0.25$^{\circ}$ from beam center. This fit is a linear least-squares fit to the natural log of the amplitude versus the square of the radius weighted by the inverse of the variance map in each pixel. The exponential of the fitted intercept is then used as the peak amplitude to normalize the aggregate amplitude and variance maps. 

Each amplitude and variance map is integrated to derive the beam solid angle and its variance, then averaged azimuthally to derive the radial beam profile. The variance at each radius is calculated from the sum of the variance of the data and the azimuthal average of the variance map. This profile is then refitted for its peak amplitude using the linear least-squares method described above and then re-normalized. The FWHM is measured by the same method as for the peak amplitude using the data point nearest to the half maximum along with three data points above it and three below it. In this case, the FWHM is given by

\begin{equation}
\mathrm{FWHM} = 2 \sqrt{\frac{\log{0.5} - a} {b}},
\end{equation}
where $b$ is the slope and $a$ is the intercept of the fitted line. Its variance is derived from the covariance of the fit. The beam window function is derived from the radial profile. All of the beam parameters, the amplitude and variance maps, the radial profile, and the window function for each sample are recorded.

Finally, the 100 sample maps are averaged using equal weights to yield the final aggregate cosmology beam for each era. In this case, the average variance map is not divided by the sum of the weights yielding the sample variance rather than the variance of the mean. The 100 sample profiles are also averaged, and their sample variance is recorded, as are the 100 sample window functions.

\subsection{Analysis: Combined Era Aggregate Beam and Window Function}

\begin{figure}
    \includegraphics[width=\linewidth]{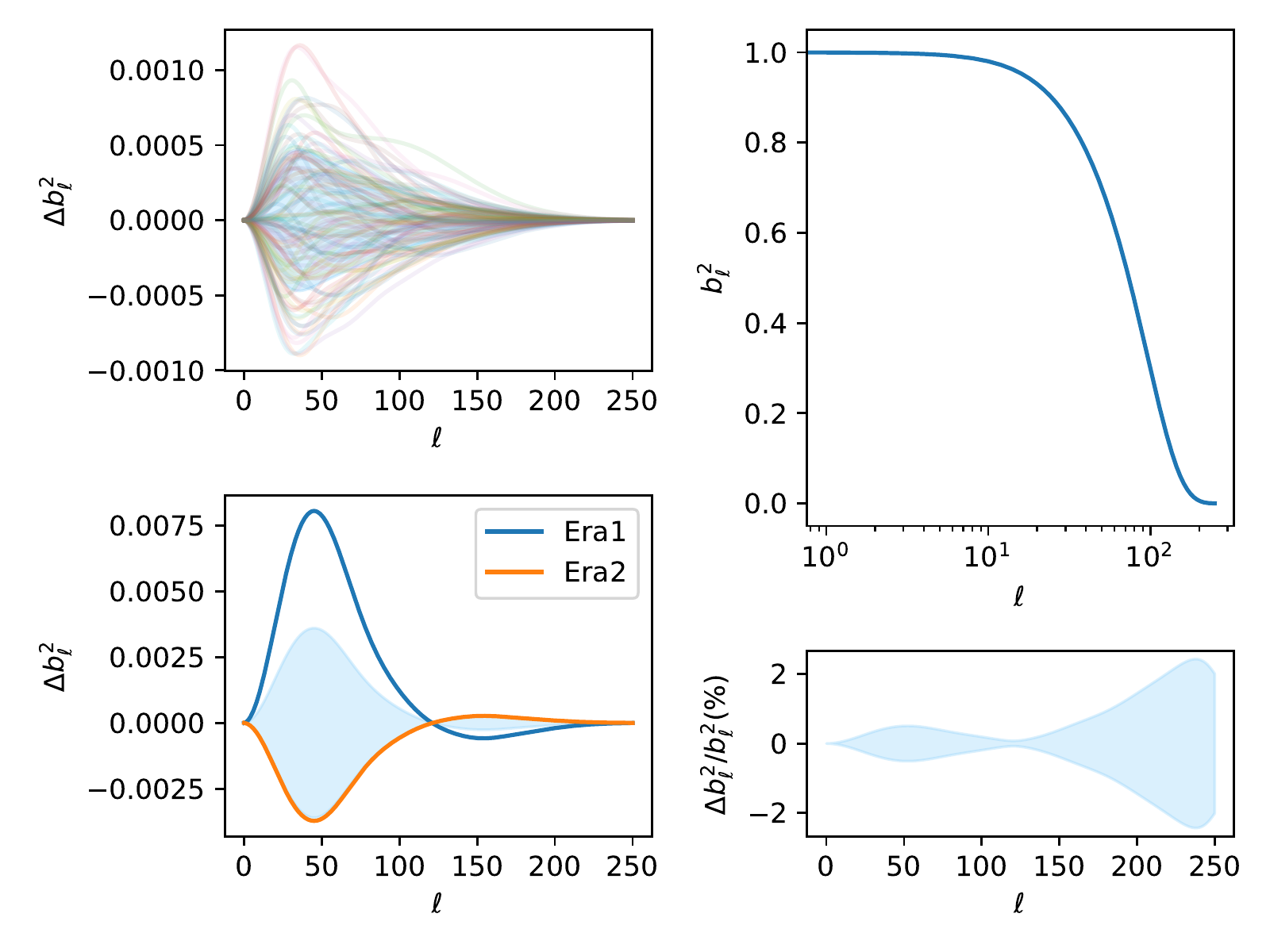}
    \caption{Combined era cosmology beam window function. Upper left panel: shading shows uncertainty due to the window function sample variance across the 100 sample radial beam profiles. Lower left panel: shading shows uncertainty due to the variance of the weighted mean of the two eras' window functions. Upper right panel: mean combined era window function. Lower right panel: combined uncertainty as a percentage of the window function. 
    \label{fig:window_func}}
\end{figure}

In the fourth step, each of the 100 sample maps for each era are scaled by the inverse of their relative efficiencies defined as the mean absolute efficiency for each era divided by their mean, then averaged weighted by the boresight angle dependent weights summed over all detectors and boresight angles for each era to yield 100 sample combined era amplitude maps, variance maps, radial beam profiles, and window functions. These were described above for the third step leaving out the deconvolution, which has already been done. 

\begin{deluxetable}{cccc}
\tablecaption{$40\,\mathrm{GHz}$ aggregate cosmology beam parameters. Values are mean (uncertainty). The solid angles are integrated measurements to a radius of 10$^{\circ}$ from beam center.}
\label{tab:beam_params}
\tablehead{
\colhead{Beam} & \colhead{FWHM} & \colhead{Solid Angle} & \colhead{Eccentricity}}
\startdata
& (deg) & ($\mu$sr) & \\ 
\hline
Era 1 & 1.559 (0.006) & 797.6 (0.3) & 0.07 (0.01) \\
Era 2 & 1.559 (0.006) & 800.2 (0.7) & 0.07 (0.01) \\
Combined & 1.559 (0.006) & 799.4 (0.5) & 0.07 (0.01) \\
\enddata
\end{deluxetable}

These 100 sample maps are then averaged using equal weights as described above, as are the 100 sample profiles and window functions, to yield the final combined era cosmology beam, radial profile, and window function. This beam and radial profile are shown in Figure \ref{fig:cos_beam} (bottom panels). For reference, the combined era beam and radial profile before cross-linking are shown in Figure \ref{fig:cos_beam} (top panels). Since the combined era window function is used in the CMB mapping for both eras and the two eras' window functions are slightly different, an additional variance term consisting of the variance of the weighted mean of the two eras' window functions is computed and added to the variance of the 100 sample window functions as shown in Figure \ref{fig:window_func}. Beam parameters for both eras and the combined beam are shown in Table \ref{tab:beam_params}.



\end{document}